\documentclass[letterpaper,12pt,hyphens,oneside]{book}
\pdfoutput=1
\usepackage{tikz}
\usepackage{array}
\usetikzlibrary{calc}
\newcommand{\tikzmark}[1]{\tikz[remember picture,overlay]\coordinate (#1);}
\newcolumntype{T}[1]{@{\hspace{\tabcolsep}}c@{\hspace{\tabcolsep}\tikzmark{#1}}}
\usepackage{fullpage}
\usepackage[T1]{fontenc}
\usepackage[utf8]{inputenc}
\usepackage{tcolorbox}
\usepackage{color}
\usepackage{graphicx}
\usepackage{framed}
\usepackage[ruled, vlined]{algorithm2e}
\usepackage{pifont}
\newcommand{\xmark}{\ding{55}}
\newcommand{\dmark}{\ding{70}}
\usepackage{standalone}
\usepackage{amsmath}
\usepackage{amsfonts}
\usepackage{amssymb}
\usepackage{paralist}
\usepackage{enumitem}
\usepackage{subcaption}
\usepackage{mdwlist}
\usepackage[mathcal]{euscript}
\usepackage{booktabs}
\usepackage{tabularx}
\usepackage[toc]{glossaries}
\usepackage{xspace}
\usepackage{balance,multirow}
\usepackage{url}
\usepackage{microtype}
\usepackage{pgf-umlsd}
\usepackage{hyperref}

\renewcommand{\paragraph}[1]{\medskip \noindent {\bf #1}.}
\long\def\symbolfootnote[#1]#2{\begingroup%
\def\thefootnote{\fnsymbol{footnote}}\footnote[#1]{#2}\endgroup}

\let\latexcite=\cite
\def\cite{\nolinebreak\latexcite}
\let\latexref=\ref
\def\ref{\nolinebreak\latexref}

\newcommand{\ignore}[1]{}

\newcommand{\mz}[1]{\textcolor{dkgreen}{MZ: #1}}
\newcommand{\tj}[1]{\textcolor{cyan}{TJ: #1}}

\newcounter{romanlistcounter}
  {\setcounter{romanlistcounter}{0}%
   \begin{list}{\textit{(\roman{romanlistcounter})}}{%
        \usecounter{romanlistcounter}%
      \setlength{\itemsep}{0pc}%
      \setlength{\itemindent}{1pc}%
      \setlength{\topsep}{0pc}%
      \setlength{\mylabelwidth}{3em}}}
  {\end{list}}

\newcounter{alphalistcounter}
  {\setcounter{alphalistcounter}{0}%
   \begin{list}{\textit{(\alph{alphalistcounter})}}{%
        \usecounter{alphalistcounter}%
      \setlength{\itemsep}{0pc}%
      \setlength{\itemindent}{1pc}%
      \setlength{\topsep}{0pc}%
      \setlength{\mylabelwidth}{3em}}}
  {\end{list}}

\def\compactify{\itemsep=0pt \topsep=0pt \partopsep=0pt \parsep=0pt}
\let\latexusecounter=\usecounter

\let\latexusecounter=\usecounter

\usepackage[
  separate-uncertainty = true,
  multi-part-units = repeat
]{siunitx}
\usepackage{listings}
\definecolor{dkgreen}{rgb}{0,0.6,0}
\definecolor{gray}{rgb}{0.5,0.5,0.5}
\definecolor{mauve}{rgb}{0.58,0,0.82}

\usepackage{float}
\usepackage{chngpage}
\usepackage{multicol}

\usepackage{adjustbox}

\newcolumntype{R}[2]{%
    >{\adjustbox{angle=#1,lap=\width-(#2)}\bgroup}%
    l%
    <{\egroup}%
}
\newcommand*\rot{\multicolumn{1}{R{65}{0.6em}}}

\makeatletter
\let\c@table\c@figure 
\let\ftype@table\ftype@figure 
\makeatother

\makeglossaries
\begin{document}


\title{Data Security on Mobile Devices: Current State of the Art, Open Problems, and Proposed Solutions} 

\author{
{\rm Maximilian Zinkus}\\
Johns Hopkins University \\
zinkus@cs.jhu.edu
\and
{\rm Tushar M. Jois}\\
Johns Hopkins University\\
jois@cs.jhu.edu
\and
{\rm Matthew Green}\\
Johns Hopkins University\\
mgreen@cs.jhu.edu
} 

\maketitle
\chapter*{Executive Summary}

In this work we present definitive evidence, analysis, and (where needed) speculation to answer the questions, $(1)$ \textit{``Which concrete security measures in mobile devices meaningfully prevent unauthorized access to user data?''} $(2)$ \textit{``In what ways are modern mobile devices accessed by unauthorized parties?''}  and finally, $(3)$ \textit{``How can we improve modern mobile devices to prevent unauthorized access?''}

We examine the two major platforms in the mobile space, iOS and Android, and for each we provide a thorough investigation of existing and historical security features, evidence-based discussion of known security bypass techniques, and concrete recommendations for remediation. In iOS we find a compelling set of security and privacy controls, empowered by strong encryption, and yet a \textit{critical lack in coverage} due to under-utilization of these tools leading to serious privacy and security concerns. In Android we find strong protections emerging in the very latest flagship devices, but simultaneously \textit{fragmented and inconsistent security and privacy controls}, not least due to disconnects between Google and Android phone manufacturers, the deeply lagging rate of Android updates reaching devices, and various software architectural considerations. We also find in both platforms exacerbating factors due to increased synchronization of data with cloud services.

The markets for exploits and forensic software tools which target these platforms are alive and well. We aggregate and analyze public records, documentation, articles, and blog postings to categorize and discuss unauthorized bypass of security features by hackers and law enforcement alike. Motivated by an accelerating number of cases since Apple v. FBI in 2016, we analyze the impact of forensic tools, and the privacy risks involved in unchecked seizure and search. Then, we provide in-depth analysis of the data potentially accessed via law enforcement methodologies from both mobile devices and associated cloud services.

Our fact-gathering and analysis allow us to make a number of recommendations for improving data security on these devices. In both iOS and Android we propose concrete improvements which mitigate or entirely address many concerns we raise, and provide analysis towards resolving the remainder. The mitigations we propose can be largely summarized as increasing coverage of sensitive data via strong encryption, but we detail various challenges and approaches towards this goal and others.

It is our hope that this work stimulates mobile device development and research towards security and privacy, provides a unique reference of information, and acts as an evidence-based argument for the importance of reliable encryption to privacy, which we believe is both a human right and integral to a functioning democracy.

\thispagestyle{empty}
\newpage

\tableofcontents
\listoffigures
\listoftables
\thispagestyle{empty}
\newpage

\setcounter{page}{1}

\newglossaryentry{AES}{
    name=AES,
    description={Advanced Encryption Standard (AES) is a cipher accepted by the National Institute of Standards and Technology, widely deployed and used to protect the confidentiality of digital data at rest and in transit}
}

\newglossaryentry{Curve25519}{
    name=Curve25519,
    description={Curve25519 is a mathematical construction used in ECDH and other cryptographic algorithms}
}

\newglossaryentry{TEE}{
    name=TEE,
    description={A Trusted Execution Environment (TEE) is a software component that runs within a dedicated isolated execution mode on a processor. It enables the storage and use of cryptographic secrets, which remain isolated from the operating system and application software.}
}

\newglossaryentry{TA}{
    name=TA,
    description={A Trusted Application is a software component that runs within a Trusted Execution Environment in order to realize a specific service.}
}

\newglossaryentry{ECDH}{
    name=ECDH,
    description={Elliptic Curve Diffie-Hellman (ECDH) is a cryptographic algorithm which enables secure two-party key agreement}
}

\newglossaryentry{NIST}{
    name=NIST,
    description={The National Institute of Standards and Technology (NIST) is a U.S. government organization responsible for standards across industries including software, hardware, and technology}
}

\newglossaryentry{VPN}{
    name={VPN},
    description={A Virtual Private Network (VPN) serves as a proxy to connect to the internet via the VPN servers. Connections to VPN servers are often encrypted and authenticated. VPNs can protect users from surveillance by their Internet Service Provider, but this protection is only as good as the VPN service}
}

\newglossaryentry{PBKDF2}{
    name={PBKDF2},
    description={Password-Based Key Derivation Function 2 (PBKDF2) \cite{nist800132} is a cryptographic algorithm which iteratively applies pseudo-random functions to an input. This can enable storage and comparison of values without storing potentially sensitive inputs in long-term storage, among other functionalities}
}

\newglossaryentry{SMS}{
    name=SMS,
    description={Short Message Service (SMS) is a text-based message service which operates over cellular phone networks}
}

\newglossaryentry{MMS}{
    name=MMS,
    description={Multimedia Messaging Service (MMS) is a media messaging protocol and service which operates over cellular phone networks to transmit photos, videos, or audio}
}

\newglossaryentry{GCM}{
    name={GCM},
    description={Galois-Counter Mode (GCM) is a cryptographic algorithm which enables authenticated encryption of a data stream using AES}
}

\newglossaryentry{CBC}{
    name={CBC},
    description={Cipher Block Chaining (CBC) is a cryptographic algorithm which enables encryption of a data stream using AES}
}

\newglossaryentry{XTS}{
    name=XTS,
    description={XOR-Encrypt-XOR-based tweaked-codebook mode with ciphertext stealing (XTS) is a cryptographic algorithm which enables encryption of a filesystem with AES, providing stronger authentication properties than CBC and some other modes, but lacking the security associated with true authenticated encryption}
}

\newglossaryentry{TouchID}{
    name=TouchID,
    description={A fingerprint biometric authentication system developed by Apple for iOS devices (and later for macOS computers).}
}

\newglossaryentry{FaceID}{
    name=FaceID,
    description={A facial-recognition biometric authentication system developed by Apple for iOS devices (mostly replacing TouchID).}
}

\newglossaryentry{TLS}{
    name=TLS,
    description={Transport Layer Security (TLS) is a cryptographic protocol for securing connections between devices on a network.}
}

\newglossaryentry{jailbreak}{
    name=jailbreak,
    description={An exploit or software package which bypasses iOS security to enable unsigned code execution and custom kernel modifications.}
}

\newglossaryentry{sandbox}{
    name=sandbox,
    description={A software-enforced set of access controls which prevent an application or service from accessing data, system resources, or hardware.}
}

\newglossaryentry{CP}{
    name=CP,
    description={CP or ``Complete Protection'' is the strongest class of Data Protection on iOS. Data which is CP is encrypted in memory, and encryption keys are evicted from memory shortly after the device is locked.}
}

\newglossaryentry{RCS}{
    name=RCS,
    description={Rich Communication Services (RCS) is a messaging protocol that is designed as a replacement for traditional SMS/MMS communications.}
}

\newglossaryentry{Siri}{
    name=Siri,
    description={Siri is the name for Apple's voice recognition system.}
}

\newglossaryentry{iMessage}{
    name=iMessage,
    description={iMessage is Apple's end-to-end encrypted messaging service. iMessage delivers text and media via signed and encrypted messages among registered Apple users.}
}

\newglossaryentry{FaceTime}{
    name=FaceTime,
    description={FaceTime is Apple's end-to-end encrypted video and audio chat service. FaceTime connects registered Apple users to enable real-time audio and video communication.}
}

\newglossaryentry{JTAG}{
    name=JTAG,
    description={Joint Test Action Group. An IEEE standard for, and implementation of, hardware debugging via on-chip pin interfaces. In practice, also a potential point of failure used to bypass low-level (bootloader, OS, etc) security systems.}
}

\newglossaryentry{SoC}{
    name=SoC,
    description={System-on-a-Chip (SoC) is a hardware configuration wherein a central processor and processor peripherals are included within a single silicon package, often appropriate for embedded or mobile devices}
}

\newglossaryentry{ARM}{
    name=ARM,
    description={Arm (previously ARM, for Advanced RISC Machine, and further prior, Acorn RISC Machine) is a family of reduced instruction set architectures primarily designed for mobile and embedded systems and SoCs}
}

\newglossaryentry{CPU}{
    name=CPU,
    description={Central Processing Unit, the main computing hardware within a computer}
}

\newglossaryentry{RNG}{
    name=RNG,
    description={A Random Number Generator (RNG) is a cryptographic algorithm for generating random or pseudo-random data for use in further cryptographic algorithms}
}

\newglossaryentry{DFU}{
    name=DFU,
    description={Device Firmware Upgrade (DFU) mode is a special device state in which firmware such as the bootloader may be upgraded or modified}
}

\newglossaryentry{USB}{
    name=USB,
    description={Universal Serial Bus (USB) is a hardware implementation and standard designed to enable universal device-to-device or client-and-host communication via a cable}
}

\newglossaryentry{Lightning}{
    name=Lightning,
    description={Lightning is a hardware implementation and proprietary standard developed by Apple for iOS devices; Lightning is used for charging and data transfer}
}

\newglossaryentry{API}
{
    name=API,
    description={Application Programming Interfaces (API) represent a set of subroutines that provide system services to application developers}
}

\newglossaryentry{2FA}
{
    name=2FA,
    description={Two-Factor Authentication (2FA) is a mechanism used to augment traditional password-based authentication to remote services, by adding an additional ``factor'' of authentication related to {\em e.g.,} biometrics or possession of a device}
}

\newglossaryentry{HSM}
{
    name=HSM,
    description={A Hardware Security Module (HSM) is a specialized computing device designed to store and operate on cryptographic secrets. These devices often include software and physical protection mechanisms intended to provide strong security for secret keys, while making these keys usable to authorized systems. Some HSMs can be provisioned with custom software}
}

\newglossaryentry{SEP}
{
    name=SEP,
    description={The Apple Secure Enclave Processor (SEP) is a specialized co-processor included in Apple iPhones since the iPhone 5S. This processor is designed to store and employ cryptographic secrets, and to handle authentication procedures using FaceID and TouchID. The SEP is designed to withstand attacks that result in a total compromise of the device operating system}
}

\newglossaryentry{AVB}
{
    name=AVB,
    description={Android Verifiable Boot is a service for verifying the integrity of the Android boot chain using a TEE.}
}

\newglossaryentry{DHS}
{
    name=DHS,
    description={The U.S. Department of Homeland Security (DHS) is one agency responsible for evaluating forensic tooling for the U.S. Federal government}
}

\newglossaryentry{operating system}
{
    name={operating system},
    description={The core software that controls access to internal systems and peripherals on a smartphone. The OS is responsible for enforcing security and access controls on software applications.}
}

\newglossaryentry{Jailbreak}
{
    name=Jailbreak,
    description={A procedure that disables security restrictions in a device operating system in order to allow the execution of third-party software}
}

\newglossaryentry{FBI}{
    name=FBI,
    description={The U.S. Federal Bureau of Investigation (FBI) is a governmental agency which serves as the principle law enforcement office, as well as a domestic security and intelligence organization}
}

\newglossaryentry{kernel}
{
    name=kernel,
    description={The core software component of a standard computer operating system, responsible for maintaining privilege separation between applications and processes on a device.}
}

\newglossaryentry{AFU}
{
    name=AFU,
    description={After First Unlock (AFU) is an Apple-specific term given to the state of a device that has been powered on, and into which the user has entered their passcode at least once. Devices remain in this state until they are manually rebooted, or some defined time period has elapsed}
}

\newglossaryentry{BFU}
{
    name=BFU,
    description={Before First Unlock (BFU) is an Apple-specific term given to the state of a device that has been powered on, but where the user has not yet entered the passcode}
}

\newglossaryentry{AOSP}
{
    name=AOSP,
    description={Android Open Source Project, an open source software project defining the non-proprietary elements of the Android operating system. AOSP is produced by the Open Handset Alliance, with sponsorship from Google.}
}

\newglossaryentry{UID key}
{
    name={UID key},
    description={The UID key is a random AES key that is stored within the device silicon at manufacture time via fuses; this key is used as an ingredient in deriving cryptographic keys for Apple's file encryption}
} 

\chapter{Introduction}\label{sec:intro}

Mobile devices have become a ubiquitous component of modern life. More than 45\% of the global population uses a smartphone~\cite{globaladoption}, while this number exceeds 80\% in the United States~\cite{pewresearch}. This widespread adoption is a double-edged sword: the smartphone vastly improves the amount of information that individuals can carry with them; at the same time, it has created new potential targets for third parties to obtain sensitive data. The portability and ease of access makes smartphones a target for malicious actors and law enforcement alike: to the former, it provides new opportunities for criminality~\cite{icloudleaks,us_cert_thefts,dhsreportphones}. To the latter it offers new avenues for investigation, monitoring, and surveillance~\cite{elcomsoft_methods,cellebrite_advanced_services,comey14}. 

Over the past decade, hardware and software manufacturers have acknowledged these concerns, in the process deploying a series of major upgrades to smartphone hardware and operating systems. These include mechanisms designed to improve software security; default use of passcodes and biometric authentication; and the incorporation of strong encryption mechanisms to protect data in motion and at rest. While these improvements have enhanced the ability of smartphones to prevent data theft, they have provoked a backlash from the law enforcement community. This reaction is best exemplified by the \Gls{FBI}'s ``Going Dark'' initiative~\cite{comey14}, which seeks to increase law enforcement's access to encrypted data via legislative and policy initiatives. These concerns have also motivated law enforcement agencies, in collaboration with industry partners, to invest in developing and acquiring technical means for bypassing smartphone security features. This dynamic broke into the public consciousness during the 2016 ``Apple v. \Gls{FBI}'' controversy~\cite{ApplevFBI_apple,apple_letter_2016,ApplevFBI_fbi,ApplevFBI_oig}, in which Apple contested an FBI demand to bypass technical security measures. However, a vigorous debate over these issues continues to this day~\cite{elcomsoft_methods,cellebrite_advanced_services,cellebrite_cloud,cellebrite_ufed,elcomsoft_nojb,oxygen_latest}. Since 2015 and in the US alone, hundreds of thousands of forensic searches of mobile devices have been executed by over 2,000 law enforcement agencies, in all 50 states and the District of Columbia, which have purchased tools implementing such bypass measures~\cite{upturn_mass_extraction}.

The tug-of-war between device manufacturers, law enforcement, and third-party vendors has an important consequence for users: at any given moment, it is difficult to know which smartphone security features are operating as intended, and which can be bypassed via technical means. The resulting confusion is made worse by the fact that manufacturers and law enforcement routinely withhold technical details from the public and from each other. What limited information is available may be distributed across many different sources, ranging from obscure technical documents to court filings. Moreover, these documents can sometimes embed important technical information that is only meaningful to an expert in the field. Finally, competing interests between law enforcement and manufacturers may result in compromises that negatively affect user security~\cite{mennicloud2020}.

The outcome of these inconsistent protections and increasing law enforcement access is the creation of massive potential for violations of privacy. More than potential, technology is \textit{already} allowing law enforcement agencies around the world to surveil people~\cite{upturn_mass_extraction,pi_deepdive,uyghurs,surveillance_of_protesters}. Technological solutions are only part of the path to remediating these issues, and while we leave the policy advocacy and work to experts in those areas, we present these contributions in pursuit of progress on the technical front.

\paragraph{Our contributions} In this work we attempt a full accounting of the current and historical status of smartphone security measures. We focus on several of the most popular device types, and present a complete description of both the available security mechanisms in these devices, as well as a summary of the known {\em public} information on the state-of-the-art in bypass techniques for each. Our goal is to provide a single periodically updated guide that serves to detail the public state of data security in modern smartphones.

\medskip \noindent 
More concretely, we make the following specific contributions:

\begin{enumerate}
    \item We provide a technical overview of the key data security features included in modern Apple and Android-based smartphones, operating systems (OSes), and cloud backup systems. We discuss which forms of data are available in these systems, and under what scenarios this data is protected. Finally, to provide context for this description, we also offer a historical timeline detailing known improvements in each feature.
    
    \item We analyze more than a decade of public information on software exploits and \Gls{DHS} forensic reports and investigative documents, with the goal of formulating an understanding of which security features are (and historically have been) bypassed by criminals and law enforcement, and which security features are currently operating as designed.
    
    \item Based on the understanding above, we suggest changes and improvements that could further harden smartphones against unauthorized access.
\end{enumerate}

\medskip \noindent
We enter this analysis with two major goals. The first is an attempt to solve a puzzle: despite substantial technological advances in systems that protect user data, law enforcement agencies appear to be accessing device data with increasing sophistication~\cite{elcomsoft_methods,dhs_forensics,graykey_news_18}. This implies that law enforcement, at least, has become adept at bypassing these security mechanisms. A major goal in our analysis is to to understand how this access is being conducted, on the theory that any vulnerabilities used by public law enforcement agencies could also be used by malicious actors.

A second and related goal of this analysis is to help provide context for the current debate about law enforcement access to smartphone encrypted data~\cite{apple_letter_2016,ApplevFBI_fbi,ApplevFBI_oig,earn_it}, by demonstrating which classes of data are already accessible to law enforcement today via known technological bypass techniques. We additionally seek to determine which security technologies are effectively securing user data, and which technologies require improvement.

\medskip \noindent
{\em Platforms examined.}
Our analysis focuses on the two most popular OS platforms currently in use by smartphone vendors: Apple's iOS and Google's Android.\footnote{An important caveat in our Android analysis is that in practice, the Android device-base is significantly more fragmented than Apple's, due to the fact that Android phones are manufactured by many different vendors. As a consequence, our analysis concentrates mainly on ``flagship'' Google devices and high-end devices that hold a significant degree of market share.}  We begin by enumerating the key security technologies available in each platform, and then we discuss the development of these technologies over time. Our primary goal in each case is to develop an understanding of the {\em current state} of the technological protection measures that protect user data on each platform.

\medskip \noindent
{\em Sources of bypass data.} Having described these technological mechanisms, we then focus our analysis on known techniques for bypassing smartphone security measures. Much of this information is well-known to those versed in smartphone technology. However, in order to gain a deeper understanding of this area, we also examined a large corpus of forensic test results published by the U.S. Department of Homeland Security~\cite{dhs_forensics}, as well as scouring public court documents for evidence of surprising new techniques. This analysis provides us with a complete picture of which security mechanisms are likely to be bypassed, and the impact of such bypasses. We provide a concise, complete summary of the contents of the \Gls{DHS} forensic tool test results in Appendix~\ref{app:forensic_tools}.

\medskip \noindent
{\em Threat Model.} In this work we focus on two sources of device data compromise: $(1)$ physical access to a device, e.g. via device seizure or theft, and $(2)$ remote access to data via cloud services. The physical access scenario assumes that the attacker has gained access to the device, and can physically or logically exploit it via the provided interfaces. Since obtaining data is relatively straightforward when the attacker has authorized access to the device, we focus primarily on unauthorized access scenarios in which the attacker does not possess the passcode or login credentials needed to access the device. 

By contrast, our cloud access scenario assumes that the attacker has gained access to cloud-stored data. This access may be obtained through credential theft (e.g. spear-phishing attack), social engineering of cloud provider employees, or via investigative requests made to cloud providers by law enforcement authorities. 
While we note that legitimate investigative procedures differ from criminal access from a legal point of view, we group these attacks together due to the fact that they leverage similar technological capabilities.

\section{Summary of Key Findings}

We now provide a list of our key findings for both Apple iOS and Google Android devices. 

\subsection{Apple iOS} Apple iOS devices (iPhones, iPads\footnote{Although our focus in this work is on smartphones, we mention other devices that share a common security platform.}) incorporate a number of security features that are intended to limit unauthorized access to user data. These include software restrictions, biometric access control sensors, and widespread use of encryption within the platform. Apple is also noteworthy for three reasons: $(1)$ the company has overall control of both the hardware and operating system software deployed on its devices, $(2)$ the company's business model closely restricts which software can be installed on the device, and $(3)$ Apple management has, in the past, expressed vocal opposition to making technical changes in their devices that would facilitate law enforcement access~\cite{apple_letter_2016}.\footnote{It should be noted that in practice, Apple has assisted law enforcement operations as required by law, rather than filing suit~\cite{apple_fbi_assist2020}, and continue to provide data to U.S. and other law enforcement agencies upon request at increasing rates, per the Apple Transparency Report~\cite{apple_transparency}.}

To determine the level of security currently provided by Apple against sophisticated attackers, we considered the full scope of Apple's public documentation, as well as published reports from the U.S. Department of Homeland Security (\Gls{DHS}), postings from mobile forensics companies, and other documents in the public record. Our main findings are as follows:

\begin{description}
\item {\bf Limited benefit of encryption for powered-on devices.} Apple advertises the broad use of encryption to protect user data stored on-device~\cite{apple_security_guides,apple_platform_security,apple_privacy}. However, we observed that a surprising amount of sensitive data maintained by built-in applications is protected using a weak ``available after first unlock'' (\Gls{AFU}) protection class, which does not evict decryption keys from memory when the phone is locked. The impact is that the vast majority of sensitive user data from Apple's built-in applications can be accessed from a phone that is captured and logically exploited while it is in a powered-on (but locked) state. We also found circumstantial evidence from a 2014 update to Apple's documentation that the company has, in the past, reduced the protection class assurances regarding certain system data, to unknown effect.

Finally, we found circumstantial evidence in both the \Gls{DHS} procedures and investigative documents that law enforcement now routinely exploits the availability of decryption keys to capture large amounts of sensitive data from locked phones. Documents acquired by Upturn, a privacy advocate organization, support these conclusions, documenting law enforcement records of passcode recovery against both powered-off and simply locked iPhones of all generations~\cite{arizona_le_records}.

\item {\bf Weaknesses of cloud backup and services.} Apple's iCloud service provides cloud-based device backup and real-time synchronization features. By default, this includes photos, email, contacts, calendars, reminders, notes, text messages (iMessage and SMS/MMS), Safari data (bookmarks, search and browsing history), Apple Home\footnote{Apple Home provides integrations for Internet of Things devices.} data, Game Center\footnote{Game Center is Apple's built-in gaming social media service.} data, and cloud storage for installed apps.

We examine the current state of data protection for iCloud, and determine (unsurprisingly) that activation of these features transmits an abundance of user data to Apple's servers, in a form that can be accessed remotely by criminals who gain unauthorized access to a user's cloud account, as well as authorized law enforcement agencies with subpoena power. More surprisingly, we identify several counter-intuitive features of iCloud that increase the vulnerability of this system. As one example, Apple's ``Messages in iCloud'' feature advertises the use of an Apple-inaccessible ``end-to-end'' encrypted container for synchronizing messages across devices~\cite{apple_icloud_security}. However, activation of iCloud Backup in tandem causes the decryption key for this container to be uploaded to Apple's servers in a form that Apple (and potential attackers, or law enforcement) can access~\cite{apple_icloud_security}. Similarly, we observe that Apple's iCloud Backup design results in the transmission of device-specific {\em file encryption keys} to Apple. Since these keys are the same keys used to encrypt data on the device, this transmission may pose a risk in the event that a device is subsequently physically compromised.\footnote{Particularly, the iCloud Backup Keybag encrypts file keys with asymmetric class keys~\cite{apple_security_guides,apple_platform_security}. Some encrypted data is accessible to Apple~\cite{apple_icloud_security}, and other data is ``end-to-end encrypted.''}

More generally, we find that the documentation and user interface of these backup and synchronization features are confusing, and may lead to users unintentionally transmitting certain classes of data to Apple's servers.

\item {\bf Evidence of past hardware (SEP) compromise.} iOS devices place strict limits on passcode guessing attacks\footnote{This describes an attack in which the attacker tests many (or all) possible passcodes in order to identify the user passcode and thus derive encryption keys.} through the assistance of a dedicated processor known as the Secure Enclave processor (\Gls{SEP}). We examined the public investigative record to review evidence that strongly indicates that as of 2018, passcode guessing attacks were feasible on \Gls{SEP}-enabled iPhones using a tool called GrayKey. To our knowledge, this most likely indicates that a software bypass of the SEP was available in-the-wild during this timeframe. We also reviewed more recent public evidence, and were not able to find dispositive evidence that this exploit is still in use for more recent phones (or whether exploits still exist for older iPhones). Given how critical the SEP is to the ongoing security of the iPhone product line, we flag this uncertainty as a serious risk to consumers.


\item {\bf Limitations of ``end-to-end encrypted'' cloud services.} Several Apple iCloud services advertise ``end-to-end'' encryption in which only the user (with knowledge of a password or passcode) can access cloud-stored data. These services are optionally provided in Apple's CloudKit containers and via the iCloud Keychain backup service. Implementation of this feature is accomplished via the use of dedicated Hardware Security Modules (\Gls{HSM}s) provisioned at Apple's data centers. These devices store encryption keys in a form that can only be accessed by a user, and are programmed by Apple such that cloud service operators cannot transfer information out of an \Gls{HSM} without user permission~\cite{apple_security_guides,apple_platform_security}. 

As noted above, our finding is that the end-to-end confidentiality of some encrypted services is undermined when used in tandem with the iCloud backup service. More critically, we observe that Apple's documentation and user settings blur the distinction between ``encrypted'' (such that Apple has access) and ``end-to-end encrypted'' in a manner that makes it difficult to understand which data is available to Apple. Finally, we observe a fundamental weakness in the system: Apple can easily cause user data to be re-provisioned to a new (and possibly compromised) \Gls{HSM} simply by presenting a single dialog on a user's phone. We discuss techniques for mitigating this vulnerability.


\end{description}

\noindent
Based on these findings, our overall conclusion is that data for iOS devices is highly available to both sophisticated criminals and law enforcement actors with either cloud or physical access. This is due to a combination of the weak protections offered by current Apple iCloud services, and weak defaults used for encrypting sensitive user data on-device. The impact of these choices is that Apple's data protection is {\em fragile}: once certain software or cloud authentication features are breached, attackers can access the vast majority of sensitive user data on device. Later in this work we propose improvements aimed at improving the resilience of Apple's security measures.

\subsection{Google Android \& Android Phones} 
Google's Android operating system, and many third-party phones that use Android, incorporates a number of security features that are analogous to those provided by Apple devices. Unlike Apple, Google does not fully control the hardware and software stack on all Android-compatible smartphones: some Google Android phones are manufactured entirely by Google, while other devices are manufactured by third parties. Moreover, device manufacturers routinely modify the Android operating system prior to deployment.

This fact makes a complete analysis of the Android smartphone ecosystem more challenging. In this work, we choose to focus on a number of high-profile phones such as Google Pixel devices and recent-model Samsung Galaxy phones, mainly because these devices are either $(1)$ representative devices designed by Google to fully encapsulate the capabilities of the Android OS, or $(2)$ best-selling Android phones, with large numbers of active devices worldwide. We additionally focus primarily on recent versions of Android (Android 10 and 11, as of this writing). We note, however, that the Android ecosystem is highly fragmented, and contains large numbers of older-model phones that no longer receive OS software updates, a diversity of manufacturers, and a subclass of phones which are built using inexpensive hardware that lacks advanced security capabilities. Our findings in this analysis are therefore necessarily incomplete, and should be viewed as an optimistic ``best case.''

To determine the level of security currently provided by these Android devices against sophisticated attackers, we considered the full scope of Google's public documentation, as well as published reports from the U.S. Department of Homeland Security (\Gls{DHS}), postings from mobile forensics companies, and other documents in the public record. Our main findings are as follows:

\begin{description}
\item {\bf Limited benefit of encryption for powered-on devices.} Like Apple iOS, Google Android provides encryption for files and data stored on disk. However, Android's encryption mechanisms provide fewer gradations of protection. In particular, Android provides no equivalent of Apple's {\em Complete Protection} (CP) encryption class, which evicts decryption keys from memory shortly after the phone is locked. As a consequence, Android decryption keys remain in memory at all times after ``first unlock,'' and user data is potentially vulnerable to forensic capture.

\item {\bf De-prioritization of end-to-end encrypted backup.} Android incorporates an end-to-end encrypted backup service based on physical hardware devices stored on Google's datacenters. The design of this system ensures that recovery of backups can only occur if initiated by a user who knows the backup passcode, an on-device key protected by the user's PIN or other authentication factor. Unfortunately, the end-to-end encrypted backup service must be opted-in to by app developers, and is paralleled by the opt-out Android Auto-Backup, which simply synchronizes app data to Google Drive, encrypted with keys held by Google.

\item {\bf Large attack surface.} Android is the composition of systems developed by various organizations and companies. The Android kernel has Linux at its core, but also contains chip vendor- and device manufacturer-specific modification. Apps, along with support libraries, integrate with system components and provide their own services to the rest of the device. Because the development of these components is not centralized, cohesively integrating security for all of Android would require significant coordination, and in many cases such efforts are lacking or nonexistent.

\item {\bf Limited use of end-to-end encryption.} End-to-end encryption for
    messages in Android is only provided by default in third-party messaging
        applications. Native Android applications do not provide end-to-end
        encryption: the only exception being Google Duo,\footnote{According to a
        recent technical draft, Google is also implementing end-to-end encryption
        for Messages via \Gls{RCS}, the replacement protocol of
        SMS/MMS~\cite{verge_googlemessages_e2esoon,google_rcs_e2e}.}
        which provides end-to-end encrypted video calls. The current lack of default
        end-to-end encryption for messages allows the service provider (for example,
        Google) to view messages and logs, potentially putting user data at risk
        from hacking, unwanted targeted advertising, subpoena, and surveillance
        systems.

\item {\bf Availability of data in services.} Android has deep integration with Google services, such as Drive, Gmail, and Photos. Android phones that utilize these services (the large majority of them~\cite{statista_android_mfg,android_certified}) send data to Google, which stores the data under keys it controls - effectively an extension of the lack of end-to-end encryption beyond just messaging services. These services accumulate rich sets of information on users that can be exfiltrated either by knowledgeable criminals (via system compromise) or by law enforcement (via subpoena power). 

\end{description}

\chapter{Technical Background}

\section{Data Security Technologies for Mobile Devices} 
Modern smartphones generate and store immense amounts of sensitive personal information. This data comes in many forms, including photographs, text messages, emails, location data, health information, biometric templates, web browsing history, social media records, passwords, and other documents. Access control for this data is maintained via several essential technologies, which we describe below. 

\begin{description}
\item {\bf Software security and isolation.} Modern smartphone operating systems are designed to enforce access control for users and application software. This includes restricting input/output access to the device, as well as ensuring that malicious applications cannot access data to which they are not entitled. Bypassing these restrictions to run arbitrary code requires making fundamental changes to the operating system, either in memory or on disk, a technique that is sometimes called ``jailbreaking'' on iOS or ``rooting'' on Android.

\item {\bf Passcodes and biometric access.} Access to an Apple or Android smartphone's user interface is, in a default installation, gated by a user-selected passcode of arbitrary strength. Many devices also deploy biometric sensors based on fingerprint or face-recognition as an alternative means to unlock the device.

\item {\bf Disk and file encryption.} Smartphone operating systems embed data encryption at either the file or disk volume level to protect access to files. This enforces access control to data even in cases where an attacker has bypassed the software mechanisms controlling access to the device. Encryption mechanisms typically derive encryption keys as a function of the user-selected passcode and device-embedded secrets, which is designed to ensure that access to the device requires both user consent and physical control of the hardware.

\item {\bf Secure device hardware.} Increasingly, smartphone manufacturers have begun to deploy secure co-processors and virtualized equivalents\footnote{Examples include the TrustZone architecture included in many \Gls{ARM} processors~\cite{arm_trustzone}.} in order to harden devices against both software attacks and physical attacks on device hardware. These devices are designed to strengthen the encryption mechanisms, and to guard critical data such as biometric templates.

\item {\bf Secure backup and cloud systems.} Most smartphone operating systems offer cloud-based data backup, as well as real-time cloud services for sharing information with other devices. Historically, access to cloud backups has been gated by access controls that are solely under the discretion of the cloud service provider,such as password authentication, making this a fruitful target for both attackers and law enforcement to gain access to data. More recently, providers have begun to deploy provider-inaccessible encrypted backup systems that enforce access controls that require user-selected passcodes, with security of the data enforced by trusted hardware at the providers' premises.
\end{description}

\section{Threat Model} In order to discuss the vulnerability of mobile devices it is pertinent to consider the threat models which underpin our analysis. In fact, in this case the threats of the traditional remote network adversary are relatively well-mitigated. It is with a particular additional capability that our threat actors, namely law enforcement forensic investigators and criminal hackers, are able to bypass the existing mitigations: protracted physical access to the mobile device. This access facilitates data extraction in that devices can be kept charging, and mitigations such as disabling physical data ports or remote lock or wipe can be evaded.

We consider deep physical analysis of hardware (particularly de-soldering and ``de-capping''\footnote{Removing the top physical layer of shielding which covers the chip.} the silicon, often done with nitric acid to gain direct access to underlying physical logic implementations) out of scope, as they seem to be prohibitively expensive and risk destroying device hardware or invalidating evidence; we see no clear evidence of this occurring at scale even in federal law enforcement offices. However, we do see evidence of some physical analysis in the form of academic~\cite{skorobogatov2016bumpy} and commercial~\cite{ios_poweroff} research, to the extent of interposing the device's connections to storage or power.

In some cases, law enforcement receive consent from the targets of investigation to access mobile devices. This consent is likely accompanied with passcodes, PINs, and/or passwords. A database of recent warrants against iOS devices shows that not only do law enforcement agents sometimes get this consent, they also seek warrants which nullify any later withdrawal of consent~\cite{vice_db} and as such can use their position and access to completely compromise the device.

In other cases, law enforcement agencies are able to execute warrants which create a ``geofence'' or physical region and period of time, such that any devices which are found to have been present therein are subject to search, usually in the form of requesting data from cloud providers (Apple for iOS devices, and Google for Android)~\cite{verge_geofence}. Such geofence warrants have massive reach and potential to violate the privacy of innocent passerby, but their use is a matter of policy and thus not in scope for our analysis.

Largely, the evidence we gather and present demonstrates that this physical access is used by law enforcement to attach commercial forensic tools which perform exploitation and data extraction on mobile devices. This approach can provide both data directly from the device, or cloud access tokens resident in device memory which can be further exploited to gain access to a target's online accounts and content. These two methods, device and cloud extraction, can provide overlapping but different categories of sensitive personal data, and, together or individually, represent a massive breach of a target's privacy.

\section{Sensitive Data on Phones}

The U.S. National Institute of Standards and Technology (\Gls{NIST}) maintains a list of data targets for mobile device acquisition forensic software. These are presented in Figure~\ref{list:nist_forensics}. The categories of data which forensic software tests attempt to extract provide us with a notion of what data is prioritized by law enforcement, and allow us to focus our examination of user data protection. The importance of these categories is corroborated by over 500 warrants against iPhones recently collected and released in the news~\cite{vice_db,vice_db_news}, and articles posted by mobile forensics companies and investigators~\cite{elcomsoft_methods,ios_significant_locations_scooped}.

While a useful resource, this list does not capture the extent of potential privacy loss due to unauthorized mobile device access, primarily falling short in two ways. First, it does not capture the long-lived nature of some identifiers, nor the potential sensitivity of each item. Second, critically, mobile devices contain information about and from ourselves but also our networks of peers, friends, and family, and so privacy loss propagates accordingly. Further, due to the emerging capabilities of machine learning and data science techniques combined with continuously increasing availability of aggregated data sets, predictions and analysis (whether correct or not) make these potential violations of privacy nearly unbounded.

\begin{figure}[H]
\begin{tcolorbox}
    \centering
    \caption{List of Targets for \Gls{NIST} Mobile Device Acquisition Forensics}
    \label{list:nist_forensics}
    \begin{itemize*}
    \item Cellular network subscriber information: IMEI, MEID/ESN
    \item Personal Information Management (PIM) data: address book/contacts, calendar, memos, etc
    \item Call logs: incoming, outgoing, missed
    \item Text messages: \Gls{SMS}, \Gls{MMS} (audio, graphic, video)
    \item Instant messages
    \item Stand-alone files:\footnote{This list has evolved over time, and certain entries (such as ``stand-alone files'') seem to imply a disconnect between this list and how modern smartphones organize and store personal information, and may be remnants of how multimedia/feature phones used to do so. Thus is is unclear if third-party app data falls into this category.} audio, documents, graphic, video
    \item E-mail
    \item Web activity: history, bookmarks
    \item GPS and geo-location data
    \item Social media data: accounts, content
    \item SIM/UICC data: provider, IMSI, MSISDN, etc
    \end{itemize*}
    \caption*{In accordance with \Gls{NIST} standards, \Gls{DHS} tests forensic software for mobile device acquisition of these categories of data~\cite{dhs_forensics,nist_forensics_spec}.}
    \caption*{Source: NIST~\cite{nist_forensics_spec}}
\end{tcolorbox}
\end{figure}

\chapter{Apple iOS}

Apple devices are ubiquitous in countries around the world. In Q4 of 2019 alone, almost 73 million iPhones and almost 16 million iPads shipped~\cite{iphonenews20,ipadnews20}. While Apple devices represent a minority of the global smartphone market share, Apple maintains approximately a 48\% share of the smartphone market in the United States~\cite{statista_ios}, with similar percentages in many western nations. Overall, Apple claims 1.4 billion active devices in the world~\cite{apple_billions}. Along with increasing usage trends, these factors make iPhones extremely valuable targets for hackers, with bug bounty programs offering up to \$2 million USD~\cite{apple_bounty,zerodium_bounty}, for law enforcement agencies executing warrants, and for governments seeking to surveil journalists, activists, or criminals~\cite{pegasus_ios}.

Apple invests heavily in restricting the operating system and application software that can run on their hardware~\cite{apple_security_guides,apple_platform_security}. As such, even users with technical expertise are limited in their ability to extend and protect Apple devices with their own modifications, and Apple software development teams represent essentially the sole \textit{technical} mitigation against novel attempts to access user data without authorization. The high value of Apple software exploits and Apple's centralized response produces a cat-and-mouse game of exploitation and patching, where users can never be fully assured that their device is not vulnerable. Apple undertakes protecting user devices through numerous and varied mitigation strategies, and while these include both technical and business approaches, the technical will be primarily and thoroughly examined in this work.

\section{Protection of User Data in iOS}\label{sec:ios_security}

In this section we provide an overview of key elements of Apple's user data protection strategy that cover the bulk of on-device and cloud interactions supported by iOS devices. This overview is largely based on information published by Apple~\cite{apple_security_guides,apple_platform_security}, and additionally on external (to Apple) research, product analyses, and security tests.

\paragraph{User authentication} Physical interaction is the primary medium of modern smartphones. In order to secure a device against unauthorized physical access, some form of user authentication is needed. iOS devices provide two mechanisms for this: $(1)$ numeric or alphanumeric passcodes and $(2)$ biometric authentication. In early iPhones, Apple suggested a default of four-digit numeric passwords, but now suggests a six-digit numeric passcode. Users may additionally opt for longer alphanumeric passphrases, or (against Apple's advice~\cite{apple_platform_security}) disable passcode authentication entirely.

\begin{figure}[H]
    \centering
    \includegraphics[width=0.8\linewidth]{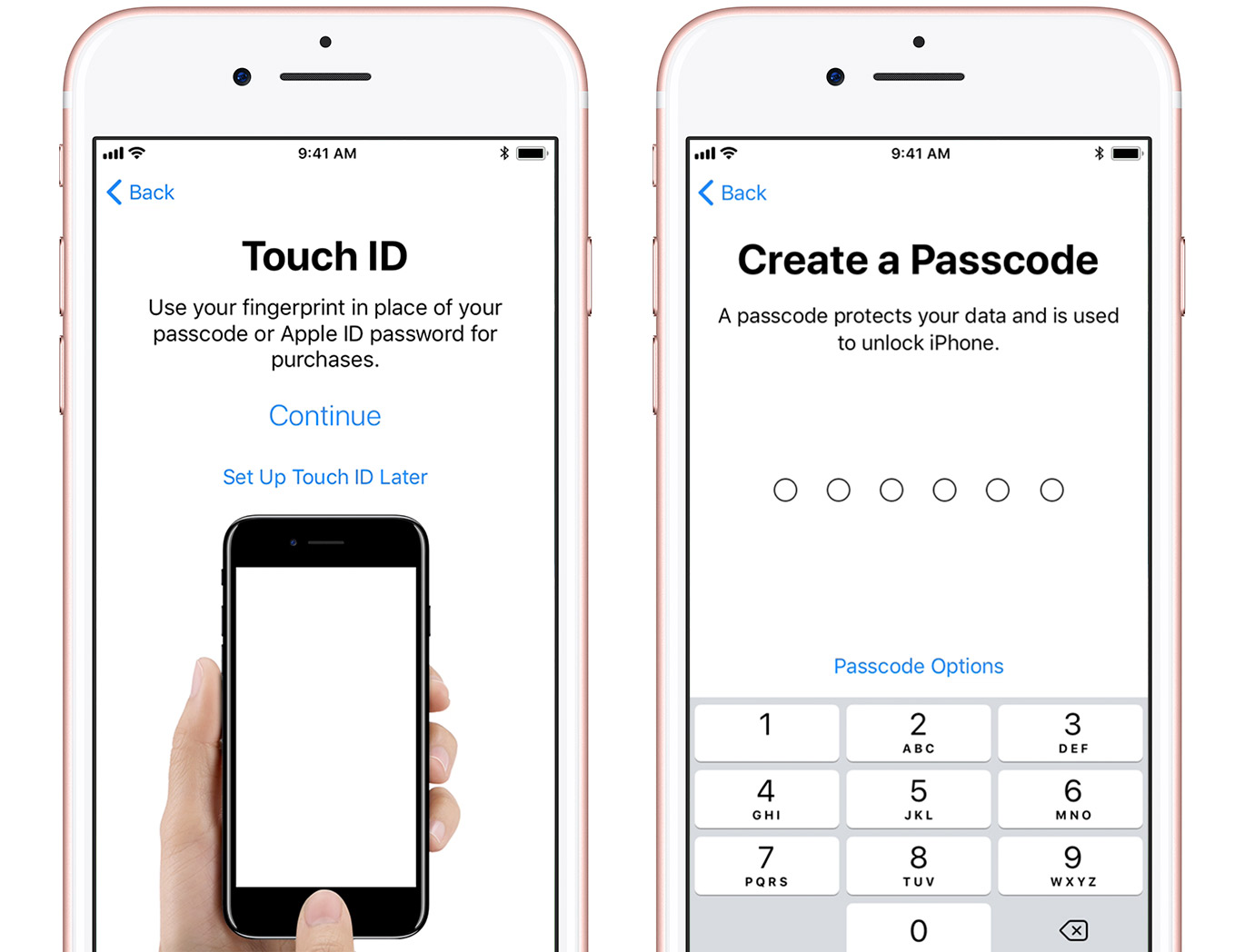}
    \caption{iPhone Passcode Setup Interface}\label{img:passcode}
    \caption*{Source: Apple~\cite{apple_passcode_image}}
\end{figure}

Because there are a relatively small number of 6-digit passcodes, iOS is designed to rate-limit passcode entry attempts in order to prevent attackers from conducting brute-force guessing attacks. In the event of an excessive number of entry attempts, device access may be temporarily locked and user data can be permanently deleted. To improve the user experience while maintaining security, Apple employs biometric access techniques in its devices: these include \Gls{TouchID}, based on a capacitive fingerprint sensor, and a more recent replacement \Gls{FaceID}, which employs face recognition using a depth-sensitive camera~\cite{apple_security_guides,apple_platform_security}. The image in Figure~\ref{img:passcode} demonstrates the \Gls{TouchID} and six-digit passcode setup interfaces on iOS.

Apple also restricts a number of passcodes that are deemed too common, or too ``easily guessed.'' For a thorough examination of this list and its effects on iOS, refer to recent works by Markert et al.~\cite{ios_user_passcodes}.

\paragraph{Code signing} iOS tightly restricts the executable code that can be run on the platform. This is enforced using digital signatures~\cite{apple_security_guides,apple_platform_security}. The mechanics of iOS require that only software written by Apple or by an approved developer can be executed on the device.

\medskip \noindent
{\em Trusted boot.} Apple implements signatures for the software which initializes the operating system, and the operating system itself, in order to verify its integrity~\cite{apple_security_guides,apple_platform_security}. These signature checks are embedded in the low-level firmware called Boot ROM. The primary purpose of this security measure, according to Apple, is to ensure that long-term encryption keys are protected and that no malicious software runs on the device.

\begin{figure}[H]
    \centering
    \includegraphics[width=0.75\linewidth]{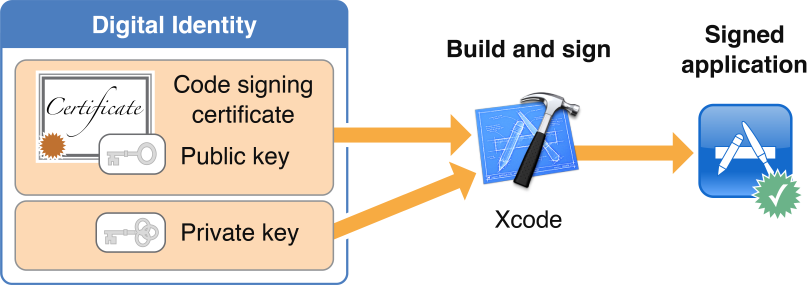}
    \caption{Apple Code Signing Process Documentation}\label{img:codesign}
    \caption*{Source: Apple Developer Documentation~\cite{apple_codesign}}
\end{figure}

\medskip \noindent
{\em App signing.} Apple authorizes developers to distribute code using a combination of Apple-controlled signatures and a public-key certificate infrastructure that allows the system to scale~\cite{apple_platform_security,apple_developer_docs}. Organizations may also apply and pay for {\em enterprise signing certificates} that allow them to authenticate software for specially-authorized iOS devices (those that have installed the organization's certificate~\cite{apple_platform_security,apple_enterprise}). This is intended to enable companies to deliver proprietary internal apps to employees, although the mechanism has been subverted many times for \gls{jailbreak}ing~\cite{pangu_bh}, for advertising or copyright fraud~\cite{enterprise_cert_misuse}, and for device compromise~\cite{wei2014apple}. The image in Figure~\ref{img:codesign} displays Apple documentation of the code signing process for developers.

\paragraph{Sandboxing and code review} iOS enforces restrictions that limit each application's access to user data and the operating system \Gls{API}s. This mechanism is designed to protect against incursions by malicious third-party applications, and to limit the damage caused by exploitation of a non-malicious application. To implement this, iOS runs each application in a ``\gls{sandbox}'' that restricts its access to the device filesystem and memory space, and carries a signed manifest that details allowed access to system services such as location services. For applications distributed via its App Store -- which is the only software installation method allowed on a device with default settings -- Apple additionally performs automated and manual code review of third-party applications~\cite{apple_app_review}. Despite these protections, malicious or privacy-violating applications have passed review~\cite{iosprivacyleaks11}.

iOS 14\footnote{Released contemporaneously with this writing.} includes additional privacy transparency and control features such as listing privacy-relevant permissions in the App Store, allowing finer-grained access to photos, an OS-supported recording indicator, Wi-Fi network identifier obfuscation, and optional approximate location services~\cite{apple_ios14_release_2020}. However, most of these features are focused on the privacy of users from app developers rather than from the phone itself, the relevant adversary under the threat model of forensics.

\begin{figure}[H]
    \centering
    \includegraphics[width=0.9\linewidth]{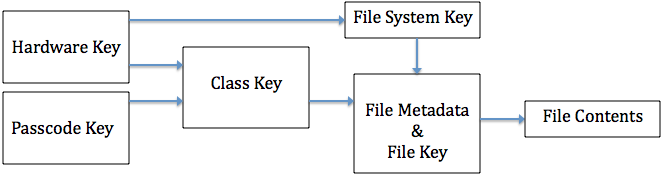}
    \caption{iOS Data Protection Key Hierarchy. Each arrow }\label{img:encryption}
    \caption*{Source: Washington University in St. Louis~\cite{wustl_ios_keys}}
\end{figure}

\paragraph{Encryption} While software protections provide a degree of security for system and application data, these security mechanisms can be bypassed by exploiting logic vulnerabilities in software or flaws in device hardware. Apple attempts to address this concern through the use of data encryption. This approach provides Apple devices with two major benefits: first, it ensures that Apple device storage can be rapidly erased, simply by destroying a single encryption key carried within the device. Second, encryption allows Apple to provide strong file-based access control to files and data objects on the device, even in the event that an attacker bypasses security controls within the operating system. The image in Figure~\ref{img:encryption} depicts the key hierarchy used in iOS Data Protection.

iOS employs industry-standard cryptography including \Gls{AES}~\cite{aes}, \Gls{ECDH} over \Gls{Curve25519}~\cite{curve25519}, and various \Gls{NIST}-approved standard constructions to encrypt files in the filesystem~\cite{apple_security_guides,apple_platform_security}.
To ensure that data access is controlled by the user and is tied to a specific device, Apple encrypts all files using a key that is derived from a combination of the user-selected passcode and a unique secret cryptographic key (called the \Gls{UID key}) that is stored within device hardware.  In order to recover the necessary decryption keys following a reboot the user must enter the device passcode. When the device is locked but has been unlocked since last boot, biometrics suffice to unlock these keys.\footnote{Except in some specific cases mostly associated with disuse timeouts~\cite{apple_passcodes_touchid_faceid}.} To prevent the user from bypassing the encryption by guessing a large number of passcodes, the system enforces guessing limits in two ways: $(1)$ by using a computationally-intensive password-based key derivation function that requires 80ms to derive a key on the device hardware~\cite{apple_security_guides,apple_platform_security}, and $(2)$ by enforcing guess limits and increasing time intervals using trusted hardware (see further below) and software~\cite{apple_security_guides,apple_platform_security}.


\medskip \noindent
{\em Data Protection Classes.} 
Apple provides interfaces to enable encryption in both first-party and third-party software, using the iOS Data Protection \Gls{API}~\cite{apple_security_guides,apple_platform_security}. Within this package, Apple specifies several encryption ``protection classes'' that application developers can select when creating new data files and objects. These classes allow developers to specify the security properties of each piece of encrypted data, including whether the keys corresponding to that data will be evicted from memory after the phone is locked (``Complete Protection'' or \Gls{CP}) or shut down (``After First Unlock'' or \Gls{AFU}).

We present a complete list of Data Protection classes in Figure~\ref{list:ios_dp_classes}. As we will discuss below {\em the selection of protection class makes an enormous practical difference in the security afforded by Apple's file encryption.} Since in practice, users reboot their phones only rarely, many phones are routinely carried in a locked-but-authenticated state (\Gls{AFU}). This means that for protection classes other than \Gls{CP}, decryption keys remain available in the device's memory. Analysis of forensic tools shows that to an attacker who obtains a phone in this state, encryption provides only a modest additional protection over the software security and authentication measures described above.

\begin{figure}[H]
\begin{tcolorbox}
    \centering
    \caption{List of iOS Data Protection Classes}
    \label{list:ios_dp_classes}
    \begin{description}
    \item \textbf{Complete Protection (\Gls{CP})}: Encryption keys for this data are evicted shortly after device lock (10 seconds).
    \item \textbf{Protected Unless Open (PUO)}: Using public-key encryption, PUO allows data files to be created and encrypted while the device is locked, but only decrypted when the device is unlocked, by keeping an ephemeral public key in memory but evicting the private key at device lock. Once the file has been created and closed, data in this class has properties similar to Complete Protection.
    \item \textbf{Protected Until First User Authentication} (a.k.a. \textbf{After First Unlock}) \textbf{(\Gls{AFU})}: Encryption keys are decrypted into memory when the user first enters the device passcode, and remain in memory even if the device is locked.
    \item \textbf{No Protection (NP)}: Encryption keys are encrypted by the hardware \Gls{UID key}s only, not the user passcode, when the device is off. These keys are always available in memory when the device is on.
    \end{description}
    \caption*{Source: Apple iOS and Platform Security Guides~\cite{apple_security_guides,apple_platform_security}}
\end{tcolorbox}
\end{figure}

A natural question arises: why not simply apply \Gls{CP} to all classes of data? This  would seriously hamper unauthorized attempts to access user data. However, the answer appears to lie in user experience. Examining the data in Table~\ref{tbl:ios_history} it seems likely that data which is useful to apps which run in the background, including while the device is locked, is kept in the \Gls{AFU} state in order to enable continuous services such as \Gls{VPN}s, email synchronization, contacts, and \gls{iMessage}. For example, if user's contacts were protected using \Gls{CP}, then a locked phone would be unable to display a name associated with a phone number for an incoming text message, and likely would just display the phone number itself.\footnote{This is indeed the behavior that phones exhibit prior to the first unlock.} This would severely impact the feature of iOS to preview sender and message content in lock screen notifications.

\paragraph{Keychain} iOS provides the system and applications with a secure key-value store \Gls{API} called the Keychain~\cite{apple_security_guides,apple_platform_security} for storing sensitive secrets such as keys and passwords. The Keychain provides encrypted storage and permissioned access to these secret items via a public \Gls{API} which third-party app developers can build around. This Keychain data is encrypted using keys that are in turn protected by device hardware keys and the user passcode, and can optionally placed into protection classes that mirror the protection classes in Figure~\ref{list:ios_dp_classes}. In addition to these protection classes, the Keychain also introduces an optional characteristic called Non-Migratory (NM), which ensures that any protected data can only be decrypted on the same device that it was encrypted under. This mechanism is enforced via the internal hardware \Gls{UID key}, which cannot be transported off of the device.\footnote{Use of this class ensures that protected data cannot be restored onto a new phone and decrypted.}

Apple selects the Data Protection classes used by built-in applications such as \gls{iMessage} and Photos, while third-party developers may choose these classes when developing applications. If they do not explicitly select a different protection class, the default class used is Protected Until First User Authentication, or \Gls{AFU}. Their documentation~\cite{apple_security_guides,apple_platform_security} claims that, at least, the following applications' data falls under some degree of Data Protection: Messages, Mail, Calendar, Contacts, Photos, and Health, in addition to all third-party apps in iOS 7 and later. Refer to Table~\ref{tbl:ios_history} for more details on Data Protection, and to Figure~\ref{fig:ios_extractables} for details on data which is necessarily \Gls{AFU} or less protected due to its availability via forensic tools.

\paragraph{Backups} iOS devices can be backed up either to iCloud or to a personal computer. When backing up to a personal computer, users may set an optional backup password. This password is used to derive an encryption key that, in turn, protects a structure called the ``Keybag.'' The Keybag contains a bundle of further encryption keys that encrypt the backup in its entirety. A limitation of this mechanism is that the user-specified backup password must be extremely strong: since this password is not entangled with device hardware secrets, stored backups may be subject to sophisticated offline dictionary attacks that can guess weak passwords, however iOS uses 10 million iterations of \Gls{PBKDF2}~\cite{pbkdf2} to significantly inhibit such password cracking~\cite{apple_security_guides,apple_platform_security}.\footnote{Since iOS 10.2, previously 10,000; compare iOS security guides for iOS 9 (May 2016) and iOS 10 (March 2017)~\cite{apple_security_guides}.} iOS devices may also be configured to backup to Apple iCloud servers. In this instance, data is encrypted asymmetrically using \Gls{Curve25519}~\cite{curve25519} (so that backups can be performed when the device is locked without exposing secret keys)~\cite{apple_security_guides}, and those keys are encrypted with ``iCloud keys'' known to Apple to create the iCloud Backup Keybag. This means that Apple itself\footnote{Acting on its own or under court order.} or a malicious actor who can guess or reset user credentials can access the contents of the backup. For both types of backups, the Keychain is additionally encrypted with a key derived from the hardware \Gls{UID key} to prevent restoring it to a new device~\cite{apple_security_guides,apple_platform_security}.

Aside from Mail, which is not encrypted at rest on the server~\cite{apple_icloud_security}, all other backup data is stored encrypted with keys that Apple has access to. This implies that such data can be accessed through an unauthorized compromise of Apple's network, a stolen credential attack, or compelled access by authorized government officials. The data classes included in an iCloud backup are listed in Figure~\ref{fig:ios_icloud_backup}.

\medskip
\begin{tcolorbox}
\begin{figure}[H]
    \caption{List of Data Included in iCloud Backup}\label{fig:ios_icloud_backup}
\begin{itemize*}
\begin{multicols}{2}
    \item App data\footnote{Unless developers opt out, refer to~\S\ref{sec:ios_security}.}
    \item Apple Watch backups
    \item Device settings
    \item Home screen and app organization
    \item \gls{iMessage}, \Gls{SMS}, and \Gls{MMS} messages\footnote{\gls{iMessage}s are stored in iCloud, rather than in backups, if iCloud for \gls{iMessage} is enabled.}
    \item Photos and videos
    \item Purchase history from Apple services
    \item Ringtones
    \item Visual Voicemail password
\end{multicols}
\end{itemize*}
\caption*{Source: Apple Support Documentation~\cite{apple_icloud_backup}}
\end{figure}
\end{tcolorbox}

\paragraph{iCloud} In addition to backups, iCloud can be used to store and synchronize various classes of data, primarily for built-in default apps such as Photos and Documents. Third-party apps' files are also included in iCloud Backups unless the developers specifically opt-out~\cite{apple_icloud_backup_optout}. Apple encrypts data in transit using Transport Layer Security (TLS) as is standard for internet traffic~\cite{apple_icloud_security}. Data at rest, however, is a more complex story: Mail is stored \textit{unencrypted on the server} (which Apple claims is an industry standard practice~\cite{apple_icloud_security}). The data classes in Figure~\ref{fig:icloud_weak} is stored encrypted with a 128-bit \Gls{AES} key known to Apple, and the data classes in Figure~\ref{fig:icloud_strong} is stored encrypted with a key derived from the user passcode\footnote{Potentially also the hardware \Gls{UID key}, although Apple claims that this data can be recovered on a new device using only the passcode~\cite{apple_icloud_security}.} and is thus protected from even Apple. There are caveats to these lists, including that Health data is only end-to-end encrypted if two-factor authentication is enabled for iCloud~\cite{apple_icloud_security}, and that Messages in iCloud, which can be enabled in the iOS settings, uses end-to-end encryption, but the key is also included in iCloud backups and thus can be accessed by Apple if iCloud backup is enabled~\cite{apple_icloud_security}.

The user experience of controlling access to iCloud data embeds relatively unpredictable aspects: for example, disabling iCloud for the default Calendar app prevents the sending of calendar invites via email on iOS 13.\footnote{This was confirmed experimentally by the authors.} A variety of exceptions and special cases, like the \gls{iMessage} example above, combined with unpredictable side-effects on user experience, makes it more difficult for users to secure a device by adjusting user-facing settings.

\medskip
\begin{tcolorbox}
\begin{figure}[H]
    \caption{List of iCloud Data Accessible by Apple}\label{fig:icloud_weak}
\begin{itemize*}
\begin{multicols}{2}
    \item Safari History \& Bookmarks\footnote{History is stored end-to-end encrypted on iOS 13 and later.}
    \item Calendars
    \item Contacts
    \item Find My\footnote{List of devices and people enrolled in Find My, an Apple service for physically locating enrolled devices~\cite{findmy}.}
    \item iCloud Drive\footnote{Used for document storage for Apple's office suite (Pages, Keynote, and Numbers).}
    \item Messages in iCloud\footnote{Unless iCloud Backup is disabled, refer to the text for additional details.}
    \item Notes
    \item Photos
    \item Reminders
    \item \Gls{Siri} Shortcuts
    \item Voice Memos
    \item Wallet Passes\footnote{Items such as ID cards, boarding passes for airlines, and other supported content.}
\end{multicols}
\end{itemize*}
\caption*{Source: Apple iCloud Security Guide~\cite{apple_icloud_security}}
\end{figure}
\end{tcolorbox}

\begin{tcolorbox}
\begin{figure}[H]
    \caption{List of iCloud Data Encrypted ``End-to-End''}\label{fig:icloud_strong}
\begin{itemize*}
\begin{multicols}{2}
    \item Apple Card transactions
    \item Home data
    \item Health data
    \item iCloud Keychain\footnote{See ``iCloud Keychain'' in this section.}
    \item Maps data\footnote{Favorite locations and other location history including searches.}
    \item Memoji\footnote{Generated emoji-style images.}
    \item Payment information
    \item Quicktype\footnote{Word suggestions above the keyboard on iOS, introduced in iOS 8~\cite{apple_quicktype}.} Keyboard learned vocabulary
    \item Safari History and iCloud Tabs
    \item Screen Time
    \item \Gls{Siri} information\footnote{It is not clear what specific data this contains, the documentation is vague.}
    \item Wi-Fi passwords
    \item W1 and H1\footnote{Apple product names for wireless-enabled integrated circuits.} Bluetooth keys
\end{multicols}
\end{itemize*}
\caption*{Source: Apple iCloud Security Guide~\cite{apple_icloud_security}}
\end{figure}
\end{tcolorbox}

\paragraph{CloudKit} Third-party developers can also integrate with iCloud in iOS applications via CloudKit, an \Gls{API} which allows applications to access cloud storage with configurable properties that allow data to be shared in real-time across multiple devices~\cite{apple_developer_docs}. CloudKit data is encrypted into one or more ``containers'' under a key hierarchy similar to Data Protection. The top-level key in this hierarchy is the ``CloudKit Service Key.'' This key is stored in the synchronized user Keychain, inaccessible to Apple, and is rotated any time the user disables iCloud Backup~\cite{apple_platform_security,apple_icloud_security,apple_cloudkit_security}.

\paragraph{iCloud Keychain} iCloud Keychain extends iCloud functionality to provide two services for Apple devices: $(1)$ Keychain synchronization across devices, and $(2)$ Keychain recovery in case of device loss or failure~\cite{apple_platform_security}. 
\begin{enumerate}
    \item Keychain syncing enables trusted devices (see below) to share Keychain data with one another using asymmetric encryption~\cite{apple_platform_security}. The data available for Keychain syncing includes Safari user data (usernames, passwords, and credit card numbers), Wi-Fi passwords, and HomeKit\footnote{Used for control of smart home devices.} encryption keys. Third-party applications may opt-in to have their Keychain data synchronized.
    \item Keychain recovery allows a user to escrow an encrypted copy of their Keychain with Apple~\cite{apple_platform_security}. The Keychain is encrypted with a ``strong passcode'' known as the iCloud Security code, discussed below.
\end{enumerate}

\noindent
Apple documentation~\cite{apple_tfa} defines a ``trusted device'' as:

\begin{quote}
``an iPhone, iPad, or iPod touch with iOS 9 and later, or Mac with OS X El Capitan and later that you've already signed in to using two-factor authentication. It’s a device we know is yours and that can be used to verify your identity by displaying a verification code from Apple when you sign in on a different device or browser.'' 
\end{quote}

In contrast with standard iCloud and iCloud backups, iCloud Keychain provides additional security guarantees for stored data. This is enforced through the use of trusted Hardware Security Modules (\Gls{HSM}s) within Apple's back-end data centers. This system is designed by Apple to ensure that even Apple itself cannot access the contents of iCloud Keychain backups without access to the iCloud Security Code. This code is generated either from the user's passcode if Two-Factor Authentication (\Gls{2FA}) is enabled, or chosen by the user (optionally generated on-device) if \Gls{2FA} is not enabled. When authenticating to the Hardware Security Modules (\Gls{HSM}s) which protect iCloud Keychain recovery, the iCloud Security Code is never transmitted, instead using the Secure Remote Password (SRP) protocol~\cite{SRP}. After the \Gls{HSM} cluster verifies that 10 failed attempts have not occurred, it sends a copy of the encrypted Keychain to the device for decryption. If 10 attempts is exceeded, the record is destroyed and the user must re-enroll in iCloud Keychain to continue using its features.\footnote{Re-enrolling appears to be a matter of toggling the selector from off to on in Settings on an iOS device.} As a final note, the software installed on Apple \Gls{HSM}s can run must be digitally signed; Apple asserts that the signing keys required to further alter the software are physically destroyed after \Gls{HSM} deployment~\cite{apple_platform_security,apple_bh_2016}, preventing the company from deliberately modifying these systems to access user data.

\begin{figure}
    \centering
    \includegraphics[width=0.9\linewidth]{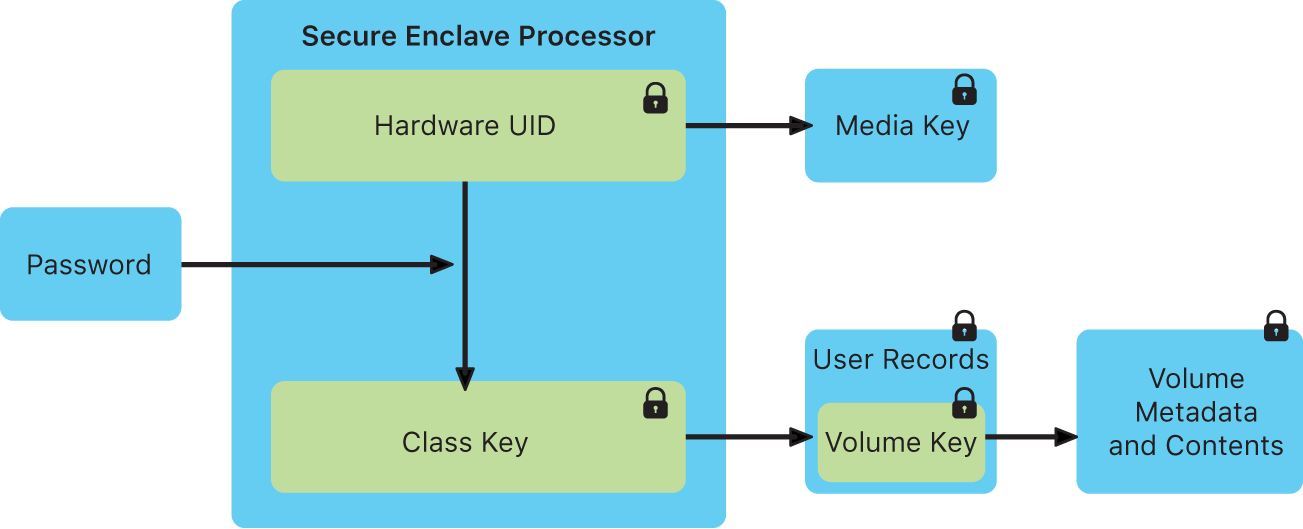}
    \caption{Secure Enclave Processor Key Derivation}\label{img:sep_encryption}
    \caption*{Source: Apple~\cite{apple_platform_security}}
\end{figure}

\paragraph{Trusted hardware} Apple provides multiple dedicated components to support encryption, biometrics and other security functions. Primary among these is the Secure Enclave Processor (\Gls{SEP}), a dedicated co-processor which uses encrypted memory and handles processes related to user authentication, secret key management, encryption, and random number generation. The \Gls{SEP} is intended to enable security against even a compromised iOS \gls{kernel} by executing cryptographic operations in separate dedicated hardware. The \Gls{SEP} negotiates keys to communicate with \Gls{TouchID} and \Gls{FaceID} subsystems (the touch sensor or the Neural Engine for facial recognition)~\cite{apple_faceid_security}, provides keys to the Crypto Engine (a cryptography accelerator), and communicates with a ``secure storage integrated circuit'' for random number generation, anti-replay counters, and tamper resistance~\cite{demystifying_sep_2016}.\footnote{The ``SEPOS'' root task of the \Gls{SEP}, as well as various drivers, services, and applications which run in the \Gls{SEP} were presented in-depth at Black Hat 2016~\cite{demystifying_sep_2016}; in this work the authors elucidate the inner workings, software security mitigations implemented (with some notably absent, including ASLR and Heap metadata protection), and attack surface of the \Gls{SEP} as implemented in iPhone 6S/iOS 9.} The image in Figure~\ref{img:sep_encryption} documents the encryption keys the \Gls{SEP} derives from the user passcode.

\begin{figure}
    \centering
    \includegraphics[width=0.9\linewidth]{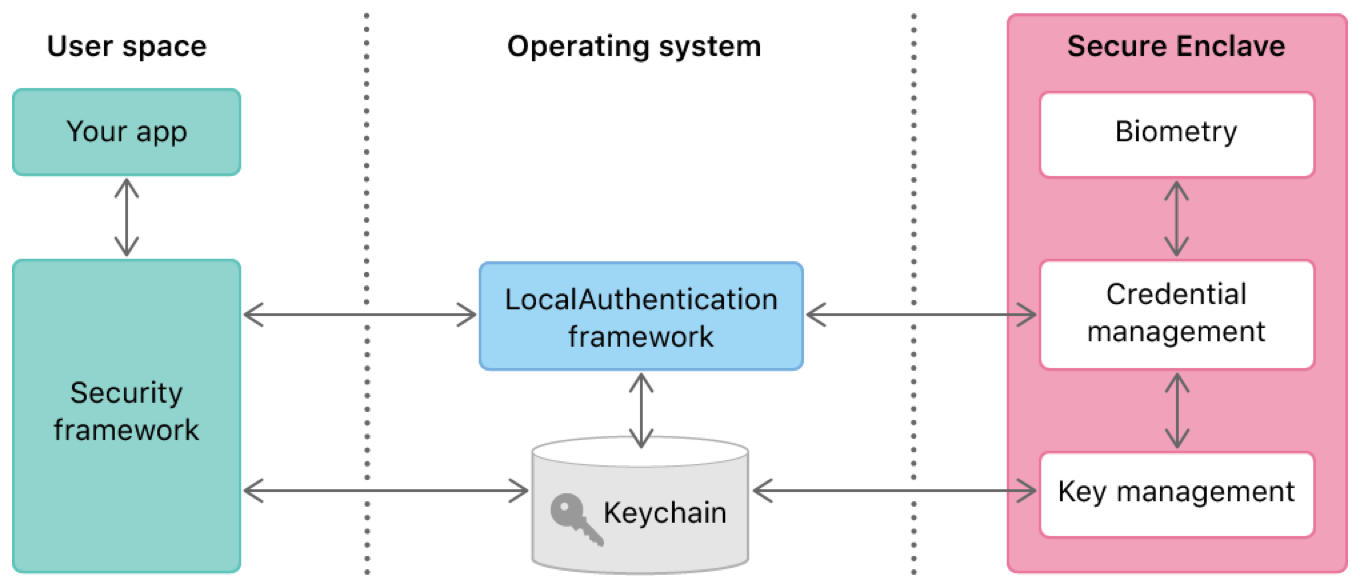}
    \caption{\texttt{LocalAuthentication} Interface for TouchID/FaceID}\label{img:faceid_sep}
    \caption*{Source: Apple~\cite{apple_localauth}}
\end{figure}

Apple documentation describes \Gls{FaceID} as leveraging the Neural Engine and the \Gls{SEP}: ``A portion of the A11 Bionic processor’s neural engine—protected within the Secure Enclave—transforms this data into a mathematical representation and compares that representation to the enrolled facial data.''~\cite{apple_faceid_security} As worded, this description fails to fully convey the features of hardware which make this possible, and as such we must speculate as to the exact method by which \Gls{FaceID} data is processed and secured. For example, it is possible that the Neural Engine simply provides data directly to the \Gls{SEP}, or via the application processor (AP). It's also possible that, similar to \Gls{TouchID}, the Neural Engine creates a shared key with the \Gls{SEP} and then passes encrypted data through the AP. Physical inspection of the iPhone X (the first generation with F\Gls{FaceID}) suggests that the Neural Engine and \Gls{SEP} are inside the A11 package~\cite{ifixit_a11}, as in the earlier design with the \Gls{SEP} and A7 \Gls{SoC} in iPhone 5s~\cite{ifixit_a7}. iPhones 6 and later (among other contemporaneous devices e.g. Apple Watches and some iPads) additionally support Apple Pay and other NFC/Suica\footnote{A Japanese contactless/wireless protocol and rough equivalent to NFC} communication of secrets via the Secure Element, a wireless-enabled chip which stores encrypted payment card and other data~\cite{apple_secureelement}. The image in Figure~\ref{img:faceid_sep} documents the \texttt{LocalAuthentication} framework through which \Gls{TouchID} and \Gls{FaceID} operate.

\paragraph{Restricting attack surface} A locked iOS device has very few methods of data ingress available: background downloads such as sync (email, cloud services, etc), Push notifications, certain network messages via Wi-Fi, Bluetooth, or the cellular network, and the physical \Gls{Lightning} port on the device. Exploits delivered to a locked device must necessarily use one of these avenues. With the introduction of \Gls{USB} Restricted Mode, Apple sought to limit the access of untrusted \Gls{USB} \Gls{Lightning} devices to access or interact with iOS, such as forensic software~\cite{elcomsoft_usb}. By reducing attack surface, iOS complicates or mitigates attacks. This protection mode simply disables \Gls{USB} communication for unknown devices after the iOS device is locked for an hour without \Gls{USB} accessory activity, or upon first use of a \Gls{USB} accessory; known devices are remembered for 30 days~\cite{elcomsoft_usb_bypass}. If no accessories are used for three days, the lockout period is reduced from an hour to immediately after device lock~\cite{elcomsoft_usb_bypass}. However, this protection is not complete, as discussed in \S\ref{sec:ios_bypass}.

\paragraph{iMessage and FaceTime} Each iPhone ships with two integrated communication packages: \gls{iMessage} and \Gls{FaceTime}. The \gls{iMessage} platform is incorporated within the Messages application as an alternative to \Gls{SMS}/\Gls{MMS}, and provides end-to-end encrypted messaging communications with other Apple devices. The FaceTime app provides end-to-end encrypted audio and videoconferencing between Apple devices. \gls{iMessage} messages are encrypted using a ``signcryption'' scheme recently analyzed by Bellare and Stepanovs~\cite{bellare2020security} with the Apple Identity Service serving as a trusted authority for user public keys~\cite{apple_platform_security}. \Gls{FaceTime} calls are encrypted using a scheme that Apple has not documented, but claims is forward-secure\footnote{Previous call content cannot be accessed even if a device or key is compromised.} based on the Secure Real-time Transport Protocol (SRTP)~\cite{srtp}. Although these protocols are end-to-end encrypted and authenticated, they rely on the Apple Identity Service to ensure that participants are authentic~\cite{apple_platform_security}.

\section{History of Apple Security Features}

The current hardware and software security capabilities of iOS devices are many and varied, ranging from access control to cryptography. These features were incrementally developed over time and delivered in new versions of iOS pre-loaded on or delivered to\footnote{First via iTunes, and then over-the-air via Wi-Fi since iOS 5 in 2011.} devices. An overwhelming percentage of iOS users update their devices: in June 2020, Apple found that over 92\% of recent\footnote{Sold new in the last four years.} iPhones ran iOS 13, and almost all the remainder (7\%) ran iOS 12~\cite{apple_ios_update_adoption}; iOS 13 was released 9 months prior. The implication of this is that while most users receive the latest mitigations, users of older devices may receive only partial security against known attacks, especially when using devices which have reached end-of-life (no longer receiving updates from Apple). To make these limitations apparent, in Appendix~\ref{app:ios_history} we provide a detailed overview of the historical deployment of new security features, as described by published Apple documents ~\cite{apple_security_guides,apple_platform_security,apple_icloud_security,apple_cloudkit_security,apple_tfa,apple_faceid_security,apple_secureelement,apple_security_updates,apple_usb}. This history is summarized in Tables~\ref{tbl:ios_history} and~\ref{tbl:ios_history_hw}.

\begin{tcolorbox}
\begin{table}[H]
\begin{center}
\begin{tabular}{p{0.1\linewidth}|p{0.34\linewidth}|p{0.50\linewidth}}
     \textbf{iOS\footnote{(Model), alternate models omitted.}} & \textbf{Data Protection}\footnote{See Figure~\ref{list:ios_dp_classes}.} & \textbf{Notes} \\ \hline
     1 (2G) & - & 4-digit passcodes \\ \hline
     2 (3G) & - & Option to erase user data after 10 failed passcode attempts introduced \\ \hline
     3 (3GS) & DP introduced &  Encrypted flash storage when device off \\ \hline
     4 (4) & Mail \& Attachments: PUO &  PUO insecure due to implementation error on iPhone 4 until iOS 7~\cite{apple_security_updates} \\ \hline
     5 (4S) & - &  \Gls{AES}-\Gls{GCM} replaces \Gls{CBC} in Keychain \\ \hline
     6 (5) & iTunes Backup: \Gls{CP}+NM\newline Location Data: \Gls{CP}\newline Mail Accounts: \Gls{AFU}\newline \gls{iMessage} keys: NP+NM &  Various other data at \Gls{AFU} or NP+NM class \\ \hline
     7 (5S) & Safari Passwords: \Gls{CP}\footnote{Possibly earlier, documentation ambiguous.}\newline Authentication tokens:\footnote{Social media accounts and iCloud.} \Gls{AFU}\newline Default and third-party apps: \Gls{AFU} &  \Gls{SEP} and \Gls{TouchID} introduced in iPhone 5S\newline  Third-party apps may opt-in to higher classes \\ \hline
     8 (6) & App Documents: \Gls{CP}\newline Location Data: \Gls{AFU}?\footnote{Documentation includes notably weakened language.} &  User passcode mixed into encryption keys\newline \Gls{XTS} replaces \Gls{CBC} for storage encryption \\ \hline
     9 (6S) & Safari Bookmarks: \Gls{CP} &  6-digit passcode default introduced \\ \hline
     10 (7) & Clipboard: ?\footnote{Documentation ambiguous.}\newline iCloud private key: NP+NM & - \\ \hline
     11 (8 \& X) & - &  \Gls{FaceID} introduced in iPhone X\newline  \Gls{SEP} memory adds an ``integrity tree'' to prevent replay attacks\newline  \Gls{USB} Restricted Mode introduced to mitigate exploits delivered over \Gls{Lightning} connector \\ \hline
     12 (XS) & - &  \Gls{SEP} enforces Data Protection in DFU (Device Firmware Upgrade) and Recovery mode to mitigate bypass via bootloader \\ \hline
     13 (11) & - & - \\ \hline
\end{tabular}
\def\arraystretch{1}
\end{center}
\caption{History of iOS Data Protection}\label{tbl:ios_history}
\caption*{Source: Apple iOS and Platform Security Guides~\cite{apple_security_guides,apple_platform_security}}
\end{table}
\end{tcolorbox}

\begin{tcolorbox}
\begin {table}[H]
\begin{center}
\begin{tabular}{p{0.2\linewidth}|p{0.26\linewidth}|p{0.48\linewidth}}
     \textbf{\Gls{SoC}\footnote{System-on-a-Chip.}\footnote{(Model), alternate models omitted.}} & \textbf{Hardware Changes} & \textbf{Notes} \\ \hline
     Samsung \Gls{ARM}-32 \Gls{CPU} (2G) & - & No dedicated hardware security components in the early phones \\ \hline
     Samsung \Gls{ARM}-32 \Gls{CPU} (3G) & - & - \\ \hline
     Samsung Cortex \Gls{SoC} (3GS) & - & Flash storage encryption driven by application processor (AP) \\ \hline
     Apple A4 (4) & - &  Shift to Apple-designed \Gls{SoC}s \\ \hline
     A5 (4S) & Crypto Engine & Dedicated cryptography accelerator documented here, potentially included in earlier generations \\ \hline
     A6 (5) & - & - \\ \hline
     A7 (5S) & \Gls{TouchID} and Secure Enclave Processor (\Gls{SEP}) & Major change, including new \Gls{UID key} inside \Gls{SEP} and shared key between \Gls{SEP} and \Gls{TouchID} sensor \\ \hline
     A8 (6) & - &  Significant software changes in this generation which rely on hardware changes of A7 \\ \hline
     A9 (6S) & Bus between flash and memory ``isolated via the Crypto Engine'' &  Interpreting this documentation implies that hardware changed to physically enforce flash storage encryption \\ \hline
     A10 Fusion (7) & - & - \\ \hline
     A11 Bionic (8/X) & \Gls{FaceID} and Neural Engine & Neural Engine somehow integrated with \Gls{SEP} to enable facial recognition secure against malicious AP \\ \hline
     A12 Bionic (XS) & ``Secure Storage Integrated Circuit'' & SSIC added to bolster \Gls{SEP} replay protection, \Gls{RNG}, and tamper detection \\ \hline
     A13 Bionic (11) & - & - \\ \hline
\end{tabular}
\def\arraystretch{1}
\caption {History of iPhone Hardware Security}\label{tbl:ios_history_hw}
\caption*{Source: Apple iOS and Platform Security Guides~\cite{apple_security_guides,apple_platform_security} and iFixit Teardowns~\cite{ifixit_a11,ifixit_a7,ifixit_a4,ifixit_a5,ifixit_a6,ifixit_a8,ifixit_a9,ifixit_a10,ifixit_a12,ifixit_a13}}
\end{center}
\end{table}
\end{tcolorbox}

\section{Known Data Security Bypass Techniques}\label{sec:ios_bypass}

As each iteration of Apple device introduces or improves on security features, the commercial exploit/forensics and \gls{jailbreak}ing communities have reacted by developing new techniques to bypass those features. Bypassing iPhone protections is an attractive goal for threat actors and law enforcement agencies alike. Hackers can receive bug bounties amounting to six or seven digits for viable exploits~\cite{apple_bounty,zerodium_bounty}; rogue governments can buy or develop and use malware to track human rights activists or political opponents~\cite{pegasus_ios}; and law enforcement agencies can create or bolster cases through investigation of phone contents, with forensic software companies signing lucrative~\cite{san_bernardino_news} contracts to help them do so~\cite{elcomsoft_methods,cellebrite_advanced_services}.

Law enforcement agencies in many jurisdictions may also provide legal requests for data. In pursuing compliance with these laws, Apple provides data to law enforcement when legal requests are made. Figure~\ref{fig:apple_legal} documents data which Apple claims to be able or not able to provide to U.S. law enforcement~\cite{apple_legal}. This method of access requires legal process, and certain information being provided to Apple, and some requests may be rejected under various circumstances~\cite{apple_transparency}. As such, this method of data collection may be supplemented or even entirely superseded by commercial data forensics methods described in \S\ref{sec:devicehacks} and \S\ref{sec:cloudhacks}.

\paragraph{Jailbreaks and software exploits} The primary means to bypass iOS security features is through the exploitation of software vulnerabilities in apps and iOS software. \Gls{jailbreak}s are a class of such exploits central to forensic analysis of iOS. The defining feature of a \gls{jailbreak} is to enable running unsigned code, such as a modified iOS \gls{kernel} or apps that have not been approved by Apple~\cite{cydia} or a trusted developer~\cite{apple_enterprise}. Other software exploits which pertain to user data privacy (speculatively) include \Gls{SEP} exploits, exploits which may enable passcode brute-force guessing and lock-screen bypasses. All of these exploits are discussed in~\S\ref{sec:jailbreaking}.

\paragraph{Local device forensic data extraction} Once a device has been made accessible using a software exploit, actors may require technological assistance to perform forensic analysis of the resulting data. This requires tools that extract data from a device and render the results in a human-readable form~\cite{nist_forensics_spec,dhs_forensics}. This latter niche has been actively filled by a variety of private companies, whose software is tested and used by the U.S. Department of Homeland Security (\Gls{DHS}) and local law enforcement, with public reporting on its effectiveness~\cite{dhs_forensics} (refer to Figure~\ref{list:nist_forensics} for the data targeted by mobile forensic software). \Gls{DHS} evaluations of forensic software tools reveal that, at least in laboratory settings, significant portions of the targeted data is successfully extracted from supported devices. Refer to~\S\ref{sec:devicehacks} for discussion of the use of such tools, and to~\S\ref{sec:ios_forensics} for further examination of the history of forensic software.

\paragraph{Cloud forensic data extraction} Cloud integrations such as Apple iCloud enable valuable features such as backup and sync. They also create a data-rich pathway for information extraction, and represent a target for search warrants themselves. The various ways these services can be leveraged by hackers or law enforcement forensics are discussed in~\S\ref{sec:cloudhacks}.

\begin{tcolorbox}
\begin{figure}[H]
\begin{center}\caption{Apple Documentation on Legal Requests for Data}\label{fig:apple_legal}\end{center}
\begin{multicols}{2}
\textbf{Data Apple makes available to law enforcement:}
\begin{center}
\begin{itemize*}
\item Product registration information\footnote{Including name, address, email address, and telephone number.}
\item Customer service records
\item iTunes subscriber information
\item iTunes connection logs (IP address)
\item iTunes purchase and download history
\item Apple online store purchase information\footnote{Linked to iCloud account.}
\item iCloud subscriber information
\item iCloud email logs (metadata)
\item iCloud email content
\item iCloud photos
\item iCloud Drive documents
\item Contacts
\item Calendars
\item Bookmarks
\item Safari browsing history
\item Maps search history
\item Messages (SMS/MMS)
\item Backups (app data, settings, and other data)
\item Find My~\cite{findmy} connections and transactions\footnote{Remote lock or erase requests}
\item Hardware address (MAC) for a given iPhone\footnote{By serial number or mobile subscriber information (IMEI, MEID, UDID).}
\item \Gls{FaceTime} call invitations\footnote{With the caveat that these do not imply any communication took place.}
\item \gls{iMessage} capability query logs\footnote{When an iOS device encounters a potential iMessage handle (phone number, email address, or Apple ID), it queries Apple to determine if the contact is able to use \gls{iMessage}.}
\end{itemize*}
\end{center}

\textbf{Data Apple claims is unavailable:}
\begin{center}
\begin{itemize*}
\item Find My location data
\item Full data extractions/user passcodes on iPhone 6/iOS 8.0 and later
\item \Gls{FaceTime} call content
\item \gls{iMessage} content
\end{itemize*}
\end{center}
\end{multicols}
\begin{center}\caption*{Source: Apple Legal~\cite{apple_legal}}
\end{center}
\end{figure}
\end{tcolorbox}

\subsection{Jailbreaking and Software Exploits}\label{sec:jailbreaking}

\paragraph{Jailbreaking} While \gls{jailbreak}s are often used for customization purposes, the underlying technology can also be used by third parties in order to bypass software protection features of a target device, for example to bypass passcode lock screens or to forensically extract data from a locked device. Indeed, many commercial forensics packages make use of public \gls{jailbreak}s released for this purpose~\cite{elcomsoft_keychain}. A commonality among these techniques is the need to deliver an ``exploit payload'' to a vulnerable software component on the device.  Viable delivery mechanisms include physical ports such as the device's \Gls{USB}/\Gls{Lightning} interface (much less viable if \Gls{USB} Restricted Mode is active~\cite{apple_usb}). Alternatively, exploits may be delivered through data messages that are received and processed by iOS and app software, for example specially-crafted text messages or web pages. In many cases, multiple separate exploits are combined to form an ``exploit chain:'' the first exploit may obtain control of one software system on the device, while further exploits may escalate control until the \gls{kernel} has been breached. Once the \gls{kernel} has been exploited, \gls{jailbreak}s usually deploy a patch to allow unsigned code to run or to initiate custom behavior such as extraction of the filesystem~\cite{elcomsoft_keychain}. 

\Gls{jailbreak}ing is a keystone of constructing bypasses to access user data due to the fact that the iOS \gls{kernel} is ultimately tasked with managing and retrieving sensitive data. As such, a \gls{kernel} compromise often allows the immediate extraction of any data not explicitly protected by encryption using keys which cannot be derived from the application processor alone. Publicly-known \gls{jailbreak}s are released by a seemingly small group of exploit developers.\footnote{As seen in Table~\ref{tbl:ios_history_jb}.} \Gls{jailbreak}s are released targeting a specific iOS version, and more rarely target specific hardware (e.g. iPhone model). Apple periodically releases software updates which patch some subset of the vulnerabilities distributed in these \gls{jailbreak}s~\cite{apple_security_updates}, and a process ensues in which exploit developers replace patched vulnerabilities with newly discovered, still-exploitable alternatives, until a major software change occurs (e.g. a new iOS major version). Table~\ref{tbl:ios_history_jb} provides the highlights of the history of \gls{jailbreak}ing in iOS, with many of these iterative updates omitted.

\Gls{jailbreak}ing was relatively popular in 2009. Exact counting is nearly impossible, but it was estimated that 10\% of iOS devices in 2009 were jailbroken~\cite{jb_analysis}.\footnote{Representing 10.5 to 12 million devices.} In 2013, roughly 23 million devices ran Cydia~\cite{cydia}, a popular software platform commonly used on jailbroken iOS devices~\cite{jb_analysis}. 150 million iPhones were sold that year~\cite{ios_sales_2013} and total iPhone sales accelerated tremendously between 2009 and 2013~\cite{apple_total_sales}, and as such it is speculatively likely, though hard to measure, that the percentage of jailbroken devices declined notably. Some analysis has been undertaken as to reasons for declining \gls{jailbreak}ing, if this even is a trend~\cite{jb_analysis}. Around 2016 (refer to Table~\ref{tbl:ios_history_jb}) there was a marked transition in \gls{jailbreak}s away from end-user usability and towards support for use by security researchers. Some of the more well-maintained \gls{jailbreak}s did include single-click functionality or re-\gls{jailbreak}ing after reboot via a sideloaded\footnote{Installed via personal computer rather than the official App Store.} app. It is possible that this transition took place in part due to the commercial viability of \gls{jailbreak} production. As the market is relatively inflexibly supplied, prices for working \gls{jailbreak}s increase directly with demand~\cite{zerodium_bounty}, and as such creators are less inclined to share exploits which are necessary for \gls{jailbreak}ing publicly. The kinds of exploits needed for forensics against a locked phone, specifically those which exploit an interface on the locked phone (commonly the \Gls{USB}/\Gls{Lightning} interface and related components), would be highly valuable to a forensics software company which at a given time did not have a working exploit of their own.

The checkm8/checkra1n \gls{jailbreak} exploits seem to be widely implemented in forensic analysis tools in 2020. These exploits work on iOS devices up to iPhone X (and any with A11 hardware iterations) \textit{regardless of iOS version} and as such are widely applicable and thus useful~\cite{checkm8}. Cellebrite Advanced Services (their bespoke law enforcement investigative service) offers Before-First-Unlock access\footnote{Among other less technically challenging offerings.} to iPhones X and earlier running up to the latest iOS~\cite{cellebrite_advanced_services}, and as such we are relatively certain they are employing checkm8.

Although they do not refer to it as \gls{jailbreak}ing, the exploits used in the Cellebrite UFED Touch and 4PC products either exploit the backup system to initiate a backup despite lacking user authorization, or exploit the bootloader to run custom Cellebrite code which extracts portions of the filesystem~\cite{cellebrite_ufed}. We categorize these as equivalent due to the practical implementation and impact similarities.

\paragraph{Passcode guessing} To access records on devices that are not in the \Gls{AFU} state, or to access data that has been protected using the \Gls{CP} class, actors may need to recover the user's passcode in order to derive the necessary decryption keys.

There are two primary obstacles to this process: first, because keys are derived from a combination of the hardware \Gls{UID key} and the user's passcode, keys must be derived {\em on the device}, or the \Gls{UID key} must be physically extracted from silicon.\footnote{Private communications from Apple engineers claim that the \Gls{UID key} cannot be obtained directly by software, and can only be used as input to the \Gls{AES} engine. This is claimed to be enforced through the silicon design, rather than through software, ensuring that only expensive physical extraction can obtain the raw UID value.} There is no public evidence that the latter strategy is economically feasible. The second obstacle is that the iPhone significantly throttles attackers' ability to conduct passcode guessing attacks on the device: this is accomplished through the use of guessing limits enforced (on more recent phones) by the dedicated \Gls{SEP} processor, as well as an approximately 80 millisecond password derivation time enforced by the use of a computationally-expensive key derivation function. 

In older iPhones that do not include a \Gls{SEP}, passcode verification and guessing limits were enforce by the application processor. Various bugs~\cite{apple_security_updates} in this implementation have enabled attacks which exploit the passcode attempt counter to prevent it from incrementing or to reset it between attempts. With four- and six-digit passcodes, and especially with users commonly selecting certain passcodes~\cite{ios_user_passcodes}, these exploits made brute-forcing the phone feasible for law enforcement. One particularly notable example of passcode brute-forcing from a technical perspective was contributed by Skorobogatov in 2016~\cite{skorobogatov2016bumpy}. In this work, the authors explore the technical feasibility of mirroring flash storage on an iPhone 5C to enable unlimited passcode attempts. Because the iPhonee 5C does not include a SEP with tamper-resistant NVRAM, the essence of the attack is to replace the storage carrying the retry counter in order to reset its value. The authors demonstrate that the attack is indeed feasible, and inexpensive to perform. Even earlier attacks include cutting power to the device when an incorrect passcode is entered to preempt the counter before it increments~\cite{ios_poweroff}, although these have largely been addressed.\footnote{A standard countermeasure introduced in non-SEP phones requires the device to increment the guess counter {\em prior} to verifying the passcode, ensuring that data must be written to storage.}

For extremely strong passwords (such as random alphanumeric passcodes, although these are relatively rarely used~\cite{ios_user_passcodes}), the 80ms guessing time may render passcode guessing attacks completely infeasible regardless of whether the SEP exists and is operating. For lower-entropy passcodes such as the default 6-digit numeric PIN, the \Gls{SEP}-enforced guessing limits represent the primary obstacle.\footnote{Without these limits, an attacker could guess all possible six-digit PINs in a matter of hours.} Bypassing these limitations requires techniques for overcoming the \Gls{SEP} guessing limits. We provide a detailed overview of the evidence for and against the in-the-wild existence of such exploits in \S\ref{sec:ios_bypass}. Refer to Table~\ref{tbl:passcode_times} for estimated passcode brute-force times under various circumstances.

\medskip
\begin{tcolorbox}
\begin{table}[H]
    \centering
    \begin{tabular}{p{0.24\linewidth}|p{0.17\linewidth}|p{0.2\linewidth}|p{0.16\linewidth}|p{0.11\linewidth}}
        \textbf{Passcode Length} & \textbf{4 (digits)} & \textbf{6 (digits)} & \textbf{10 (digits)} & \textbf{10 (all)}\footnote{The iOS US QWERTY keyboard allows around 114 character inputs.} \\\hline
        \textbf{Total Passcodes} & $10^4 = 10,000$ & $10^6 = 1,000,000$ & $10^{10}$ & $3.7 \times 10^{20}$\\\hline
        \textbf{Total Allowed}\footnote{iOS prevents users from choosing certain 4- or 6-digit passcodes considered too ``easily guessed.'' For more information, refer to~\cite{ios_user_passcodes}.} & 9,276 & 997,090 & - & -\\\hline
        \textbf{80 ms/attempt}\newline \textit{in expectation} & 12.37 minutes\newline \textit{6.19 minutes} & 22.16 hours\newline \textit{11.08 hours} & $\sim$25 years\newline\textit{$\sim$12 years} & \footnote{Roughly 68 times the age of the universe.} \\\hline
        \textbf{10 mins/attempt}\footnote{From an Elcomsoft article which claimed this rate using GrayKey on a before-first-unlock device~\cite{elcomsoft_knownpasscode}.}\newline \textit{in expectation} & $\sim$70 days\newline $\sim$\textit{35 days} & $\sim$20 years\newline $\sim$\textit{10 years} & $\sim$200,000 yrs. & \footnote{Inestimable.} \\\hline
    \end{tabular}
    \caption{Passcode Brute-Force Time Estimates}\label{tbl:passcode_times}
    \caption*{A more intelligent passcode guessing strategy could succeed much more quickly.}
\end{table}
\end{tcolorbox}

\paragraph{Exploiting the \Gls{SEP}} The Secure Enclave Processor is a separate device that runs with unrestricted access to pre-configured regions of memory of the iOS device~\cite{demystifying_sep_2016} and as such a \Gls{SEP} exploit provides an even greater level of access than an OS \gls{jailbreak}. Moreover, exploitation of the \Gls{SEP} is the most likely means to bypass security mechanisms such as passcode guessing limits. In order to interface with the \Gls{SEP} with sufficient flexibility, \gls{kernel} privileges are required~\cite{demystifying_sep_2016}, and thus a \gls{jailbreak} is likely a prerequisite for \Gls{SEP} exploitation. In the case that such an exploit chain could be executed, an attacker or forensic analyst might gain unfettered access to the encryption keys and functionality used to secure the device, and thus would be able to completely extract the filesystem.

In 2018, Grayshift,\footnote{\url{https://www.grayshift.com/}} a forensics company based in Atlanta, Georgia, advertised and sold a device they called GrayKey, which was purportedly able to unlock a locked or disabled iPhone by brute-forcing the passcode. Modern iPhones allegedly exploited in leak GrayKey demonstration photos~\cite{graykey_news_18} included a \Gls{SEP}, meaning that for these phones, brute-force protections should have prevented such an attack. As such, we speculate that GrayKey may have embedded not only a \gls{jailbreak} but also a \Gls{SEP} exploit in order to enable this functionality. Other comparable forensic tools seemed only able to access a subset of the data which GrayKey promised at the time~\cite{cellebrite_advanced_services,dhs_forensics,elcomsoft_keychain} which provides additional circumstantial evidence that \Gls{SEP} exploits, if they existed, are rare. 

\begin{figure}[H]
    \centering
    \includegraphics[height=0.24\textheight]{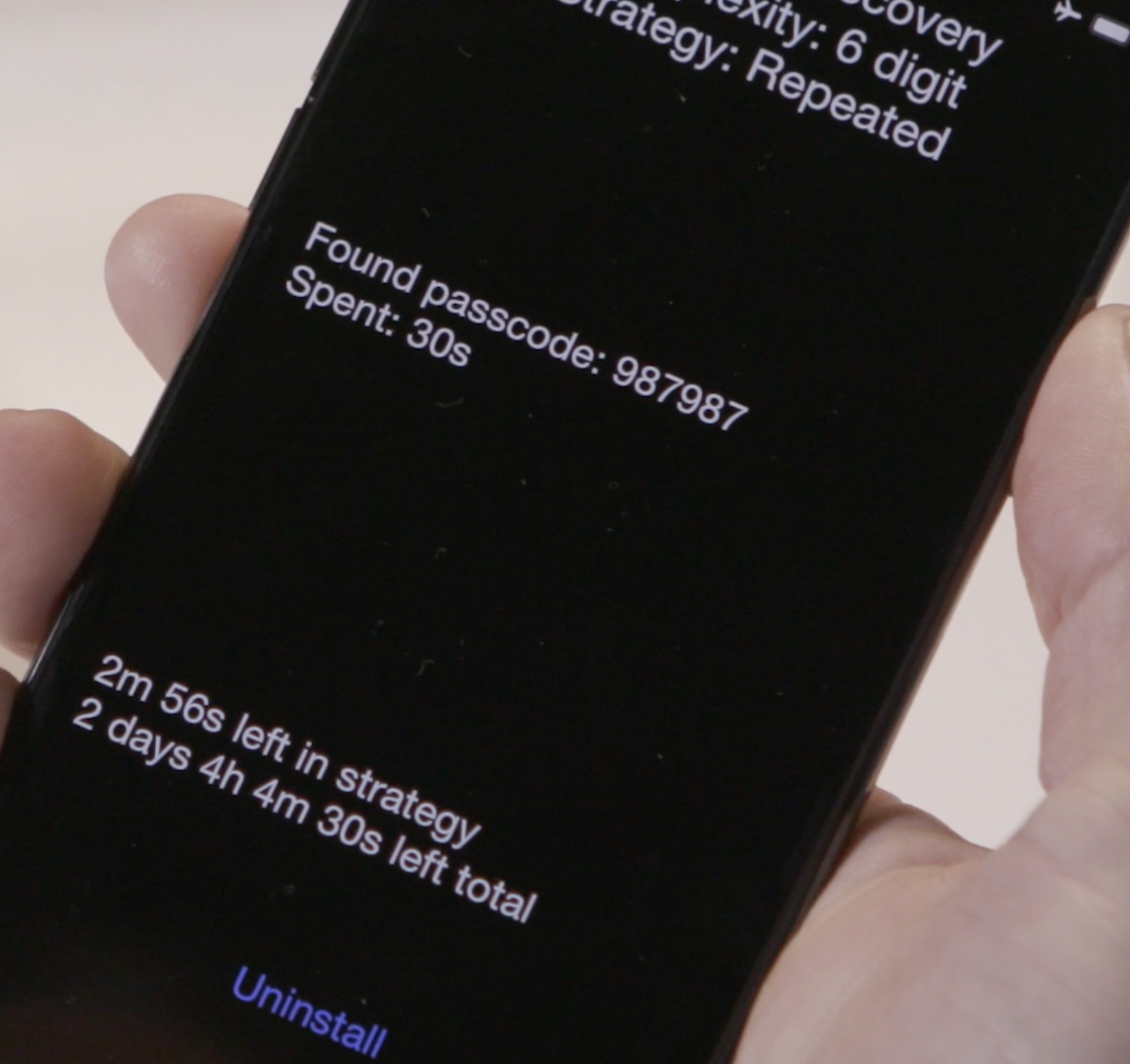}
    \includegraphics[height=0.24\textheight]{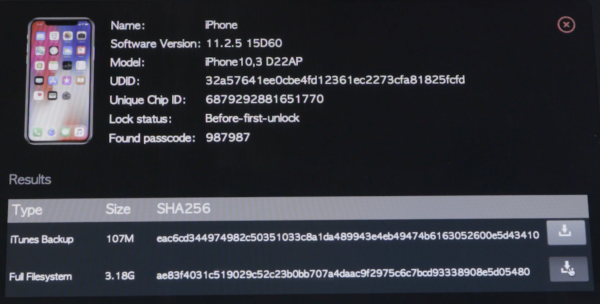}
    \caption{Alleged Leaked Images of the GrayKey Passcode Guessing Interface}\label{img:graykey_passcode}
    \caption*{Left: passcode guessing on an iPhone X. Right: forensic extraction. Source: MalwareBytes SecurityWorld~\cite{reed2018graykey}}
\end{figure}

A 2018 article from the company MalwareBytes~\cite{graykey_news_18} provided an alleged screenshot of an iPhone X (containing a \Gls{SEP}) running iOS 11.2.5 (latest at the time) in a before-first-unlock state (all data in the \Gls{AFU} and \Gls{CP} protection classes encrypted, with keys evicted from memory and thus unavailable to the \gls{kernel}). The images indicate that the GrayKey exploit had successfully executed a guessing attack on a 6-digit passcode, with an estimated time-to-completion of approximately 2 days, 4 hours.
The images also show a full filesystem image and an iTunes backup extracted from it, which should only be possible if the passcode was known~\cite{apple_security_guides,apple_platform_security} or somehow extracted. As the time to unlock increases with passcode complexity, we presume that GrayKey is able to launch a brute-force passcode guessing attack from within the exploited iOS, necessitating a bypass of \Gls{SEP} features which should otherwise prevent such an attack. Figure~\ref{img:graykey_passcode} depicts the Graykey passcode guessing and extraction interfaces.

In August 2018, Grayshift unlocked an iPhone X with an unknown 6-digit passcode given to them by the Seattle Police Department; Grayshift was reportedly able to break the passcode in just over two weeks~\cite{upturn_mass_extraction}. Further documents~\cite{arizona_le_records} imply extensive use of Graykey to recover passcodes against iOS devices.

In January 2020, an FBI warrant application indicates that GrayKey was used to access a visibly locked iPhone 11 Pro Max~\cite{koch_iphone_news,iphone11_warrant}. The significance of this access is twofold: $(1)$ the iPhone 11 Pro Max, released with iOS 13, is not vulnerable to the checkm8 jailbreak exploit~\cite{checkm8}, and $(2)$ the iPhone 11 includes a SEP. The apparent success of this extraction indicates that Grayshift possessed an exploit capable of compromising iOS 13. There is simply not enough information in these warrants to know if this exploit simply performed a jailbreak and logical forensic extraction of non-CP encrypted data, or if it involved a complete compromise of the SEP and a successful passcode bypass. 

Also in 2020, Grayshift filed with the U.S. Federal Communications Commission (FCC) ID program for certification of the GrayKey device~\cite{graykey_fcc,graykey_fcc_news}. This filing initiated the public release of documents detailing the Radio Frequency (RF) hardware and capabilities of GrayKey, as well as an image of the internals. Analysis\footnote{Including on Twitter: \url{https://twitter.com/parallel_beings/status/1305554386828038146?s=20}} of these reports and images has induced speculation that the GrayKey device first exploits iOS then gains additional access to software systems through post-exploitation tools, potentially including by interfacing with \Gls{JTAG}~\cite{jtag}, a hardware debugging interface. The GrayKey hardware also includes Wi-Fi and Bluetooth capabilities which could be used for updates, data exfiltration, or exploitation of iOS devices.

\begin{tcolorbox}
\begin {table}[H]
\begin{center}
\begin{tabular}{p{0.1\linewidth}|p{0.08\linewidth}|p{0.2\linewidth}|p{0.14\linewidth}|p{0.36\linewidth}}
     \textbf{iOS}\footnote{(Model), alternate models omitted.} & \textbf{Year} & \textbf{\Gls{jailbreak}}\footnote{Not an exhaustive list.} & \textbf{Author(s)} & \textbf{Notes}\\ \hline
     1.1.1-4 (2G) & 2007 & JailbreakMe & iPhone Dev Team\footnote{Not affiliated with Apple.} & - \\ \hline
     2.2-5 (3G-4) & 2008- & QuickPwn, redsn0w & iPhone Dev Team & Supported for multiple versions, renamed redsn0w in 2009 \\ \hline
     3 (3GS) & 2009 & purplera1n, blackra1n~\cite{purplera1n,purplera1n_news} & geohot & First ``unlock,'' meaning the cell network carrier can be changed \\ \hline
     5 (4S) & 2012 & Absinthe & Multiple & Notable collaboration of iPhone Dev Team, Chronic Dev Team, and others \\ \hline
     6-7 (4S) & 2013- & evasi0n, evasi0n7 & evad3rs & -\\ \hline
     7.1-9.3.3 (4S-6S) & 2014- & pangu, pangu8, pangu9~\cite{pangu_news,pangu_bh} & Pangu Team & Extensively presented at hacker conferences such as BlackHat\\ \hline
     8.1-8.4 (6) & 2014 & TaiG & TaiG & -\\ \hline
     10.1-10.2 (6S) & 2016-2017 & yalu + mach\_portal & Luca Todesco, Marco Grassi, Ian Beer\footnote{Developed exploits used, not directly affiliated.} & Not intended for end-user use, usable by experts\\ \hline
     11-11.1.2 & 2017 & LiberiOS & Jonathan Levin & Designed for security experts, lacking an interface for general usability\\ \hline
     11-13.5 & 2018 & unc0ver & pwn20wned, Sam Bingner & Extensively maintained with many bug fixes, features, performance improvements, and counter-patches against Apple fixes\\ \hline
     12-14 & 2019 & checkm8 and {checkra1n}~\cite{checkm8} & Luca Todesco \textit{et al} & Also extensively maintained, working for iOS 14 beta even before its full release\\ \hline
\end{tabular}
\def\arraystretch{1}
\end{center}
\caption{History of \Gls{jailbreak}s on iPhone}\label{tbl:ios_history_jb}
\end{table}
\end{tcolorbox}

\paragraph{Lock screen bypass} Lock screen bypasses tend to be induced exceptional cases discovered seemingly often by end users who explore the lock screen interface beyond developers' intent. These bypasses are characterized by unexpected user inputs to some lock screen-available application (often a camera, clock, or weather application), and tend to require some care to be performed consistently. Table~\ref{tab:ios_ls_bypass} catalogs a history of some notable lock screen bypasses.

Based on our evaluation, there is no evidence these bypasses are used by law enforcement to access personal data. However, as the phone is in a locked state, these bypasses serve as concrete indications of data which could be recovered without breaking encryption.

\begin{table}[H]
\begin{tcolorbox}
\centering
    \begin{tabular}{p{0.1\textwidth}|p{0.7\textwidth}}
        \textbf{iOS} & \textbf{Lock Screen Bypass} \\ \hline
         4.1 & Access to contacts and photos\footnote{\url{https://www.forbes.com/sites/andygreenberg/2010/10/25/careful-iphone-owners-simple-backdoor-lets-anyone-bypass-password-protection/}}  \\ \hline
         6.1 & Access to contacts and photos\footnote{\url{https://www.forbes.com/sites/adriankingsleyhughes/2013/02/14/ios-6-1-bug-allows-snoopers-access-to-your-photos-and-contacts/}} \\ \hline
         6.1.2 & Access to contacts and photos, initiate iTunes sync\footnote{\url{https://www.forbes.com/sites/adriankingsleyhughes/2013/02/18/new-ios-lock-screen-vulnerability-uncovered/}} \\ \hline
         7 (beta) & Trivially access photos\footnote{\url{https://www.forbes.com/sites/andygreenberg/2013/06/12/bug-in-ios-7-beta-lets-anyone-bypass-iphone-lockscreen-to-access-photos/}} \\ \hline
         7.0.1 & Access and share photos\footnote{\url{https://www.forbes.com/sites/andygreenberg/2013/09/19/ios-7-bug-lets-anyone-bypass-iphones-lockscreen-to-hijack-photos-email-or-twitter/}} \\ \hline
         9.0.1 & Access to contacts and photos\footnote{\url{https://appleinsider.com/articles/15/09/23/ios-9-security-flaw-grants-unrestricted-access-to-photos-and-contacts}}\\ \hline
         9.2.1 & Multiple bypasses to access the home screen, extent of compromise unclear\footnote{\url{https://seclists.org/fulldisclosure/2016/Mar/15}} \\ \hline
         12.0.1 & Access to photos\footnote{\url{https://appleinsider.com/articles/18/10/12/voiceover-bug-lets-hackers-view-iphone-photos-send-them-to-another-device}}\\ \hline
         12.1 & Access to contacts\footnote{\url{https://www.theverge.com/2018/11/1/18051186/ios-12-1-exploit-lockscreen-bypass-security}}\\ \hline
         13 & Access to contacts\footnote{\url{https://www.theverge.com/2019/9/13/20863993/ios-13-exploit-lockscreen-bypass-security}}\\ \hline
    \end{tabular}
    \caption{History of iOS Lock Screen Bypasses}\label{tab:ios_ls_bypass}
    \caption*{This list is not exhaustive. Drawn from various sources, primarily~\cite{ios_lockscreen_bypasses}}
\end{tcolorbox}
\end{table}

\subsection{Local Device Data Extraction}\label{sec:devicehacks}

Seizure of iOS devices has occurred in high-profile and common criminal cases alike. Motherboard (the technology branch of Vice news) published a database of over 500 cases involving iPhone unlocking, of which 295 included executed unlocking of iPhones either directly or through federal law enforcement or commercial partners~\cite{vice_db,vice_db_news}. The San Bernardino shooting in 2015 famously sparked the ``Apple v. FBI'' case when the FBI sought access to a locked iPhone 5C~\cite{ApplevFBI_apple,ApplevFBI_fbi}, and similar cases have occurred since~\cite{iphone11_warrant}.

In order to use forensic tools to the greatest extent, law enforcement officials seize iPhones along with any other devices which may aid in accessing those phones such as laptops and/or other Apple products~\cite{elcomsoft_methods,nist_forensics}. They isolate these devices, for example by placing them in Faraday bags, preventing remote wipe or control, and keep them charged to prevent shutdown (and thus eviction of encryption keys from memory), and in some cases to prevent iOS \Gls{USB} Restricted Mode~\cite{apple_usb} from coming into effect~\cite{elcomsoft_methods,elcomsoft_usb,nist_forensics} using inexpensive adapters. \Gls{USB} Restricted Mode can also be bypassed in certain circumstances by booting the iOS device to \Gls{DFU} mode and delivering an exploit~\cite{elcomsoft_usb_bypass}. Devices are often examined by officials trained in the use of third-party forensic tools, or devices are sent to those third parties for examination~\cite{elcomsoft_methods,cellebrite_advanced_services}.

In the field, many devices are likely to be seized in the After First Unlock (\Gls{AFU}) state, having been unlocked once since their last boot but currently locked. This is immediately clear when considering the alternatives: if the device has been power cycled, it is very likely to have been used shortly after being powered on. In order to use most functionality, unlocking is required, and thus the device transitions to \Gls{AFU}. In this state, many decryption keys for data objects belonging to the \Gls{AFU} (or weaker) protection class will remain resident in memory, having been decrypted by the \Gls{SEP}~\cite{apple_security_guides,apple_platform_security}. iOS will automatically use these keys to decrypt data requested by the iOS \gls{kernel} on behalf of itself or of apps, including any forensic tools that can be installed without causing a device reboot.

Through companies including Elcomsoft~\cite{elcomsoft_methods}, Cellebrite~\cite{cellebrite_advanced_services}, and Oxygen~\cite{oxygen_latest}, law enforcement agents can obtain forensic extractions of even the latest devices. These companies are naturally vague in the details they make public as they rely on exploits which must be unmitigated by Apple, however they make some information available in the form of advertisements and blogs~\cite{elcomsoft_methods,cellebrite_advanced_services,oxygen_latest}. Known exploits are also commonly used on older devices which no longer receive updates or which cannot be patched, such as with hardware/firmware vulnerabilities like checkm8~\cite{oxygen_latest}.\footnote{For more information on the checkm8 \gls{jailbreak}, refer to \S\ref{sec:jailbreaking}} The common goal of extraction tools is to run an extraction agent on the device which reads any available files and exfiltrates them to an analysts device or computer. Sometimes, this takes the form of inducing a backup of the iOS device to the investigator's computer. This approach generally requires either a \gls{kernel} compromise or running unsigned code, and thus a \gls{jailbreak}, or exploiting/bypassing Apple's signing infrastructure (potentially among other exploits) to sign and execute such an agent~\cite{elcomsoft_nojb}.

In total, depending on the device (iPhone model), device state (e.g. \Gls{AFU}), types of seized devices, iCloud settings (e.g. \Gls{2FA} enabled), and operational security (e.g. passwords on sticky notes next to laptop) of the target, law enforcement may obtain partial or complete access to extensive categories of data as listed in Figure~\ref{fig:ios_extractables}. It is also clear that extraction of this depth and magnitude (all or most data categories listed) is historically the norm due to the extent of data accessed by forensic tools for over a decade as tested by \Gls{DHS} following \Gls{NIST} standards for forensic software testing~\cite{dhs_forensics,nist_forensics_spec}.

\begin{figure}[H]
\begin{tcolorbox}
    \caption{List of Data Categories Obtainable via Device Forensic Software}\label{fig:ios_extractables}
\begin{multicols}{2}
\begin{itemize*}
    \item Contacts
    \item Call metadata\footnote{Participant phone numbers, call duration, etc.}
    \item \Gls{SMS}/\Gls{MMS} messages
    \item Stored files
    \item App data\footnote{Particularly, data from applications which are not designed to opt into higher protection classes than \Gls{AFU}.}
    \item Location data
    \item Wi-Fi networks\footnote{Which can be used to determine location history~\cite{wifi_location}.}
    \item Keychain data\footnote{Some Keychain data which is configured into higher protection classes may be unavailable for extraction.} (authentication tokens, encryption keys, and passwords)
    \item Deleted data\footnote{Only in some cases when the full filesystem was extracted}
    \item iCloud authentication token(s)
\end{itemize*}
\end{multicols}
\caption*{Source: Elcomsoft~\cite{elcomsoft_methods} and Cellebrite~\cite{cellebrite_advanced_services} blogs and documentation, among others}
\end{tcolorbox}
\end{figure}

\begin{figure}[H]
    \centering
    \includegraphics[width=0.9\linewidth]{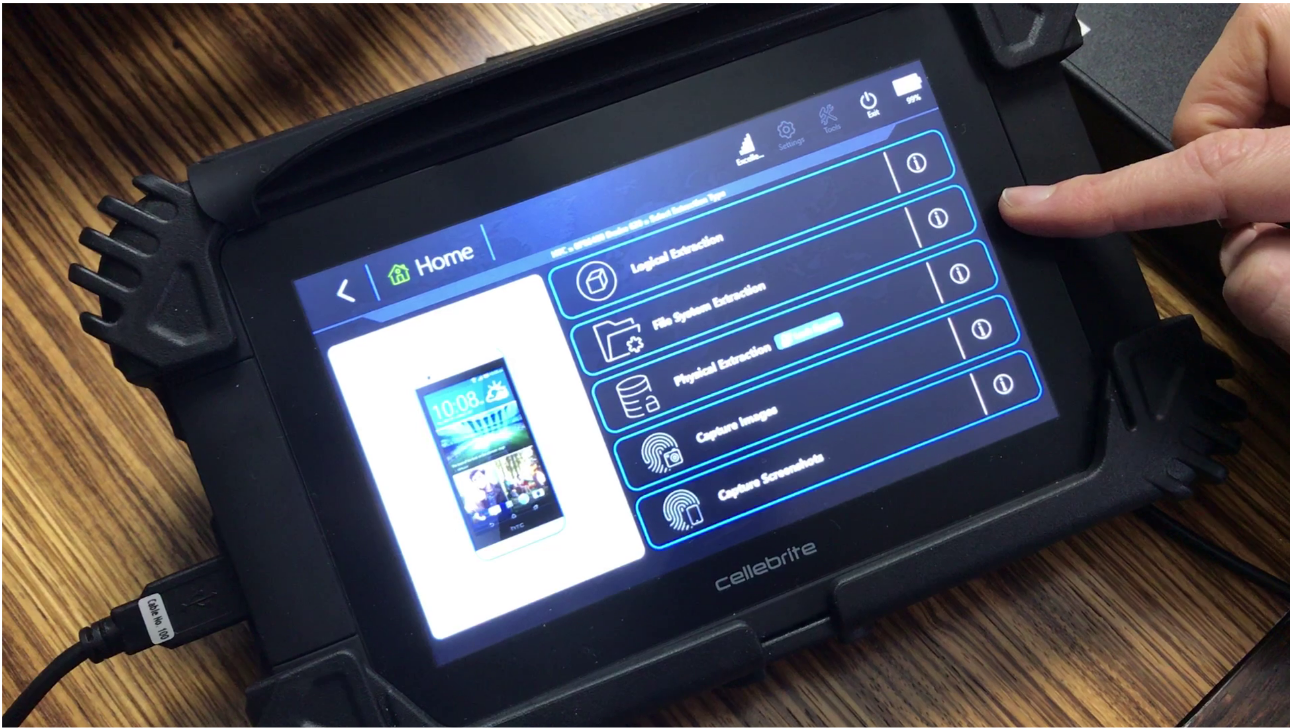}
    \caption{Cellebrite UFED Touch 2}\label{img:cellebrite_ufed_touch}
    \caption*{Source: Privacy International~\cite{pi_deepdive}}
\end{figure}

\begin{figure}[H]
    \centering
    \includegraphics[height=0.8\textheight]{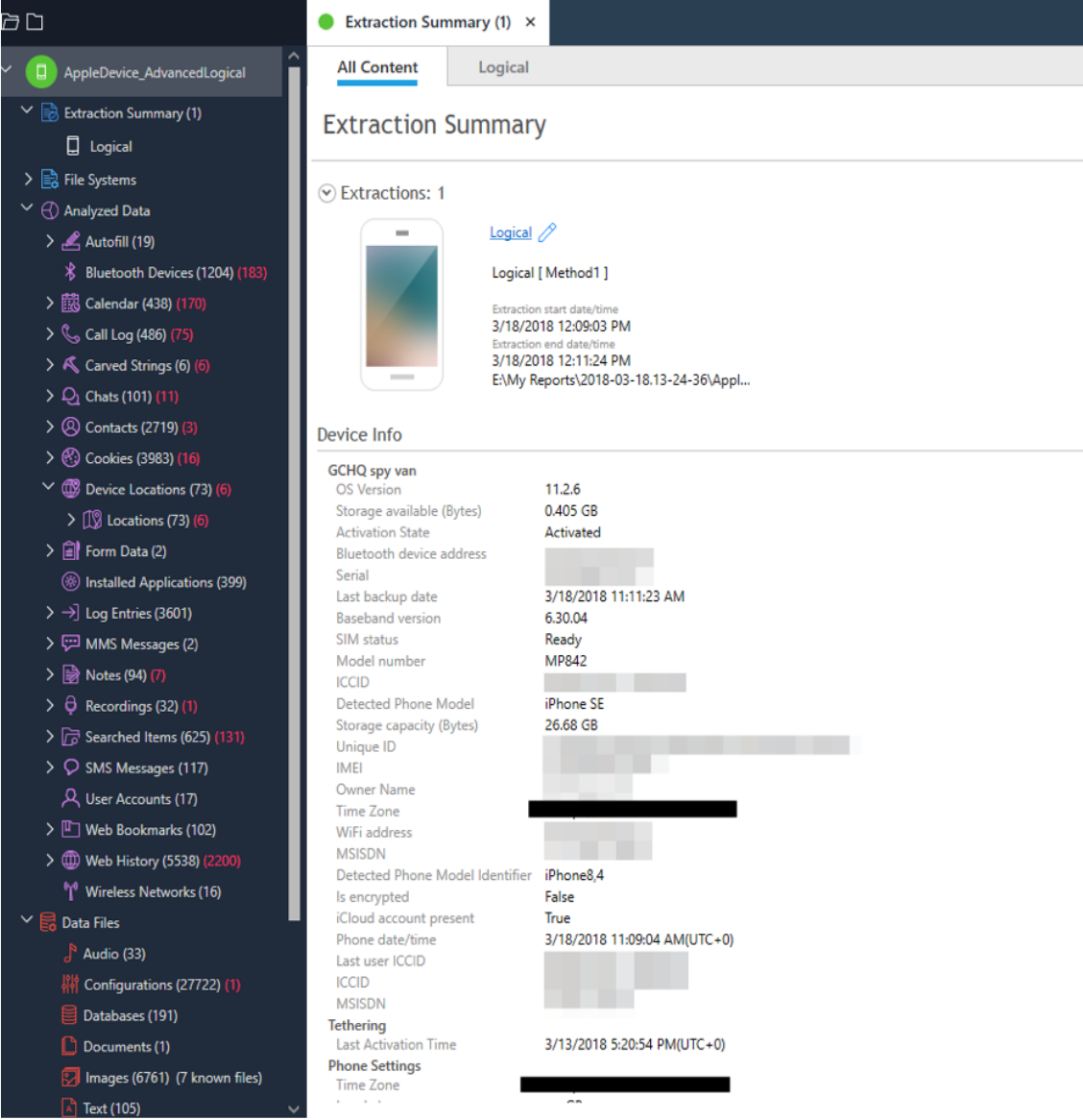}
    \caption{Cellebrite UFED Interface During Extraction of an iPhone}\label{img:cellebrite_ufed}
    \caption*{Source: Privacy International~\cite{pi_deepdive}}
\end{figure}

\paragraph{Bypassing Data Protection} The encryption implemented in Data
Protection is the last layer of defense against unauthorized device access,
allowing devices to maintain data security even in the event that an attacker
compromises the OS running on the device. It is up to app developers to opt-in
to Complete Protection for sensitive data. However, this is not always done,
even when critical for user privacy or confidentiality. For example, Elcomsoft
discovered in 2020 that the iOS app for \texttt{mail.ru} (a Russian email provider and internet company) elected to put email authentication tokens into the iOS Keychain at the ``Always Available'' level of protection, meaning that these keys were entirely unprotected against physical or logical extraction~\cite{elcomsoft_keychain}. Because there are no known practical attacks against modern encryption itself, current data extraction techniques (when the user is not available or willing to cooperate) tend to follow one of the following three approaches:

\medskip \noindent
{\em Accessing devices with keys available.} The most straightforward approach to bypassing encryption is to simply obtain the device in a state where the necessary decryption keys are loaded into the device memory. Forensic tools such as the Cellebrite UFED (see images in Figures~\ref{img:cellebrite_ufed_touch} and~\ref{img:cellebrite_ufed}) or XRY Logical are able to connect to iOS devices and either initiate a backup or otherwise request files and data over Bluetooth and/or a physical link via the \Gls{Lightning} port~\cite{pi_deepdive}. Data which is accessible by the iOS \gls{kernel} (data for which keys have been decrypted and are available in memory) can be requested and exfiltrated directly. These forensic tools seem to work through a combination of proper use of iOS \Gls{API}s and exploitation or circumvention of access controls in iOS. Forensic software companies sell devices which are as easy-to-use as possible~\cite{oxygen_latest} and then offer bespoke consultations for devices which are inaccessible using these more basic methods~\cite{cellebrite_advanced_services}. Refer to Figure~\ref{fig:ios_extractables} for a complete list of data categories seemingly regularly~\cite{vice_db} extracted via forensic software.

\begin{figure}[H]
    \centering
    \includegraphics[width=0.9\linewidth]{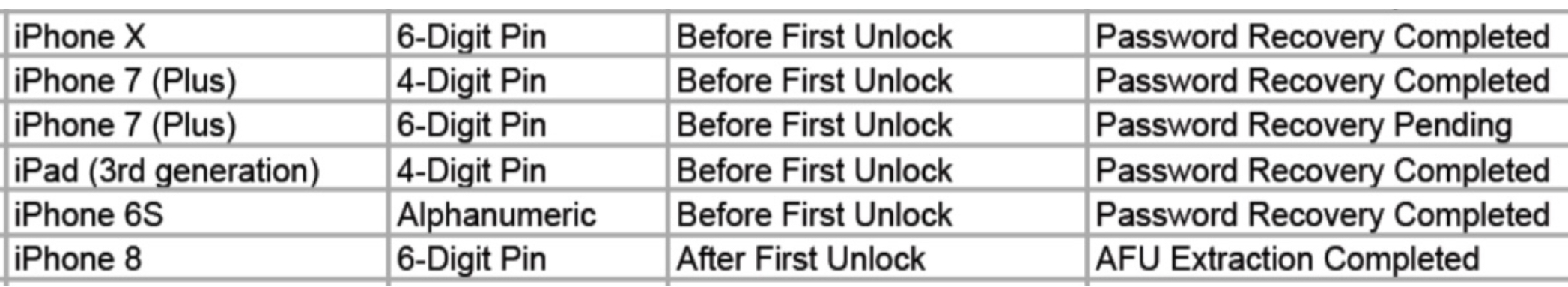}
    \caption{Records from Arizona Law Enforcement Agencies Documenting Passcode Recovery on iOS}
    \caption*{Note passcode recovery on AFU and BFU iPhones up to X. Many rows omitted.}
    \caption*{Source: Upturn~\cite{arizona_le_records}}
    \label{fig:arizona_le_logs}
\end{figure}

\medskip \noindent
{\em Bypassing protections.} Data which is protected under Data Protection and not available for immediate extraction may still be accessed in some cases. GrayKey, for example, seems to have been able to extract user passcodes by exploiting the \Gls{SEP} to enable passcode brute-force guessing~\cite{graykey_news_18,koch_iphone_news}. Documents obtained by Upturn~\cite{arizona_le_records}\footnote{For policy analysis of these documents and far more, see Upturn's recent document entitled ``Mass Extraction.''~\cite{upturn_mass_extraction}.} display records of law enforcement agencies gaining AFU and even BFU access to iPhones, presumably using Graykey. Figure~\ref{fig:arizona_le_logs} displays a small selection of these records.

\begin{figure}[H]
    \centering
    \includegraphics[width=0.45\linewidth]{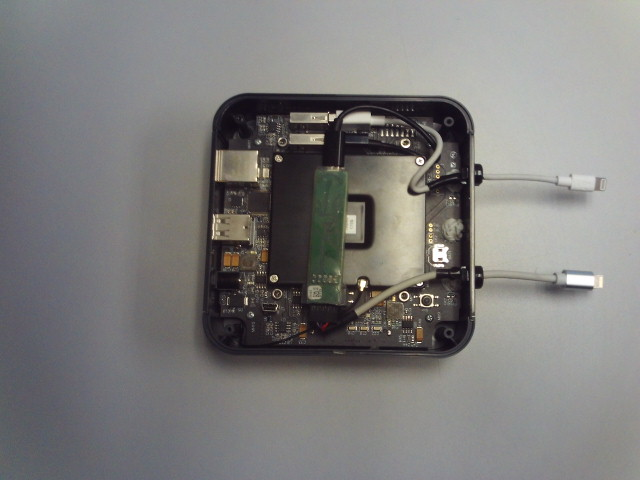}
    \includegraphics[width=0.45\linewidth]{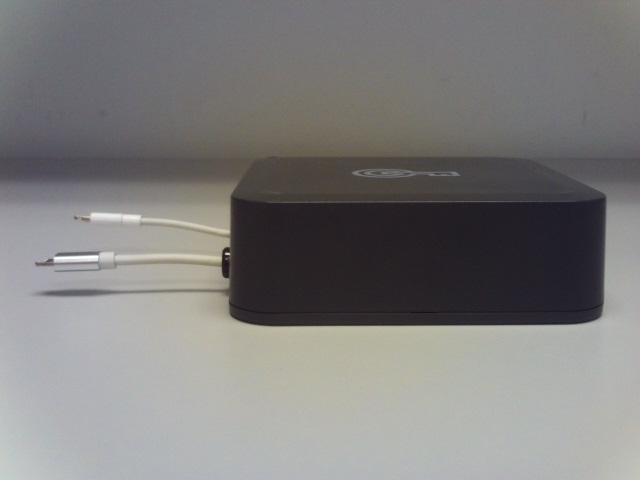}
    \caption{GrayKey by Grayshift}\label{img:graykey}
    \caption*{Left: Interior view. Right: Exterior view. Source: FCC ID~\cite{graykey_fcc}}
\end{figure}

An Elcomsoft blog article alleges that a brute-force passcode guessing strategy can only be conducted rapidly (many passcode attempts per second) if the device is in an \Gls{AFU} state, and that otherwise such guessing attacks require upwards of 70 days to brute-force a 4-digit passcode~\cite{elcomsoft_knownpasscode}. Once the user passcode is known, law enforcement access to user data is relatively unbounded~\cite{elcomsoft_knownpasscode}, as the entire iOS filesystem, Keychain, and iCloud contents can be extracted, among other capabilities. For example, ``Significant Locations''~\cite{apple_location_privacy} (GPS locations which the iOS device detects are commonly occupied) could be extracted directly from the device using GrayKey in addition to the Cellebrite Physical Analyzer product in 2018~\cite{ios_significant_locations_scooped}.\footnote{Significant Locations are end-to-end encrypted when syncing to iCloud~\cite{apple_icloud_security} but this protection is irrelevant for data at rest.} GrayKey in 2019 could even reportedly bypass \Gls{USB} Restricted Mode~\cite{pi_deepdive}. See also Figure~\ref{img:graykey} for images of the GrayKey device from the FCC ID filing~\cite{graykey_fcc}.

\medskip \noindent
{\em Alternate data sources.} In the case that forensic tools and even privileged (law enforcement only) consulting services such as those offered by Cellebrite and GrayKey fail to provide desired results, it is possible for investigators (or equivalently, hackers) to access iOS device data by other means. If a target's laptop can be seized in a search and accessed (perhaps with the password written nearby), access to iOS backups can render device extraction redundant. Even when encrypted on a local computer, iOS backups have been accessible to law enforcement due to a weak password storage mechanism enabling brute-force decryption via password cracking~\cite{apple_security_updates}. Further, forensic software companies offer cloud extraction services detailed in \S\ref{sec:cloudhacks}.

\subsection{Cloud Data Extraction}\label{sec:cloudhacks}

iOS devices alone contain extensive personal data which law enforcement agents may seek to extract. While devices themselves are increasingly protected from such analysis (refer to \S\ref{sec:ios_security}), data is also increasingly stored or synchronized to the cloud~\cite{cellebrite_cloud,elcomsoft_keychain,elcomsoft_tokens,pi_cloud_extraction}. Thus, cloud service data extraction can replace or even surpass the value of mobile devices from the perspective of law enforcement. Apple complies with warrants to the extent they claim to be able to, and offer law enforcement access to various categories of data, particularly those stored in Apple's iCloud. For details, see Figure~\ref{fig:apple_legal}. However, law enforcement agents also use warrants to replace consent to access devices and services~\cite{vice_db} and cut Apple out of the loop. Speculatively, the reasons for this could be reduced legal exposure, saving cost/time, or preventing potential notifications of targets.

Forensic software companies have followed this market and branched into developing cloud extraction solutions~\cite{cellebrite_cloud,pi_cloud_extraction}. While a full \Gls{SEP} exploit could extract the entire Keychain and thus authentication credentials for many cloud services, these tools commonly exploit iOS devices to extract cached authentication tokens only protected at the \Gls{AFU} class or worse, for services such as iCloud~\cite{apple_security_guides,apple_platform_security}, Dropbox, Slack, Instagram, Twitter, Facebook, Google services, and Uber~\cite{pi_cloud_extraction}. These tokens seem to be stored in \Gls{AFU} to prevent interruptions of service upon device lock, but this convenience poses a huge risk as it removes \Gls{SEP} protection, relying only on the security of the iOS \gls{kernel}.

If other Apple devices are seized (laptops, Apple TVs, etc), they may be used to aid in authentication to iCloud, but commonly usernames and passwords can be acquired directly during search and seizure either through technical measures such as software tools~\cite{elcomsoft_methods,elcomsoft_keychain} or simply by searching for written passwords. Essentially, access can be incrementally increased in combination with other access in a compounding fashion. Figure~\ref{fig:cloud_extractables} lists data categories potentially obtainable via cloud forensic extraction.

\begin{figure}[H]
\begin{tcolorbox}
    \caption{List of Data Categories Obtainable via Cloud Forensic Software}\label{fig:cloud_extractables}
\begin{multicols}{2}
\begin{itemize*}
    \item iOS backups
    \item Search history on Safari or Chrome\footnote{Prior to iOS 13~\cite{elcomsoft_methods}. Private communications additionally indicate that these records are cleared monthly, but the authors cannot verify this claim.}
    \item iCloud documents and app data
    \item Contacts
    \item Call metadata\footnote{Participant phone numbers, call duration, etc.}
    \item Calendars
    \item Photos and videos
    \item Notes
    \item Reminders
    \item Find Friends and Find My data\footnote{These services enable location tracking of Apple devices, whether the user's own or their friends~\cite{findmy}.}
    \item Device information
    \item Payment card information
    \item Dropbox content
    \item Social media accounts and content (Twitter, Facebook, Instagram)
    \item Google services content
    \item Uber account and activity
\end{itemize*}
\end{multicols}
\caption*{Source: Cellebrite~\cite{cellebrite_cloud}, Elcomsoft~\cite{elcomsoft_tokens}, and Privacy International~\cite{pi_cloud_extraction}}
\end{tcolorbox}
\end{figure}

\subsection{Conclusions from Bypasses}

The evidence above indicates the availability of current and historic bypass mechanisms for iOS protection measures. This strongly indicates is that, with sufficient time, money, and fortunate circumstance (e.g. capturing a phone in an \Gls{AFU} state), law enforcement agents can typically extract significant (if not all) personal data from modern iOS devices, despite Apple's claims around user privacy~\cite{apple_letter_2016,apple_security_guides,apple_platform_security}. This is exacerbated by Apple's failure to widely deploy Complete Protection over user data, and its failure to more broadly secure cloud services (particularly, the decision to store cloud authentication tokens in \Gls{AFU}). These facts combine to offer extensive access to law enforcement agents, rogue governments, and criminals. 

\section{Forensic Software for iOS}\label{sec:ios_forensics}

For nearly as long as there have been iPhones, there have been forensic tools designed to circumvent the protection measures on those phones to enable law enforcement agents to access sensitive personal information in pursuing a case. Phone forensics is not new~\cite{old_forensics2005}, but with the introduction of the iPhone in 2007 the amount of personal data aggregated onto one device that so many people began carrying every day increased massively. Additionally, it is critical to realize the accessibility of professional forensic tools such as Cellebrite's UFED~\cite{cellebrite_ufed}, and even of individualized consulting services such as Cellebrite's Advanced Services~\cite{cellebrite_advanced_services} for unlocking phones. Law enforcement agencies, including local departments, can unlock devices with Advanced Services for as cheap as \$2,000 USD per phone, and even less in bulk~\cite{upturn_mass_extraction}, and commonly do so~\cite{upturn_mass_extraction,vice_db_news,pi_deepdive}.

For a complete list of the forensic tools tested by \Gls{DHS}, as determined by publicly-available reports as of this writing, and the data forensically extracted by those tools, see Appendix~\ref{app:forensic_tools} and archived \Gls{DHS} reports~\cite{dhs_forensics}. Unfortunately, the NIST standard for testing devices is unclear as to whether the device should be in a locked state during testing~\cite{nist_forensics_spec}; we note, however, that certain categories of data seem inaccessible to forensic software in various cases, and as such we assume that this was caused by Data Protection. If false, then these tests simply document the extent to which the tested forensic tools support iOS data transmission and formatting. If true, then in summary these reports imply the following:

\medskip \noindent
{\em Successful extraction.} In most tests, most or all of the targeted data (see Figure~\ref{list:nist_forensics}) is successfully extracted against the latest iOS devices. However, there are exceptions, and reports after January 2016 are less clear as discussed below. Simultaneously, documents acquired by Upturn demonstrate records of law enforcement access -- as extensive as Before-First-Unlock passcode recovery -- to all generations of iPhones~\cite{upturn_mass_extraction,arizona_le_records}.

\medskip \noindent
{\em Relevance of Data Protection.} The major category of such exceptions are tests against iOS versions which include updates to Data Protection (see Table~\ref{tbl:ios_history}). In some cases, forensic tools were unable to access certain data. Particularly, app data and files seem to have been successfully protected against forensics for a time in 2015-16, coinciding with an expansion of DP.

\medskip \noindent
{\em Limited software diversity.} A small number of forensic software companies frequently iterate their products. This is demonstrated in the \Gls{DHS} tests as many Cellebrite, XRY, Lantern, Oxygen, MOBILedit, Secure View, and Lantern devices being tested between 2009 and 2019 (see Appendix~\ref{app:forensic_tools}), and each generally successfully extracting data from contemporaneous generations of iOS devices over time~\cite{dhs_forensics}.

\medskip \noindent
{\em Reporting inconsistencies.} Starting in February 2016, the quality of the reports degrades notably, and it is unclear in many cases whether data was extracted from iOS and not displayed properly by the forensic software, or simply not extracted at all. Some reports\footnote{For example see April 2017 ``Electronic Evidence Examiner - Device Seizure v1.0.9466.18457.''} showed inconsistencies between the analysts notes on forensic performance and the summary tables in the final report (e.g. claimed extracted data in the notes, but this success not indicated in the summary table).

\medskip \noindent
{\em GrayKey.} One notable exception to the vagueness of post-January 2016 reports was the GrayKey test in June 2019. GrayKey definitively extracted data covered by Data Protection on iPhones 8 and X according to the report.

Based on this information and a report of over 500 forensic access warrant filings and executions against iOS devices~\cite{vice_db}, it seems that law enforcement agencies are generally able to pay for or collaborate to gain access to iOS device, even of the latest generations. Particularly, federal agencies (\Gls{FBI}, \Gls{DHS}) are able to consistently extract data, and are consulted by local law enforcement agencies for such services if consent to access a device is not attained.

\section{Proposed Improvements to iOS}\label{sec:iOS-improvements}

iPhone users entrust the privacy of their communications and activities and the security of their accounts and data to Apple. They rely on Apple's responsiveness to vulnerabilities and mitigation of potential attacks to keep them safe from malicious actors, including hackers, potentially overreaching agencies, or rogue governments and corporations. However, as the history and current landscape of iOS protection bypasses show, Apple devices are not immune to compromise and their users are not completely protected. Impenetrable security is of course an impossible goal, but based on our analysis we have formulated the following recommendations for Apple software and hardware to mitigate channels of data extraction and improve iOS security.

\paragraph{Leveraging Data Protection} Data Protection on iOS is a powerful tool to ensure the encrypted storage of sensitive data. However, the \Gls{AFU} class is used as the default for third-party apps, and for many built-in data categories, notably including cloud service authentication tokens. iPhones are naturally mostly carried in an \Gls{AFU} state, and so this class of protection is in fact a liability given the availability of iOS forensic software devices. There are benefits to the \Gls{AFU} state: for example, keeping VPN authentication secrets available during lock~\cite{apple_security_guides,apple_platform_security} prevents an interruption of connectivity, and keeping contacts accessible means that iMessages and SMS/MMS received while the phone is locked can display the sender identity (name) rather than just a phone number or email address.\footnote{When the device is before first unlock and a text is received, the sending phone number is displayed, as tested by the authors with iOS 13.} Apple provides these features in iOS and therefore justifies their \Gls{AFU} classification. However, other potentially sensitive data, most notably cloud service authentication tokens, are also classed as \Gls{AFU}, putting them at risk even when not in use. We suggest two complementary approaches: first, that Apple thoroughly review the need for \Gls{AFU} in each case it is applied, and default more data into the \Gls{CP} class; and second, that Apple develop and deploy a system for runtime classification of data accesses in order to determine if and how often data is needed while locked. In such a system, data could be automatically promoted to CP if unused during device lock. Dynamically increasing security based on app and user behavior would result in a system more widely protected without interrupting the user experience, which based on the amount of data currently classed as \Gls{AFU} is an important point for Apple in the design of iOS security. Apple has developed a powerful framework for protecting user data; the only limitation of Data Protection is that it is not applied in the strictest fashion possible.

\paragraph{Dynamic Data Protection} In addition to strengthening usage of Data Protection and particularly the Complete Protection class, we additionally recommend an intelligent system which learns from user and app behavior to predict when certain file keys will be used, and evict the encryption keys from memory of files which are unlikely to be  imminently used. The advantages of this system would be minimal user experience interruption (with a good predictor, but with Apple's significant developments in machine learning hardware on iOS devices we believe this is feasible) in combination with increased protection against \gls{jailbreak} access to values in memory. Such a system could fetch keys as needed, and potentially re-authenticate the user though \Gls{FaceID} if an access was considered suspect (anomalous), or a \Gls{USB} device had been attached. As the user will likely be looking at their screen during use, such \Gls{FaceID} re-authentication could be seamlessly woven into the user experience, or even be opt-in to prevent user frustration or confusion.

\paragraph{End-to-end encrypted backups} Apple continues to hold keys which are able to decrypt iCloud Backup data. However, there seems to be little value in maintaining Apple's ability to decrypt this data aside from recovery of certain data in case of complete device and account loss~\cite{apple_icloud_backup,apple_icloud_security}. Google offers and end-to-end encrypted backup service~\cite{kensinger2018google} for Android which, while imperfect (refer to Chapter~\ref{chapter:android}), vastly improves the security of Android backups. Why Apple, a company which markets itself around user privacy~\cite{apple_privacy}, has not implemented a competing solution is an issue of curiosity and concern. Apple already has the infrastructure and design in place to implement such a system~\cite{apple_icloud_security}. Keys could be held in iCloud Keychain, and backup data would thus be end-to-end encrypted, inaccessible to Apple but available to any trusted device the user has authenticated.

\paragraph{End-to-end encrypted iCloud content} Apple maintains access to photos, documents, contacts, calendars, and other highly personal data in iCloud. Similarly to backups, the infrastructure to place these data into CloudKit containers and allow them to be end-to-end encrypted amongst trusted devices (which can access iCloud Keychain to share encryption keys) would massively reduce the efficacy of cloud extraction techniques. If user data loss is a concern, recommend that users create both local and iCloud backups regularly, or automate this process. SMS and MMS messages are immediate prime candidates for this as the iMessage app (which manages SMS, MMS, and iMessage) already integrates with CloudKit.

\paragraph{Avoid special cases which bypass encryption} iMessage in iCloud uses an end-to-end encrypted container to prevent Apple from accessing iMessage content. However, this security is rendered moot when the encryption key is also included in an iCloud Backup to which Apple has access. Based on Apple documentation, it seems that this feature was included to provide an additional avenue for recovery of iMessages, but the security implications are significant and as such this loophole should be closed. We thus recommend that Apple continue transitioning data into end-to-end encrypted iCloud CloudKit containers rather than holding encrypted data and the relevant keys.

\paragraph{Local backup passwords} Although Apple increased the iteration count for \Gls{PBKDF2}, the one-way cryptographic function which protects the password in a backup, to the point that most brute-force attacks are infeasible, local backups have two limitations. First, because of the nature of a local backup, guessing limits cannot be enforced. Thus, sufficiently high-entropy passwords must be chosen by users, which may not always be the case. User education and interface design to encourage strong backup passwords, or a system which involves an on-device secret could strengthen this mechanism. Additionally, in iOS 11, Apple removed the requirement to input the old backup password if a new one is to be set~\cite{elcomsoft_ios11_horror}, and as such the passcode is sufficient to initiate a new backup. Although Apple goes to great length to protect the passcode, this represents a single point of failure where multiple layers of security could be used instead.

\paragraph{Strengthening iCloud Keychain} iCloud Keychain represented a significant step forward for the security of Apple's cloud services, enabling strong encryption with keys isolated even from Apple, assuming trust of the \Gls{HSM}s. That trust, however, creates a point of failure for the system: users have no way of ensuring that the \Gls{HSM} cluster they are backing encrypted data up to behaves in the way they expect. That is, as the device interaction with the \Gls{HSM} is limited to authentication via the SRP protocol and sending encrypted data, there is no authentication of the security or correctness of the \Gls{HSM}. Even if the user is enrolled with trustworthy \Gls{HSM}s initially, if their iCloud Keychain record is invalidated the user can simply re-enroll~\cite{apple_icloud_security}; this is relevant as a targeted user could be intentionally invalidated, and re-enroll with a compromised server rather than an \Gls{HSM}. As Apple seeks to provide options for recovery of this data, it can't be encrypted with keys on-device (as these would be lost with device loss/failure) or with keys derived from the user passcode or passphrase (in case of forgetting those). iCloud Keychain data could potentially be encrypted with keys derived directly from the user biometrics, but this isn't without risk either. Another plausible solution for authenticating Apple \Gls{HSM}s to users could entail a Certificate Transparency-like~\cite{certificate_transparency} solution wherein peers validate the addresses and correctness of the \Gls{HSM}s and publicly broadcast or peer-to-peer share this information.\footnote{This solution was informally proposed by Saleem Rashid, and is reminiscent of TACK, a previous Internet Draft by Moxie Marlinspike and Trevor Perrin~\cite{tack}.}

\paragraph{Abstract control of the \Gls{USB} interface} \Gls{USB} Restricted Mode~\cite{apple_usb} was a step towards securing iOS devices against invasive forensic devices which operate by attaching to the \Gls{Lightning} port and transmitting exploits, data, and/or commands over the \Gls{USB} to \Gls{Lightning} interface. However, we observe that forensic software tools are continuing to exploit this interface~\cite{elcomsoft_usb} and further that the checkm8 \gls{jailbreak} is unpatchable on iPhone X and earlier~\cite{checkm8}. In order to address these issues, the iOS kernel should be able to interpose and manage the \Gls{USB} interface on iOS devices such that security controls can be enabled, cryptographically-secured device authentication can occur, and perhaps even intelligent systems which recognize commonly-used \Gls{Lightning} devices and evict encryption keys or authenticate the user when anomalies are detected. As part of increasing \Gls{USB} protections, debugging and recovery interfaces must be hardened as well, particularly to mitigate exploits such as checkm8.

\paragraph{Further restrict debugging and recovery interfaces such as DFU and JTAG} DFU mode, short for Device Firmware Upgrade, is a low-level bare-bones operating system which enables directly installing firmware patches. The cases in which this is required are limited, and ideally uncommon as this mode is intended entirely for remediating software problems. Users can place their own device in DFU mode, and exploits such as checkm8 can potentially exploit the capabilities of this mode~\cite{checkm8}. In order to prevent such unpatchable exploits, DFU mode could require cryptographic authentication to access, for example using secure hardware on a trusted Apple computer owned by the user or even by Apple. The inconvenience of such a requirement would be offset by the rarity of its use.

Additionally, per the 2020 FCC filing by Grayshift of the GrayKey device~\cite{graykey_fcc}, we observe what may be \Gls{JTAG} hardware in the device (as shown by the interior view images, refer to Figure~\ref{img:graykey}). If this hardware debugging interface is vulnerable to unauthorized access, then further (ideally cryptographic, if possible) measures should be taken to secure this interface, particularly as end users will almost certainly have no need of it.

\paragraph{Increase transparency} Apple invests significant marketing effort into demonstrating their commitment to user privacy~\cite{ApplevFBI_apple,apple_letter_2016,apple_transparency}. However, inconsistencies in Apple's approach to practically implementing privacy controls disempower users of iOS device. For example, user control over which apps can access contacts, calendars, and other built-in app data is a powerful yet understandable tool for managing privacy, but enabling or disabling the ``iCloud'' toggle in iOS settings can have dramatic privacy and even functionality implications which are relatively opaque. Increasing transparency and empowering users through safe default settings and informative, consistent interfaces would improve the usability and practical privacy experienced by iOS users.

Apple could integrate the extensive work of user interface researchers, particularly those who design for user empowerment and follow egalitarian principles. Design Justice~\cite{designjustice,costanza2018design} is an ongoing area of research, industrial collaboration, and outreach which promote principles to foster equality and access in design. Opaque interfaces, such as the discussed controls surrounding iCloud and their pertinence to privacy, could be redeveloped with the Design Justice Principles in mind.

\paragraph{Leverage research constructions} In certain cases, functionality and security can create contradictory requirements for a system. For example, it could be a requirement that location data remain private and on-device, never synchronized to iCloud, but Apple could still seek to provide an iCloud service which allows location-based alerts. In such cases, where user privacy and features collide, constructions such as Secure Multiparty Computation (SMC/MPC) could be leveraged to, for example, allow iCloud and the device to privately compute a shared set intersection of user locations on-device to enable such alerts, or to enable some other such functionality. The research literature has provided extensive options in the form of cryptographic constructions which can enable cloud functionalities without risking user privacy. 

\paragraph{Leverage the community} Billions of Apple devices exist in the world, and so iOS is extremely widely used~\cite{apple_billions}. This worldwide community includes academics and professionals, students and experts, journalists and activists. Apple, then, has potential access to immeasurable knowledge of the experiences and needs of their users. With the bug bounty program~\cite{apple_bounty}, and through accepting externally-reported vulnerabilities in general, Apple began connecting with this community to improve the security and privacy of their products, and is continuing to do so with the Security Research Device program~\cite{apple_srd}. Leveraging the wealth of ongoing research, and embracing and implementing academic constructions can improve the security and privacy of Apple products as far as the state of the art allows. To facilitate this research, Apple could open source critical components (filing patents to protect their intellectual property if needed/applicable) such as the \Gls{SEP} and reap the benefits of allowing this network of academic and professional researchers unfettered access to help improve the devices they use and rely on every day.

\chapter{Android}\label{chapter:android}

Android is the most popular smartphone platform in the world, with  74.6\% of global smartphone market share as of May 2020~\cite{statista_android} spread across over a dozen major device manufacturers~\cite{pathak2010nokia}. The technical and logistical challenges in securing Android phones and protecting their users' privacy are numerous, and do not fully overlap with other segments of personal computing. hhe Android operating system is based on the open source Linux codebase, and so derives the benefits and risks of this underlying OS as well as new risks created by the mobile-specific features included in Android. Android's dominant market share also means that the impact of a new security vulnerability is acutely felt around the world. 

The base platform of Android, the Android Open Source Project (\Gls{AOSP}), is a collection of open-source software developed by the Open Handset Alliance~\cite{openhandset}, an entity that is commercially sponsored by Google. \Gls{AOSP} defines the baseline functionality of the Android operating system. However, the Android ecosystem is in practice more complex, due to the fact that most commercial Android phones also incorporate proprietary Google software known as Google Mobile Services (GMS)~\cite{google_gms}. GMS includes proprietary \Gls{API}s and software services, along with core Google apps such as Chrome, Drive, and Maps, and the Google Play Store, which is used for app distribution \cite{android_dev_playstore}. There exist non-Google forks of Android such as Amazon's FireOS~\cite{amazon_fireos}, but most of the Android devices in the United States~\cite{statista_android_mfg} are manufactured by Google's partners~\cite{android_certified} and use GMS. For the purposes of this chapter, we will consider ``Android'' to be the combination of the Android Open Source Project and Google Mobile Services.

\section{Protection of User Data in Android}\label{sec:android_prot}
 
Android devices employ an array of precautions to protect user data: user authentication, runtime verification, data encryption, and application sandboxing. This overview is based on documentation released by the Android Open Source Project~\cite{aosp_adoptable,aosp_appsandbox,aosp_appsigning,aosp_authentication,aosp_devicestate,aosp_dmverity,aosp_filebased,aosp_fingerprint,aosp_fsverity,aosp_fulldisk,aosp_keystore,aosp_metadata,aosp_release9,aosp_scoped,aosp_selinux,aosp_trusty,aosp_verified,aosp_verifying,aosp_gatekeeper}.

\begin{figure}
    \centering
    \includegraphics[width=0.3\linewidth]{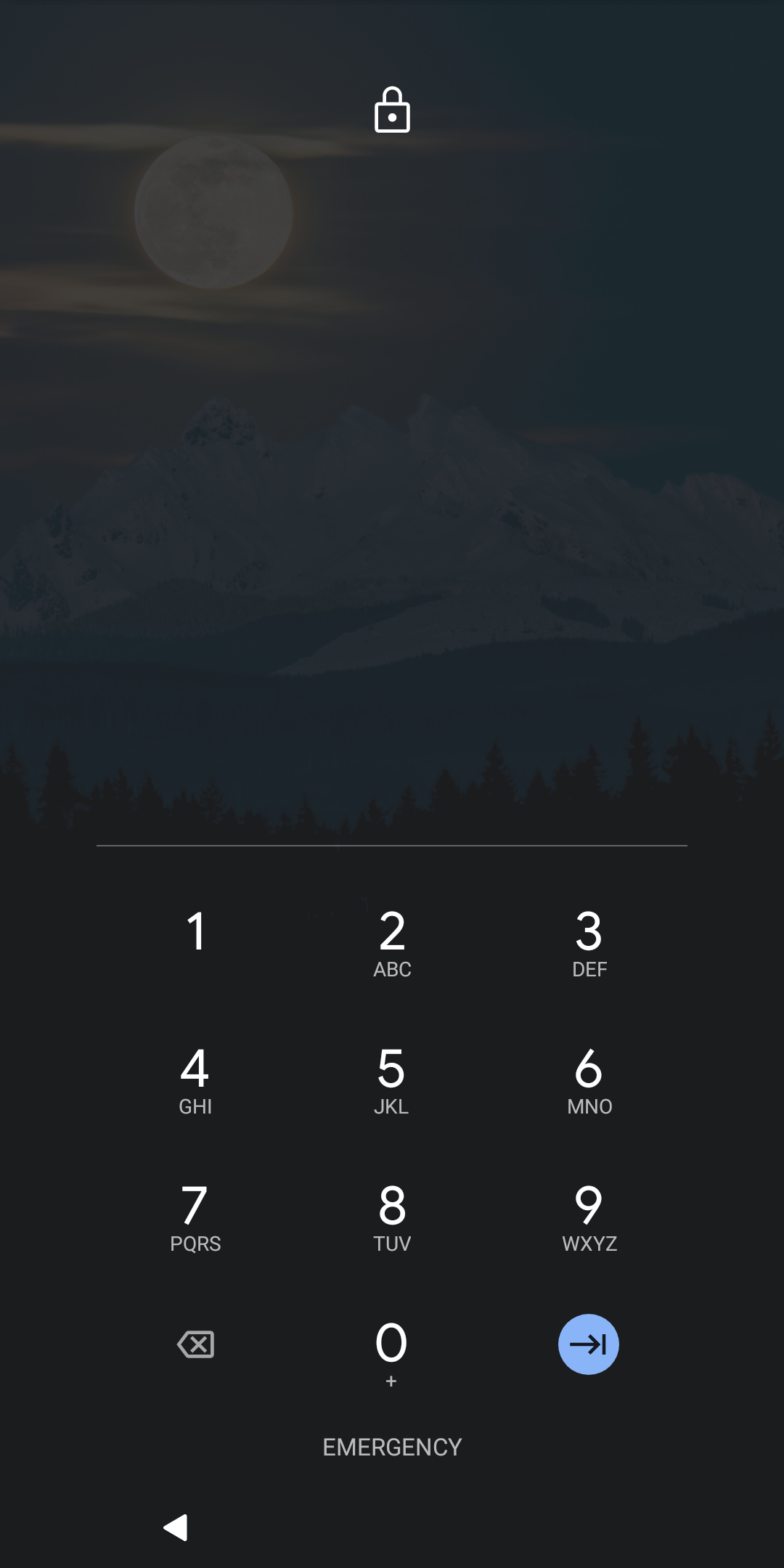}
    \caption{PIN Unlock on Android 11}\label{img:android_pin_unlock}
\end{figure}

\paragraph{User authentication} Android provides numeric passcode and alphanumeric passphrase authentication options for users, using an interface illustrated in Figure~\ref{img:android_pin_unlock}. Android additionally provides an authentication mode known as {\em pattern unlock}, wherein a user authenticates by drawing a pattern on a grid of points~\cite{aosp_authentication}. The security of the latter scheme has been called into question~\cite{aviv2010smudge}, with one academic study noting that effective security may be equivalent to a 2 or 3-digit numeric passcode~\cite{aviv2015bigger}. Many Android devices also deploy biometric authentication sensors; fingerprint authentication in particular is natively supported by the \Gls{AOSP} operating system~\cite{aosp_fingerprint}.\footnote{Most fingerprint readers are implemented either as external capacitive sensors or within the screen of the device.} 
A new biometric authentication feature, Face Authentication, is currently under development in \Gls{AOSP} \cite{aosp_faceauth}. Android biometric authentication is deployed by manufacturers, and is not required by the Android OS.

Android devices support an additional set of authentication features known collectively as {\em Smart Lock}. These use information about a phone's environment as criteria to keep it unlocked under certain conditions. Critically, this is a method of \emph{keeping devices unlocked}, which is distinct from a method of \emph{unlocking a locked device}, which can only be done with a PIN/password/pattern or a biometric. For example, Smart Lock will keep a device unlocked while it is on-body or in a specific location (such as at home)~\cite{androidsupport_smartlock}. One of the Smart Lock features, Trusted Face (formerly known as Face Unlock), allowed users to use their face (via the device camera) to unlock a device. This feature was removed in Android 10~\cite{androidpolice_faceunlock}, as it was easy to fool using static images \cite{register_faceunlock}, and is being replaced with the more general Face Authentication biometric framework~\cite{aosp_faceauth}.

\paragraph{Application sandboxing} The Android systems utilizes application sandboxing as a mechanism to prevent apps from compromising each other or the OS as a whole. The basic sandboxing mechanism is provided through discretionary access control (DAC): files have permissions based on Linux user IDs, which are enforced in the Linux \gls{kernel}. Each Android app has its own Linux user ID, and by default only has access to its own files, so any attempted direct access to the files of other apps or the core OS will result in an error. DAC is enforced by the Linux \gls{kernel}~\cite{aosp_appsandbox}. Theoretically, bypassing this mechanism would require a \gls{kernel} exploit in order to escalate privileges. However, the Android OS provides many other avenues for inter-application communication, which creates further opportunities for privilege escalation~\cite{smalley2013security}. 

To enhance the limited security provided by DAC, Android also employs Security Enhanced Linux (SELinux), which allows granular control of the privileges granted to an application~\cite{aosp_selinux,smalley2013security}. For example, in the traditional DAC model, a compromise of a root-owned process\footnote{``Root'' in the context of traditional Linux/Unix refers to a special user account that enjoys super-user privileges over the entire system.} would be tantamount to compromise of the whole system. With SELinux, polices are configured so that a process is allocated only the specific privileges it needs to function; this is known as mandatory access control (MAC). In this setting, compromise of a root-owned process would not necessarily lead to full system compromise, since the \gls{kernel} would prevent the process from performing actions outside of its capability set~\cite{aosp_selinux_concepts}. In Android, SELinux is enforced between the system and individual apps: apps are confined to the permissions available in their domains (process groups), and any violations result in an error at the \gls{kernel} level~\cite{aosp_selinux}. 

\paragraph{Encryption} There are two forms of user data encryption on Android: full-disk, a direct parallel to Linux disk encryption, and file-based, which enables more granular categorizations of encrypted data.

\medskip \noindent
{\em Full-disk encryption} is the original encryption mechanism on Android, introduced in Android 4.4~\cite{aosp_fulldisk}. It is based on the Linux \gls{kernel}'s {\tt dm-crypt} module~\cite{dmcrypt_cryptsetup}, and uses a master key to encrypt block devices with \Gls{AES}-128 in \Gls{CBC} mode with ESSIV~\cite{aosp_fulldisk}. The master key in turn is encrypted using a key derived from the user authentication credential: a numeric passcode, alphanumeric passphrase, or pattern. However, because the same key is used for the entire disk partition, Android requires the master encryption key to be available on boot in order to activate core system functionality: specifically, subsystems such as telephony, alarms, and accessibility services remain unavailable until the user unlocks their phone. Once entered, the master key is never evicted from device memory, as it is needed for \emph{all} block device operations on the data partition~\cite{aosp_fulldisk}.

\medskip \noindent
{\em File-based encryption} is a more recent development, and provides more granular control than full-disk encryption. The underlying mechanism for this encryption is the Linux \gls{kernel}'s {\tt fscrypt} module~\cite{kernel_fscrypt}, and uses \Gls{AES}-256 in \Gls{XTS} 
mode for file data encryption, and \Gls{CBC} mode to encrypt file metadata such as names~\cite{aosp_fulldisk}. File-based encryption segregates user files into two categories: {\em Credential Encrypted} (CE) and {\em Device Encrypted} (DE). CE data is encrypted using a key derived from the user's passcode, and is thus available only after the device is unlocked. DE data is based only on device-specific secrets and is available both during the initial boot and after device unlock~\cite{android_dev_directboot}. Because of this separation, file-based encryption allows the user to access important functionality prior to user  authentication. CE is the default for all applications, and DE is reserved for specific system applications (Dialer, Clock, Keyboard, etc). Filesystem metadata (such as directory structure and DAC permissions) is not covered by this mechanism; a different encryption subsystem, metadata encryption~\cite{aosp_metadata} encrypts this information. The combination of file-based encryption and metadata encryption protects all content on the device as a result.

Outside of the CE and DE storage in file-based encryption, Android has no analogue to the iOS {\em Complete Protection} protection class: keys remain in memory at all time following the first device unlock~\cite{android_dev_directboot}.\footnote{This makes Android CE protection equivalent to Apple's \Gls{AFU} protection class, whereas DE protection is analogous to Apple's not protected (NP) class.} Thus, an attacker who can logically gain access to the memory of a device following boot and first unlock can directly access encrypted data and keys. 

Mandatory application of an iOS-style Complete Protection would be beneficial to privacy: this would reduce the possibility of live acquisition of encryption keys for a device. On the other hand, it would create large development and usability problems for Android phones and existing applications. Currently, the Android application lifecycle~\cite{android_dev_lifecycle} distinguishes processes in a hierarchy of four categories based on importance: from high to low importance, those in the foreground (actively shown), those visible but in the background, those handling a background service, and those that are not needed anymore and can be killed. The current lifecycle does not incorporate the device's ``locked'' state into the paradigm, which means that an encryption scheme that evicts keys on lock may require a retooling of the Android process lifecycle.  Enabling CE/DE storage itself required such a modification to application logic~\cite{android_dev_directboot}, and it is unclear how complex the additional changes would be in order to achieve complete protection as an option. 

\paragraph{Removable storage} Android has first-class support for expandable storage, such as SD cards; Android allows applications to install to and hold data on an SD card. Media, such as photos and videos, can also be stored on the SD card. Originally, the ``external storage'' directory was considered to be removable storage -- Android would have limited control over the access to files on the SD card (since the card could be removed anyway). Android 6 introduced adoptable storage~\cite{aosp_adoptable}, which allows the system to consider the SD card as a part of its own internal storage, which in turn allows the system to perform encryption and access control on the external medium. The downside of using this mode is that the SD card becomes tied to a specific device; since encryption keys are stored on the device, the SD card cannot be trivially moved to a different device~\cite{aosp_adoptable}.

\paragraph{Trusted hardware} Many security features are enforced by hardware mechanisms in modern Android phones. This is frequently implemented by through the use of a Trusted Execution Environment (\Gls{TEE}), which is a dedicated software component that runs in a special isolated mode on the application processor. This component enables, among other features, the secure storage and processing of cryptographic secrets in a form that is logically isolated from the operating system and application software.

Because the Android operating system is developed separately from the actual hardware on which it runs, Android cannot guarantee the presence of hardware support for the protection features it implements. In practice, a large majority of Android phones use the \Gls{ARM} architecture~\cite{arm_in_android}, and these system-on-chip (\Gls{SoC}) models (with ARMv7-A, ARMv8-A, and ARMv8-M~\cite{arm_trustzone}) include a mechanism called \Gls{ARM} TrustZone. Generically, \Gls{ARM} TrustZone partitions the system into two ``worlds'', each backed by a virtualized processor: a \emph{secure world} for trusted applications, and a \emph{normal world} for all remaining functionality. The secure world portion is used as a trusted execution environment for computing on sensitive data, such as key material~\cite{pinto2019demystifying}. Unlike Apple's SEP, TrustZone makes use of the same processor hardware on the SoC as is used by the Android OS and applications, and does not utilize a co-processor for trusted operations, but still provides a number of security properties for the Android system. 

\begin{figure}
    \centering
    \includegraphics[width=0.8\linewidth]{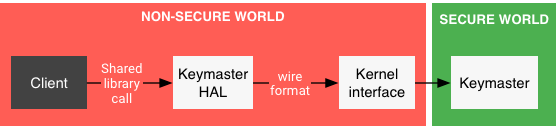}
    \caption{Flow Chart of an Android Keymaster Access Request}\label{img:android_keymaster_flow}
    \caption*{Source: Android Open Source Project Documentation~\cite{aosp_keystore}}
\end{figure}

In particular, three standard trusted applications (\Gls{TA}s) are defined by \Gls{AOSP} to run in an Android TEE: Keymaster, Gatekeeper, and Fingerprint. The Keymaster TA is the backing application for Android's Keystore functionality, which allows apps to load, store, and use cryptographic keys. Clients outside of the TEE communicate with Keymaster through a hardware abstraction layer (HAL), which then passes the request to the TEE via the \gls{kernel}; this flow is seen in Figure~\ref{img:android_keymaster_flow}. The Gatekeeper and Fingerprint TAs are used to authenticate to the Keymaster, showing that a user has enter the correct passcode, password, or pattern (Gatekeeper) or has provided a correct fingerprint biometric (Fingerprint) before a Keymaster request is serviced~\cite{aosp_keystore}.

Some Android manufacturers have begun to include additional dedicated security processors. For example, Samsung includes a {Secure Processor} in its Galaxy S line~\cite{samsung_secproc} and Google's includes a Titan M chip in its Pixel line~\cite{google_titanm}. This type of module, known as a \emph{secure element}, is an analogue to the Secure Enclave Processor in iOS: a dedicated cryptographic co-processor, separate from the primary processor, for computation on secret data. In these systems, Keymaster keys are offloaded to the secure element, and all key-related actions occur off the main processor~\cite{aosp_release9}. This offloading is in an effort to improve the security of key material, thus reducing the risk of e.g. side channel attacks on a TrustZone TEE~\cite{cho2018prime+,zhang2016truspy,qiu2019voltjockey}. 

\begin{figure}
    \centering
    \includegraphics[width=0.8\linewidth]{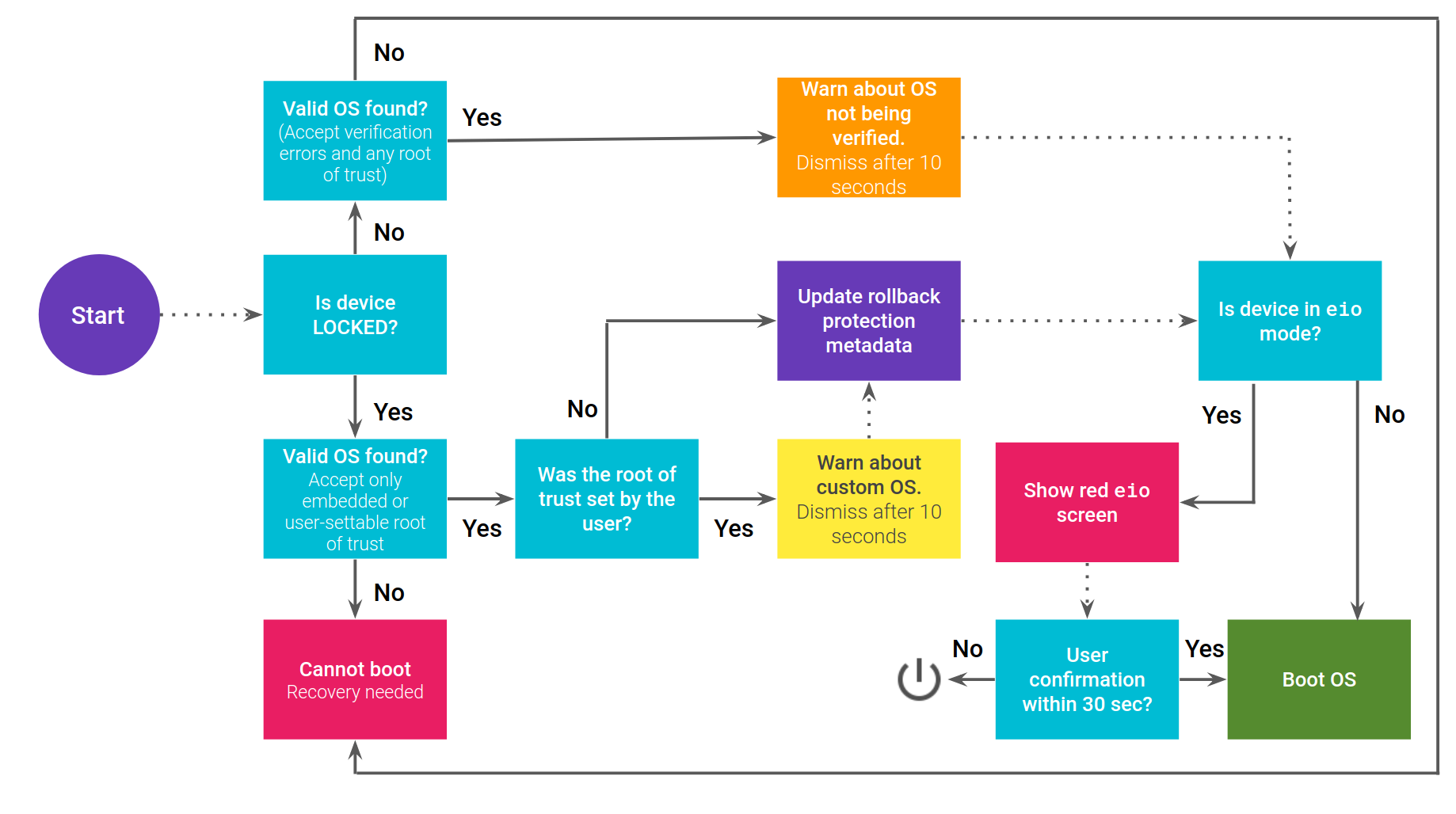}
    \caption{Android Boot Process Using Verified Boot}\label{img:android_verified}
    \caption*{Source: Android Open Source Project Documentation~\cite{aosp_bootflow}}
\end{figure}

\paragraph{Android Verified Boot} Android can optionally use the TEE to validate the integrity of the initial boot of the system, via its Android Verified Boot (\Gls{AVB}) process. The verification module, known as \texttt{dm-verity}, is designed to verify block devices using a cryptographic hash tree. Each hash node is computed as the device mapper loads in block data from a partition, with a final tree computed~\cite{aosp_dmverity}. If this root hash value does not match an expected value stored in trusted hardware, verification fails and the device is placed into a non-functional error state. \Gls{AVB} additionally uses tamper-evident storage to prevent software rollbacks i.e. use of a prior version of Android that has vulnerabilities but is otherwise valid and trusted by the hardware root~\cite{aosp_verifying}. Figure~\ref{img:android_verified} depicts the \Gls{AVB} boot process.

\Gls{AVB} can be disabled via a process known as ``unlocking.'' An unlocked bootloader does not perform \texttt{dm-verity} checks on the \texttt{boot} partition, allowing untrusted code to execute on device boot~\cite{aosp_devicestate}. This has legitimate use cases: for instance, unlocking can be useful for developers of Android modifications, and for users who wish to gain root access to their phones for additional customization and control. More information about rooting and bootloader unlocking can be found in \S\ref{sec:android_root}.

\paragraph{Google Mobile Services} Thus far we have described options and security features that are directly available in the \Gls{AOSP}. However, many additional security features are tied to Google Mobile Services (GMS), which requires a software license from Google and is not part of the \Gls{AOSP}. GMS includes a number of Google-branded apps, including Drive for storage, Gmail for email, Duo for video calls, Photos for image storage, and Maps for navigation, as well as the Google Play Services background task, which implements additional \Gls{API}s and security features. \Gls{AOSP} has some equivalents of these apps -- such as generic ``Email'' and ``Gallery'' apps -- but the GMS apps are generally higher quality~\cite{arstechnica_aospcreep} and are directly supported by Google, with updates delivered through the Play Store \cite{google_gms}. A device that is approved to use GMS and passes compatibility tests is known as ``Play Protect certified''~\cite{play_protect}. Several manufacturers, like Samsung, LG, and HTC, partner with Google to have GMS included~\cite{android_certified}; the few exceptions include those for business (Amazon FireOS~\cite{amazon_fireos}) and geopolitical (Huawei~\cite{huawei_fork}) reasons.

Google services utilize the Google Cloud, which enforces network encryption (\Gls{TLS}) for communications between a client and the server. All data stored in the Google Cloud is encrypted at rest using keys known to Google~\cite{gcloud_encryption,gcloud_suite}. Core Google services such as Gmail (for email), Drive (for files), Photos (for pictures), and Calendar (for scheduling) also store data on the Google Cloud. This implies that data backed up using Android Auto-Backups (which use Drive) along with some forensically-relevant pieces of user data (as defined in Figure~\ref{list:nist_forensics}) are available to Google. Currently, the only GMS app that is known to use end-to-end encryption is Google Duo~\cite{duosupport_e2e,omara2020duo}, which is used for real-time video communication, though Google Messages may add end-to-end encryption soon~\cite{verge_googlemessages_e2esoon}.

\begin{figure}[H]
    \centering
    \includegraphics[width=0.6\linewidth]{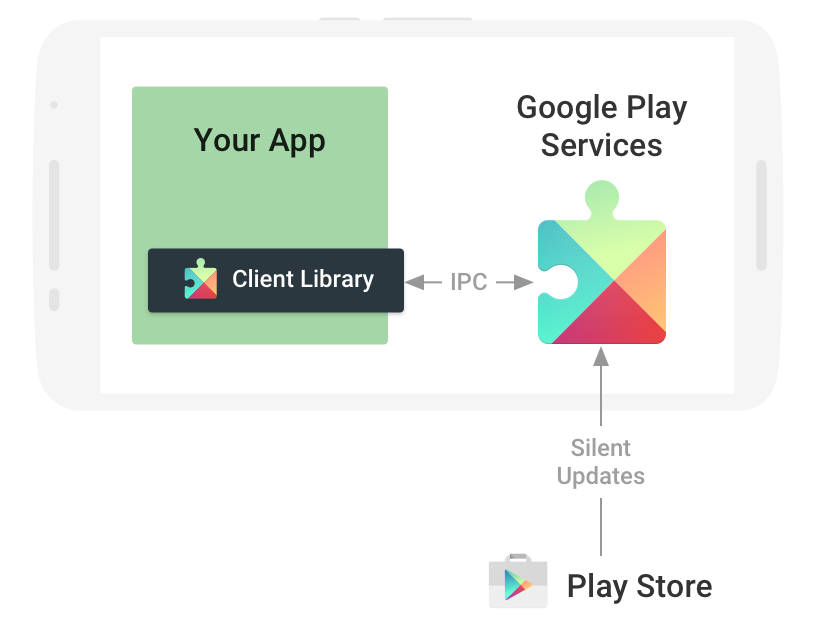}
    \caption{Relationship Between Google Play Services and an Android App}\label{img:android_playservices}
    \caption*{Source: Google \Gls{API}s for Android~\cite{googleapi_playservices}}
\end{figure}

Application developers integrate with GMS via the Play Services \Gls{API}s. Any Android app that wishes to integrate with the GMS on Android must do so through Play Services, which persistently runs in the background~\cite{googleapi_playservices}. App access to Google Account information, Mobile Ads, and Location information, among others, are all handled through Play Services~\cite{googleapi_playservices_setup}, as seen in Figure~\ref{img:android_playservices}. Of particular note is the SafetyNet \Gls{API}~\cite{googleapi_playservices_safetynet}, which is a validation service for Android phones. Play Services monitors a device, and records whether an Android system is rooted or potentially compromised by malware; developers can then request this information for a device. Applications can chose to disable access to certain features based on the result of the SafetyNet check~\cite{android_dev_safetynet,androidpolice_safetynet}. For example, Google Pay, an electronic payment platform akin to Apple Pay, is disabled on devices that do not pass SafetyNet validation~\cite{xda_googlepay,9to5_googlepay}. An app for a non-GMS Android device (e.g. for Amazon Fire OS) cannot use these features, as Play Services is only available through a GMS license with Google~\cite{google_gms}. 
 
\begin{figure}
    \centering
    \includegraphics[width=0.85\linewidth]{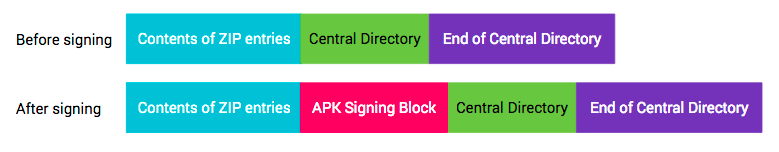}
    \caption{Signature location in an Android APK}\label{img:android_signing}
    \caption*{Source: Android Open Source Project Documentation~\cite{aosp_apksigv2}}
\end{figure}

\paragraph{Package signing, app review, and sideloading} \Gls{AOSP} Android requires that any application that runs is signed by its developer~\cite{android_dev_sign}. The actual mechanism involves verifying that the signature over the APK (Android PacKage) binary matches the public key distributed with the binary. The APK Signing Block contains all of the information required to validate the binary: a certificate chain and tuples of (algorithm, digest, signature). This block is embedded in the binary distributed to devices, as in Figure~\ref{img:android_signing}. Starting with Android 11, Android also provides support for \texttt{fs-verity}, Linux \gls{kernel} feature that allows continuous verification of read-only files on a filesystem at the \gls{kernel} level~\cite{kernel_fsverity}. In Android, \texttt{fs-verity} is used to continuously validate APK files beyond the initial signing check~\cite{aosp_fsverity}. 
 
\begin{figure}[H]
    \centering
    \includegraphics[width=0.6\linewidth]{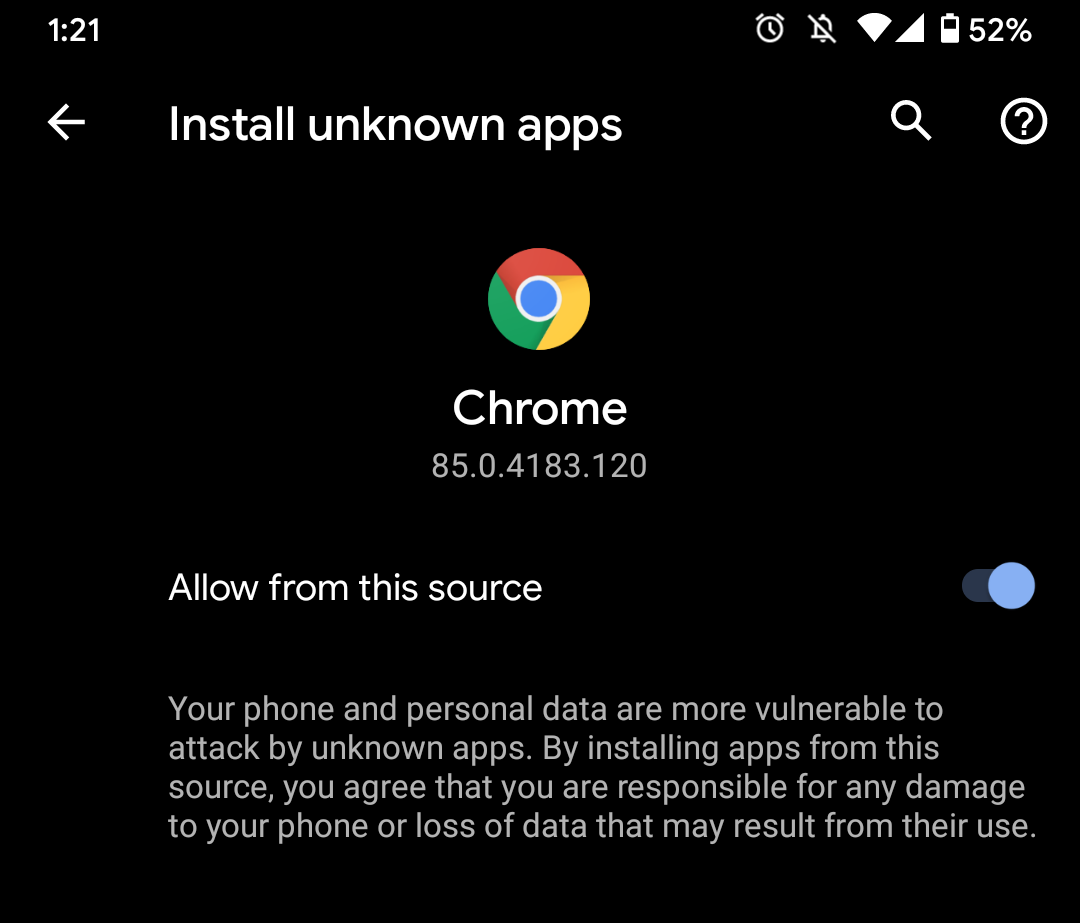}
    \caption{Installing Unknown Apps on Android}\label{img:android_sideload}
    \caption*{Allowing Chrome to install apps from unknown sources, also known as ``sideloading.''}
\end{figure}

The \Gls{AOSP} signature checks do not verify if the developer is legitimate or approved by Google, but simply acts as an authenticity check on the distributed binary itself. Google reviews apps submitted to the Play Store~\cite{play_review}, and Android devices that use GMS incorporate an additional check to ensure that an application is downloaded from the Google Play Store~\cite{play_protect,play_apps}. GMS-enabled Android phones have additional the option of installing apps from ``unknown sources'' by modifying a setting~\cite{hildenbrand2020sideloading}. This allows users to ``sideload'', or install apps that are not distributed on the Play Store. Sideloading is required for the installation of applications from alternative (non-Google approved) developers or storefronts \cite{fdroid,amazon_appstore}. Figure~\ref{img:android_sideload} illustrates an example of this setting, allowing Chrome to download and install unknown apps. This feature is off by default, and enabling it comes with a stern warning in the system, as applications distributed outside of Google's approved channels can be a vector for malware~\cite{collier2019mobile,android_sec_guide_2017,android_sec_guide_2018}.

\begin{figure}
    \centering
    \includegraphics[width=0.5\linewidth,trim={0 0.25cm 0 0.25cm},clip]{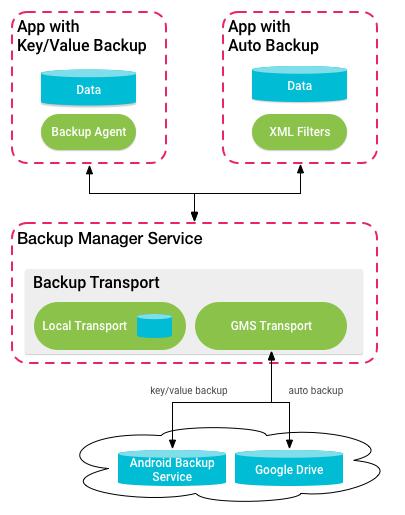}
    \caption{Android Backup Flow Diagram for App Data}\label{img:android_backup}
    \caption*{Source: Android Developer Documentation~\cite{android_dev_backup_test}}
\end{figure}

\paragraph{Backups} GMS provides transport for application data backups for Android devices. Application developers can choose to have their application's data backed up in either of two ways. The original method, available since the GMS in Android 2.2, is a Key/Value Backup (also known as the Android Backup Service)~\cite{android_dev_keyvaluebackup}. As the name implies, it allows apps to backup key and value pairs corresponding to a user's configuration. Use of Key/Value Backup is opt-in, and must be manually configured by the application developer~\cite{android_dev_databackup}. The GMS in Android 6 introduced Android Auto-Backup~\cite{android_dev_autobackup}, which automatically synchronizes app data to a user's Google Drive. This service requires no configuration from the application developer; developers must opt-out of the service on behalf of their users~\cite{android_dev_databackup}. Both types of backups -- Key/Value Backup and Auto-Backup -- are stored on the Google Cloud~\cite{android_dev_backup_test}. The transfer of a backup from a device occurs via GMS~\cite{android_dev_backup_test}. Data flows for these two methods are summarized in Figure~\ref{img:android_backup}. 

\begin{figure}
    \centering
    \includegraphics[width=0.5\linewidth,trim={0 0.25cm 0 0.25cm},clip]{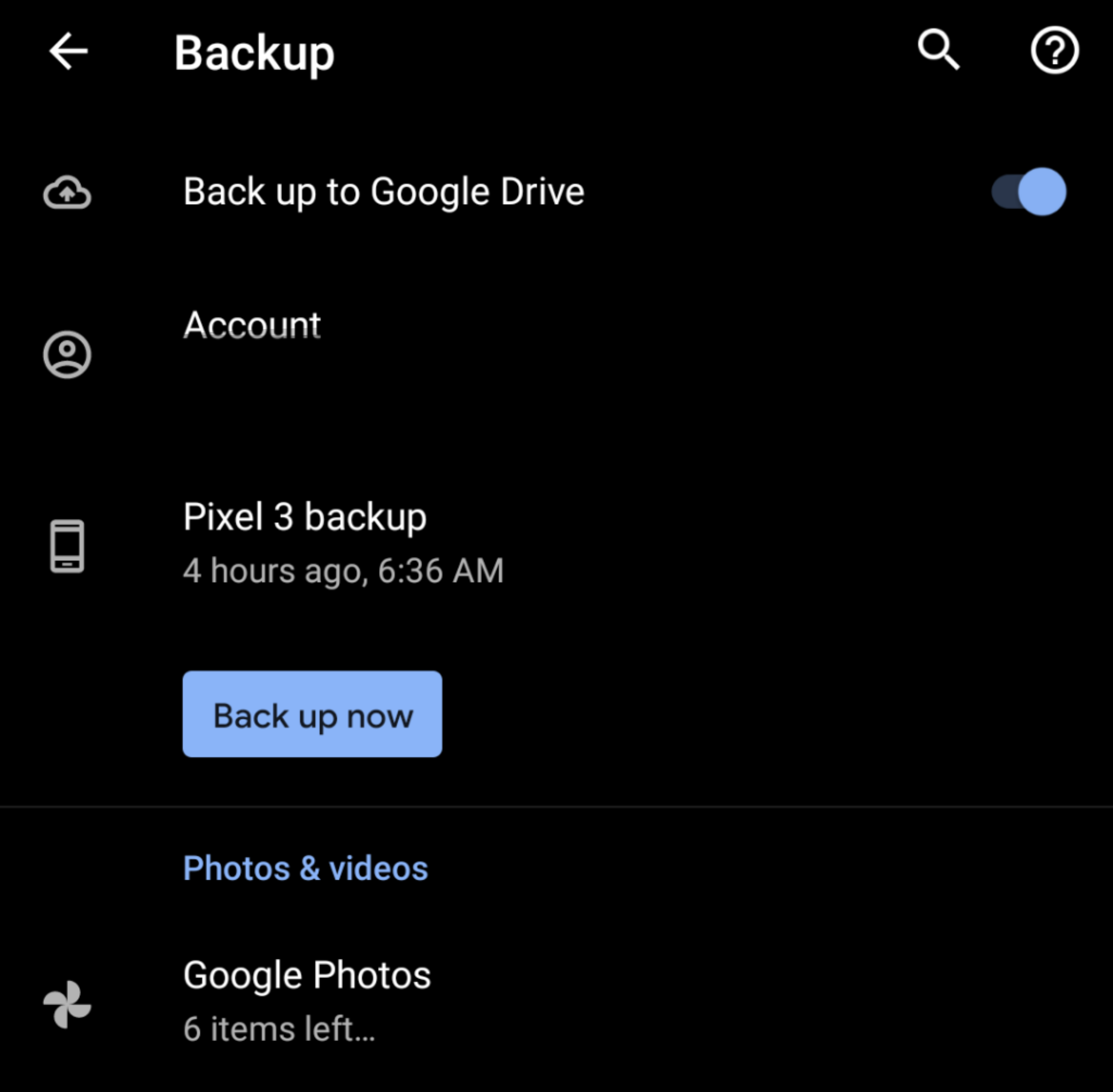}
    \caption{Android Backup Interface}\label{img:android_backup_scrot}
\end{figure}

Since 2018, Google has provided end-to-end encryption for application data backups stored on the Android Backup Service (Key/Value Backup data)~\cite{kensinger2018google}. Devices can generate an decryption key on-device for its backups, and use a user's authentication factor (PIN, password, or pattern) to encrypt this key. To protect this key, Google deploys a Google Titan \Gls{HSM}~\cite{titan_in_depth} as part of Google's Cloud Key Vault service~\cite{ncc_cloud_key}, and device-generated backup encryption keys are held by this device. The Titan \Gls{HSM} is configured to reveal the backup key when presented with the correct authentication from the client (in the form of user's passcode). It also provides brute-force attack prevention and software rollback prevention. However, this end-to-end encryption is only provided for the opt-in Key-Value Backup service~\cite{kensinger2018google}. We find no indication that the automatic Android Auto-Backup service is covered by these guarantees, and it is additionally unclear how many developers manually opt-in to the Key-Value service rather than simply using Android Auto-Backup.

These backup services operate over application-local data only; other forms of personal data on an Android device are separately backed-up to Google services, as shown in Figure~\ref{fig:google_acct_backup}. This data is stored on the Google Account associated with a device, synchronized to Google Drive (similar to the data in Android Auto-Backup). Android support documentation~\cite{androidsupport_backuprestore} mentions that this data is encrypted, either with Google Account password information or with the user authentication factor (PIN/password/pattern). Google has access to encryption keys derived from the user password. However, the documentation is not specific as to what data is or is not covered by the different encryption methods. Refer to Figure~\ref{img:android_backup_scrot} for an example backup configuration for a Pixel 3 running Android 11.

Some device manufacturers implement their own backup services \cite{samsung_backup,lg_backup,htc_backup}. \Gls{AOSP} provides password-protected backups~\cite{passwordbackups_adb} to a local computer via the Android Debug Bridge~\cite{man_adb}, Android's \Gls{USB} debugging tool. There are also a number of third-party backup tools \cite{backupsw_drfone,backupsw_syncdroid,backupsw_syncios}.

\begin{tcolorbox}
\begin{figure}[H]
    \caption{List of Data Categories Included in Google Account Backup}\label{fig:google_acct_backup}
\begin{itemize*}
\begin{multicols}{2}
\item Contacts
\item Google Calendar events and settings
\item \Gls{SMS} text messages (not MMS)
\item Wi-Fi networks and passwords
\item Wallpapers\footnote{In general, photos and videos can be backed up to Google Photos, a separate Google offering.}
\item Gmail settings
\item Apps
\item Display settings (brightness and sleep\footnote{It is not clear from the documentation what other display settings are stored.})
\item Language and input settings
\item Date and time\footnote{The documentation does not specify what exactly this category entails.}
\item Settings and data for apps not made by Google (varies by app)
\end{multicols}
\end{itemize*}
\caption*{Source: Android Support Documentation~\cite{androidsupport_backuprestore}}
\end{figure}
\end{tcolorbox}

\begin{figure}
    \centering
    \includegraphics[width=0.45\linewidth]{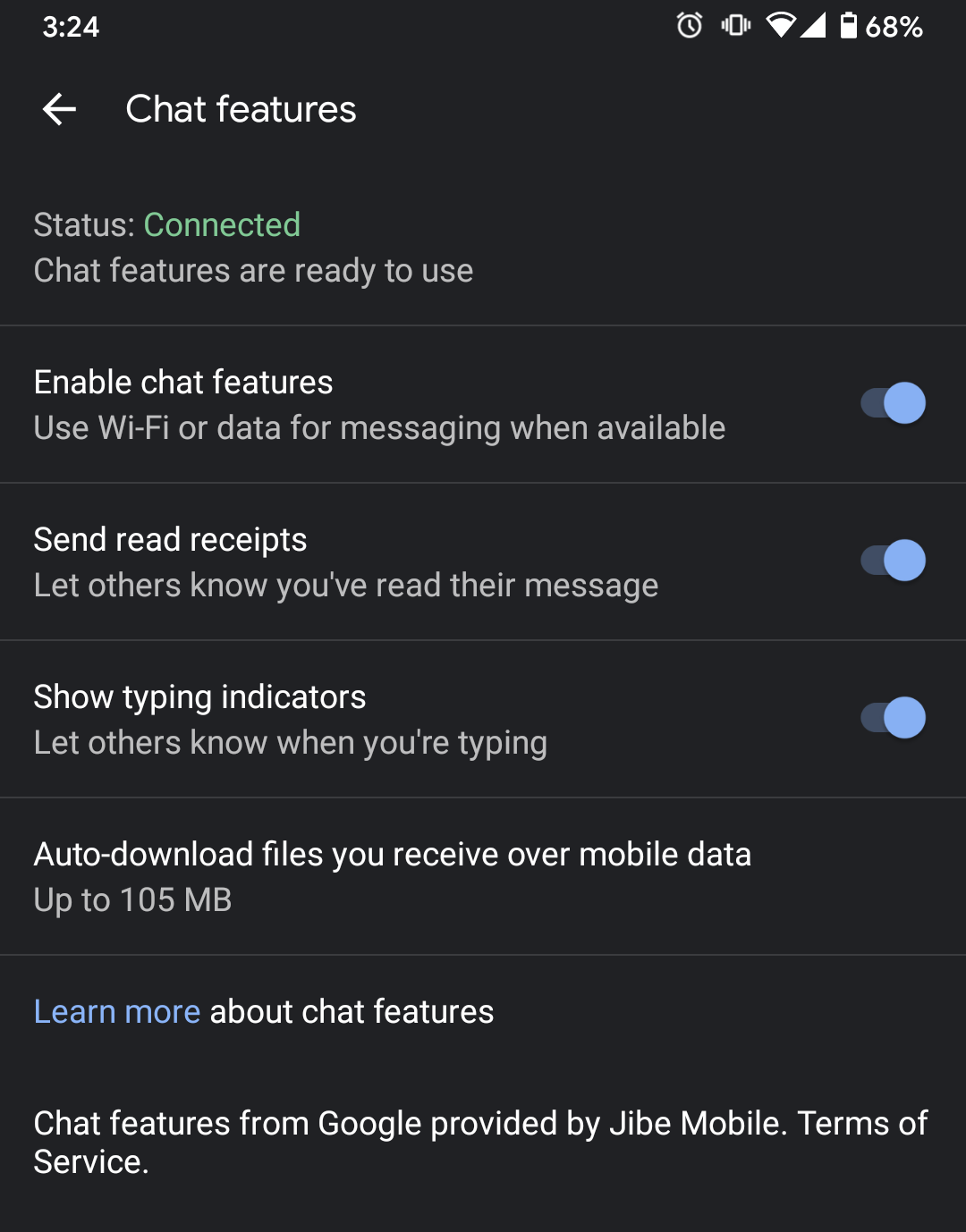}
    \caption{Google Messages \Gls{RCS} Chat Features}\label{img:android_chat_features}
\end{figure}

\paragraph{Messages \& Duo} Android provides support for standard carrier-based
\Gls{SMS} and MMS messaging with a front-end provided from \Gls{AOSP}. While
information is encrypted on-device~\cite{aosp_fulldisk,aosp_filebased}, these
messages are not end-to-end encrypted in transit and are thus vulnerable to both
eavesdropping and spoofing attacks~\cite{verge_cracksms,messagetap2019}.
Moreover, carriers maintain copies of both metadata and content on their
back-ends~\cite{cell_metadata}. The GSMA carrier associate has developed the
more modern Rich Communication Services (\Gls{RCS})~\cite{gsma_rcs} as an
alternative to legacy \Gls{SMS}/MMS. \Gls{RCS} is a communication protocol that
adds features such as typing indicators and read receipts to standard text
messages. Individual providers (such as a carrier) deliver messages to each
other on behalf of their customers~\cite{verge_google_rcs}. Google Messages, a
part of the GMS, supports \Gls{RCS} on Android under the term ``Chat features'',
as seen in Figure~\ref{img:android_chat_features}. Google also provides Jibe
Cloud, which allows users that do not have an carrier support for \Gls{RCS} to
use it~\cite{jibe_cloud}. \Gls{RCS} support is currently unavailable in
\Gls{AOSP} alone, although work is being done to provide system-level support
for it~\cite{git_rcscommit}. \Gls{RCS} communication is secure between
individual service providers~\cite{verge_google_rcs} but is not end-to-end
encrypted~\cite{rcs_not_e2e,verge_google_rcs}. Developer builds of Google
Messages provide evidence that Google is preparing for such an
enhancement~\cite{verge_googlemessages_e2esoon}, and a recent technical draft
from Google confirms this~\cite{google_rcs_e2e}.

\begin{figure}
    \centering
    \includegraphics[width=0.9\linewidth]{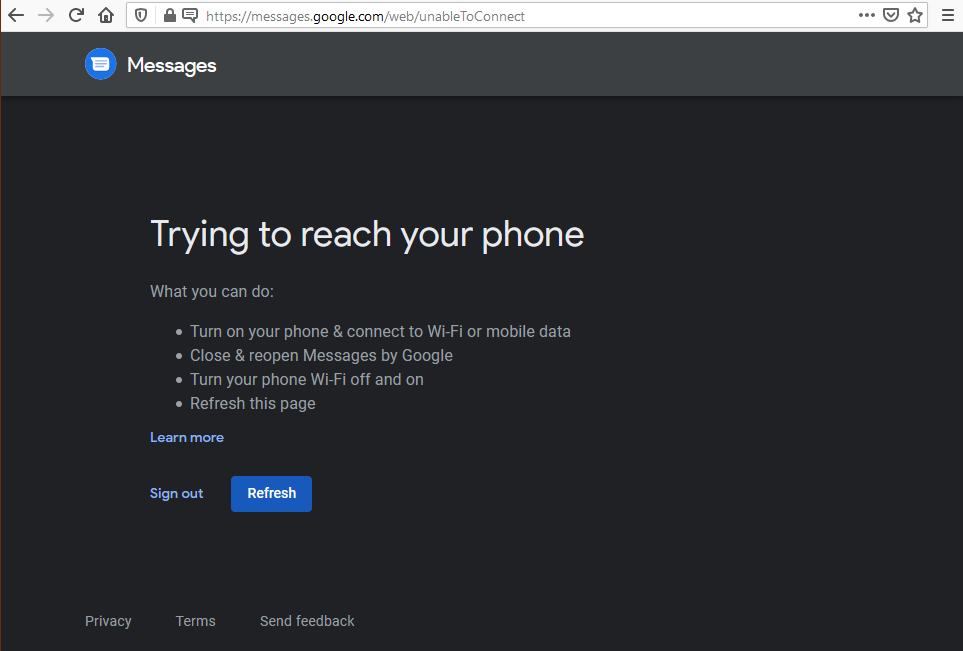}
    \caption{Google Messages Web Client Connection Error}\label{img:android_messages}
\end{figure}

Google Messages also has a web client, wherein messages pass through from the browser through Google's servers to the user's device for sending~\cite{messages_computer}. Our  testing shows that the web client for Messages does not work without a connection to a phone. We turned on Airplane Mode on a Pixel 3 device running Android 11, and noted that the web client was unable to send messages. The error message shown on the web client is found in Figure~\ref{img:android_messages}. There is no mention of end-to-end encryption between the browser and the device in the documentation, which implies that Google is able to see a user's messages as they pass through Google's infrastructure. Google claims to not keep any \Gls{RCS} messages on their backend~\cite{verge_google_rcs} via data retention policies.

\begin{figure}
    \centering
    \includegraphics[width=0.7\linewidth]{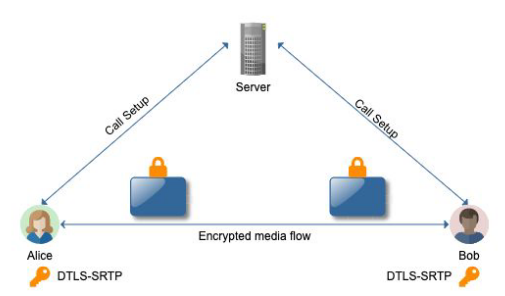}
    \caption{Google Duo Communication Flow Diagram}\label{img:android_duoflow}
    \caption*{Source: Google Duo End-to-End Encryption Whitepaper~\cite{omara2020duo}}
\end{figure}

Duo is a GMS-provided video calling app that allows for end-to-end encryption by default. Devices participating in a Duo call perform key-exchange with each other to create a shared secret, and use that to encrypt the audio and video streams of the data~\cite{omara2020duo}. In a one-on-one call, calls are peer-to-peer, i.e., audio and video data from one device is sent to the other without passing through a server in the middle; only call setup is routed through a Google server initially. This flow is shown in Figure~\ref{img:android_duoflow}. For group communications (and for one-on-one calls that block a peer-to-peer connection), calls are routed through a Google server. Audio and video is still end-to-end encrypted (via pairwise key-exchange between each of the participants), but the server learns metadata about the participants as it sets up the public keys and routes the encrypted communication between the participants~\cite{duosupport_e2e}.

\section{History of Android Security Features}

The previous section highlights key concepts in the current state of Android device security. Android has evolved significantly, however, from its original release in 2008, with major revisions of the operating system adding additional security features. Table~\ref{tbl:aosp_history} highlights the security features added in each release of the \Gls{AOSP}. Appendix~\ref{app:android_history} goes into detail about each item. 

\paragraph{Update adoption} Operating system software updates have been historically slow to propagate to end devices~\cite{malchev2017treble}. Although Android is developed by Google, manufacturers often customize the \Gls{AOSP} source code~\cite{aosp_genkernel}, both to better match their target hardware and to add vendor-specific features. For example, Samsung modifies Android to provide its Bixby virtual assistant on its Samsung Galaxy series of Android devices~\cite{westenberg2018bixby}. These modifications introduce added complexity to the development of update images, which slows update adoption rates~\cite{triggs2019are}. Compounding this is the shorter support lifecycle for Android. There is a broad array of Android phones, offered at different price levels. Those phones in the entry-level and mid-market segments rarely receive updates, if at all~\cite{pathak2010nokia}.

In an effort to reduce this additional hurdle to update deployment, Android 8.0 (2017) introduced Project Treble~\cite{malchev2017treble}, which allows for updates to the Android OS without changing the underlying vendor implementation. Treble decouples the Android framework from the device-specific software that runs on a smartphone. This has improved update adoption rates~\cite{triggs2019are}, especially in the United States~\cite{android_market_share_us}. Update adoption still lags worldwide; as of May 2019, 42.1\% of users worldwide were running version 5.0 (2014) or earlier, translating to around one billion Android devices in use that do not receive security updates~\cite{which_noupdates}.

\begin{table}
\begin{tcolorbox}
\centering
\begin{tabular}{p{0.15\linewidth}|p{0.8\linewidth}}
     \textbf{Android} & \textbf{Highlights of New Security Features} \\ \hline
     2.1 (2010) & APK signing v1 (JAR) \\ \hline
     4.0 (2011) & Keychain credential storage for non-system apps \\ \hline
     4.2 (2012) & Application verification (outside of \Gls{AOSP}) \\ \hline
     4.3 (2012) & SELinux logging policies \\ \hline
     4.4 (2013) & Verified Boot 1 (warning only)\newline SELinux enforcing (blocking) policies\newline Full-disk encryption\\ \hline
     5.0 (2014) & SELinux MAC between system and apps\newline Full-disk encryption by default with TEE-backed storage \\ \hline
     6.0 (2015) & Runtime permissions (instead of install-time permissions)\newline Trusted keystore for symmetric keys\newline APK validation (manifest-based) \\ \hline
     7.0 (2016) &  Stricter DAC file permissions on app storage (751 $\rightarrow$ 700)\newline Verified Boot 1 (stronly enforced)\newline Keymaster 2 (key attestation \& version binding)\newline File-based encryption\newline Uniform system CAs\newline StrictMode file access (removal of \texttt{file://})   \\ \hline
     8.0 (2017) & Verified Boot 2 (\Gls{AVB})\newline \texttt{seccomp-bpf} system call filtering\newline Project Treble/HIDL abstraction layer\newline Google Play Protect (anti-malware)\newline Keymaster 3 (ID attestation)\\ \hline
     9.0 (2018) & APK signing (v3)\newline All ``non-privileged'' apps run in individual SELinux sandboxes\newline Keymaster 4 (key import)\newline Metadata encryption (things not encrypted by File-based)\newline Stricter seccomp \\ \hline
     10.0 (2019) & Limited raw view of FS (no access to \texttt{/sdcard})\newline Removal of full-disk encryption\newline Location access only in foreground\newline Block access to device identifiers\newline Mandatated storage encryption \\ \hline
     11.0 (2020) & File-based encryption enhancements\newline Package-visibility filtering\newline Scoped storage enforcement \\ \hline
\end{tabular}
\def\arraystretch{1}
\caption {History of Android (\Gls{AOSP}) Security Features}\label{tbl:aosp_history}
\caption*{This is not an exhaustive list of features. Some Android versions have been omitted.}
\end{tcolorbox}
\end{table}

\section{Known Data Security Bypass Techniques}

Android's large user base has created a market for exploits that bypass its security features. Even though Android's security has improved over the years (see Table~\ref{tbl:aosp_history}), bypasses continue to be developed both in public and by private groups. The payoff for these bypasses can be substantial: for individuals and groups, large sums in the forms of bug bounties (both from Google~\cite{android_bounty} and others~\cite{zerodium_bounty}), and for governments and law-enforcement agencies, significant access into the lives of individuals under investigation. 

\paragraph{Rooting and software exploitation} Software exploits form the core of bypass techniques on Android, especially given the open-source nature of the \Gls{AOSP}. The usual end-goal of this category of exploits is to attain superuser access, a process known as ``rooting''. A rooted device can be used for non-malicious customization or development~\cite{root_propaganda}, but access to superuser capabilities allows for near-total control of a device, making it attractive for forensics~\cite{android_self_forensics}. Non-rooting software exploits include compromises of individual applications and capabilities that bypass the trusted hardware capabilities on a device, among others. These exploits are discussed in Sections~\ref{sec:android_root} and~\ref{sec:android_other}.

\paragraph{Data extraction} Forensic examiners have several strategies to perform security bypasses. If an examiner has access to a user's password already, none of the Android protections matter and they can typically extract all of the data from both the device and the Google cloud. An examiner that actually has to perform a local device bypass can either do so manually, using available exploits on the internet, or they can do so by contracting the work to a forensics company. For cloud data, law enforcement can either procure information through a legal process with Google or by accessing Google services with tokens extracted from a local device. Local data extraction techniques are discussed in Section~\ref{sec:android_local_extract} and cloud extraction in Section~\ref{sec:android_cloud_extract}.

\subsection{Rooting}\label{sec:android_root}

Rooting on Android allows a standard user to gain superuser permission on a device, with ``a root'' being the term for a process that does so successfully. This is analogous to jailbreaking on iOS, as it allows modification of the underlying operating system. Unlike iOS, however, installation of unauthorized (non-Google Play) Android applications is supported (see Section~\ref{sec:android_prot}), making rooting for this reason less likely. Moreover, some devices explicitly support rooting and bootloader modification~\cite{ok_to_bootloader_unlock} for non-malicious development purposes~\cite{aosp_devicestate,aosp_bootloaderunlock}.

\begin{figure}
    \centering
    \includegraphics[width=0.3\linewidth]{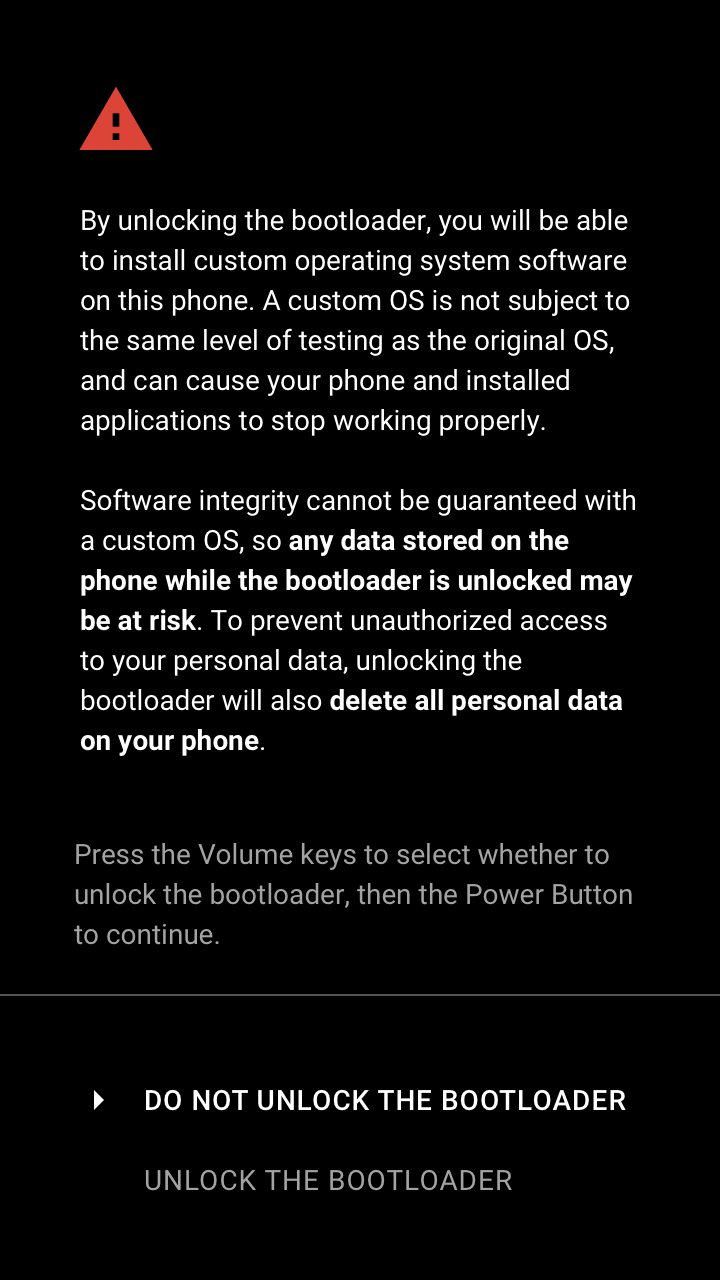}
    \caption{Legitimate Bootloader Unlock on Android}\label{img:android_bootloader}
    \caption*{Source: Android Open Source Project Documentation~\cite{aosp_bootflow}}
\end{figure}

Most roots on Android are related to state of the device bootloader. In locked mode, the bootloader prevents modification to the boot partition, preventing unverified code from running~\cite{aosp_verified,aosp_verifying}, but when in unlocked mode, boot images can be modified, allowing for the application of a root or even the installation of a custom operating system~\cite{aosp_devicestate,aosp_bootflow,aosp_bootloaderunlock}. As a security measure, the process of unlocking the bootloader deletes all personal data on an Android device; the system can no longer trust the bootloader, which means it can no longer assure the data in its storage~\cite{aosp_verified,aosp_bootloaderunlock}. Figure~\ref{img:android_bootloader} shows the default message in \Gls{AOSP} when a bootloader is unlocked through this method. Common roots~\cite{magisk_xda,firmware_xda} operate under this assumption, and do not attempt to circumvent such a data wipe. However, there are some roots that do not require any kind of bootloader unlock, loading the rooting code via an exploit in the operation system. Some of these roots are publicly available~\cite{root_list_xda}, and can be directly sideloaded into an Android device. Other roots must be manually developed for individual devices, which we discuss below. 

Android rooting has been around since the earliest days of the operating system: the first Android device, the HTC Dream (or T-Mobile G1), had a jailbreak found months after its release~\cite{dream_root,dream_release}. At the time, manufacturers and carriers would restrict features (such as \Gls{USB} tethering~\cite{tether_root}) behind payment, or install unwanted applications~\cite{bloat_root}.  Users circumvented disabled these features by rooting their devices. However, as manufacturers started to improve software, rooting became less critical~\cite{no_more_root_opinion}. Roots continue to be developed, however~\cite{magisk_xda,firmware_xda}.

Manufacturers have also moved to reduce the incentives surrounding rooting for non-malicious purposes. Samsung includes the Knox warranty bit, a one-time programmable electronic fuse, in its devices starting in 2013~\cite{samsung_knox_wp}. This fuse ``trips''~\cite{samsung_knox_help} when it detects a unapproved \gls{kernel}, and can only be ``replaced'' with a total teardown of the system board~\cite{samsung_knox_help}. Moreover, Google's SafetyNet attestation service, which validates if a system can be trusted or not, fails on rooted devices~\cite{android_dev_safetynet}. Passing a SafetyNet check is required for certain services such as Google Pay~\cite{xda_googlepay,9to5_googlepay}. Anecdotally~\cite{magisk_xda}, Google SafteyNet has been difficult to circumvent. As of 2017, SafetyNet data shows that only 5.6\% of active Android devices are rooted (either by consent or maliciously)~\cite{android_sec_guide_2017}. Google is currently testing hardware attestation for SafetyNet in some devices, which could further disincentivize legitimate rooting~\cite{androidpolice_hardware_safetynet},

\begin{figure}
    \centering
    \includegraphics[width=0.6\linewidth]{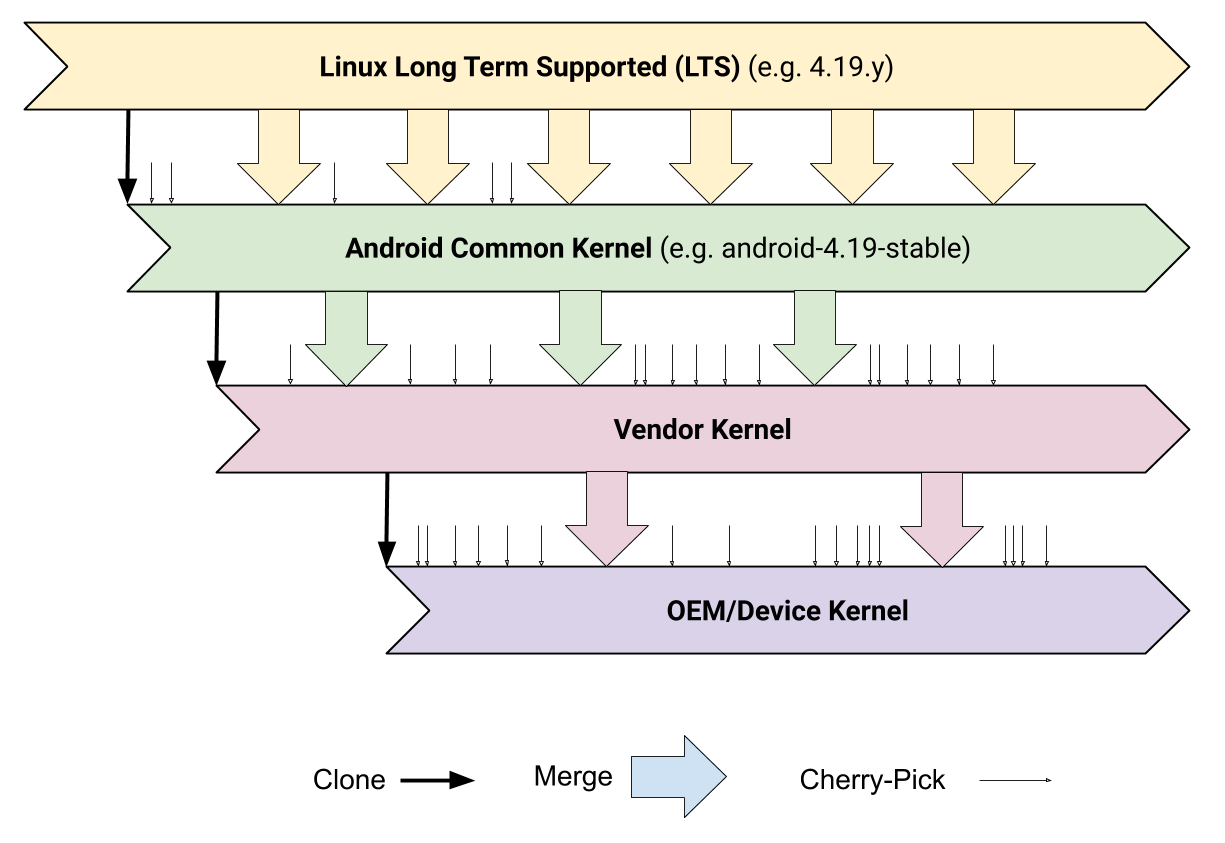}
    \caption{\Gls{kernel} Hierarchy on Android}\label{img:android_kernels}
    \caption*{Source: Android Open Source Project Documentation~\cite{aosp_genkernel}}
\end{figure}

For the purposes of this discussion, we focus on manual, exploit-based roots. These are roots that are achieved as a result of a vulnerability in Android, and do not trigger an erase of data on the phone, and are potentially more valuable in forensics, as a result. We consider exploitation at two levels: the core operating system (\gls{kernel}) and device-specific code (SoC vendor and OEM code).

\paragraph{Exploiting the core operating system} An Android device \gls{kernel} is the combination of several patches and drivers from various sources, as shown in Figure~\ref{img:android_kernels}. Underlying the entire system is a \gls{kernel} from the long-term support branch of Linux with Android-specific changes (the Android Common \Gls{kernel}). On top of this, the system-on-chip vendor can make modifications to better support its hardware, and the final OEM manufacturer of the device can add even more patches~\cite{aosp_genkernel}. Software vulnerabilities can appear at any of these levels and compromise the entire system. 
Although Google and others make several patches to the Android \gls{kernel}, the \gls{kernel} is still Linux at its core. From a security standpoint, this means that exploits that provide superuser access on Linux may have impact on Android as well. For example, Dirty CoW was a general Linux vulnerability that used a race condition to write to system files~\cite{nvd_cve_dirtycow}; this vulnerability was then applied to Android, and developers created a root exploit based around it~\cite{android_dirtycow_root}. There are also vulnerabilities that were originally discovered on Android that actually impacted the Linux \gls{kernel} at-large. An example of this is PingPongRoot, a use-after-free vulnerability used in an Android root that required a \gls{kernel} patch upon disclosure~\cite{pingpongroot191942}.

Zero-day Android root vulnerabilities are still being actively found and exploited~\cite{samsung_zday,TiYunZong,quickroot}. These operating system-level vulnerabilities are attractive for forensics because they are more or less device-agnostic. The fragmented state of Android hardware means that a common operating system exploit can be applied to more devices. 

\paragraph{Device-specific bypasses} This class of vulnerabilities is specific to properties associated with a particular device. These vulnerabilities may require more expertise and time to properly exploit, but can completely ignore the protections provided by the Android system. Generally, these bypasses look at a lower level of abstraction, such as system-on-chip firmware.

A high-profile~\cite{nvd_cve_qcomm_edl1,nvd_cve_qcomm_edl2} example of this is Qualcomm EDL. Qualcomm devices have an ``Emergency Download Mode'' (EDL)~\cite{aleph_qcomm_edl_1} that allow OEM-signed programmers to flash changes onto the system. When placed into EDL mode (usually by hardware circuit shorts~\cite{aleph_qcomm_edl_1}), an Android device boots into a debug interface rather than the usual Android Bootloader. Then, using the tools of a programmer, an attacker can flash data onto the Android partition. These programmers are ostensibly guarded by an OEM, but are often publicly available~\cite{aleph_qcomm_edl_1}. By flashing the partition with such a programmer, attackers have been able to attain root and exfiltrate data, even on encrypted devices. For example, EDL can be used to flip the bootloader lock bit on some devices, which can then boot into a custom image with a root shell, which can in turn be used to dump the data from a device~\cite{aleph_qcomm_edl_2}. 

Qualcomm EDL is not the only low-level bypass on Android. Certain MediaTek-based devices can be rewritten (or ``flashed'') without a bootloader unlock, using the publicly-available SP-Flash-Tool~\cite{xda_sp_flash}. There is also work on using AT modem commands to circumvent protections on Android~\cite{tian2018attention}. This is not an exhaustive list of all possible bypasses, but a reflection of the security issues that arise due to the complexity and fragmentation in the Android space.

\subsection{Alternative Techniques}\label{sec:android_other}

Root-level access for an Android device, while a sufficient condition to bypass Android permissions checks, is not a necessary one. We briefly discuss alternative methods of bypassing Android security controls that do not strictly rely on rooting the phone as an attack strategy.

\begin{figure}
    \centering
    \includegraphics[width=0.9\linewidth]{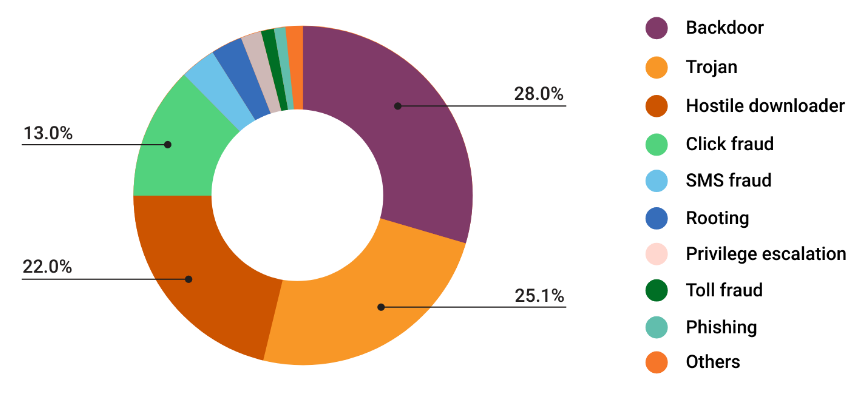}
    \caption{Distribution of PHAs for Apps Installed \emph{Outside} of the Google Play Store}\label{img:android_pha}
    \caption*{Source: Android Year-in-Review 2018~\cite{android_sec_guide_2018}}
\end{figure}

\paragraph{Malware and app compromise} Android malware is rare, but has not been eradicated. According to Google-released data, 0.68\% of devices that allowed installations from unknown sources had some form of PHA, compared to 0.08\% of devices that allowed installation from the Google Play Store exclusively~\cite{android_sec_guide_2018}. Google's PHA definition is broad: categories include ``click fraud'', a form of adware that generates clicks on ads surreptitiously~\cite{android_sec_guide_2018} -- not a form of PHA that would lead directly to data bypass -- as well as rooting apps, even if they were user-wanted. Figure~\ref{img:android_pha} is a chart of all PHAs on Android found outside of the Google Play Store. In general, an attacker who has spyware or another form of malware installed already on a device has a clear advantage from a bypass standpoint one that does not. It is non-trivial to have an app installed before forensic analysis, though; an examiner would have to either fool the user into installing a PHA or coerce them to do so.

A second, but related, bypass technique is to leverage insecurities in existing app to bypass protections. An app already on the device may have some kind of open interface or remote code execution vulnerability that can be leveraged to bypass security protections and retrieve information. For example, a double-free vulnerability in WhatsApp could be used to remotely gain a shell and steal from the WhatsApp message database~\cite{whatsapp_app_rce}. Compromising an app only allows a limited exploit context, as the regular Android sandboxing protections are still in place. But, it does provide access all of the information of an app and an opportunity to pivot into the rest of the system. 

\paragraph{Trusted hardware attacks} Exploitation of trusted components on Android either target the trusted hardware itself or on the TEEs that run on them. Researchers have found several issues with TrustZone and TrustZone-enabled TEEs over the years~\cite{cerdeira2020sok}. Some TrustZone exploits (as described in ~\cite{cerdeira2020sok}) require existing access to the device, in the form of an app installed on the device already. These exploits are less useful for forensics, as the threat model there involves data recovery for systems without prior access. 
Still, these kinds of attacks could be used as a part of an exploit chain instead, gaining a foothold in the system first, and then attacking TrustZone and the TEE. As of this writing, there are no public security exploits for the Samsung Secure Processor and Google Titan M secure element co-processors, making them valuable targets: Google in particular offers \$1,000,000 for a successful exploit of Titan M~\cite{android_bounty}.

\paragraph{Passcode guessing} If an Android device using file-based encryption or full-disk encryption is rebooted, it will request a passcode for decryption. A six-digit PIN or simple pattern can be brute-forced in a manner of minutes on commodity hardware, so Android employs Gatekeeper as a protection mechanism. There are two components of Gatekeeper: a hardware-abstraction layer (HAL), which is an \Gls{API} for clients to enroll and verify passwords, and a trusted application (TA), which actually runs on a TEE. Gatekeeper creates an HMAC for every passcode that it has approved; this HMAC also includes a User Secure Identifier (SID), which is generated by a secure random number generator upon enrollment with Gatekeeper. It validates that HMAC of passcode attempts matches that of the approved passcode. The HMAC key that Gatekeeper uses for this purpose does not leave the TEE. As a brute-force prevention mechanism, Gatekeeper enables request throttling: the TA does not respond to service requests if a timeout is in place~\cite{aosp_gatekeeper}. Circumventing Gatekeeper would involve breaking the Gatekeeper TA as implemented in a specific TEE OS, exploiting the entire TEE/TrustZone architecture as discussed above, or finding a loophole that skips the Gatekeeper check entirely. We discuss evidence of such circumvention in Section~\ref{sec:android_local_extract}.

\paragraph{Lock screen bypasses} Lock screen bypasses are cases wherein a user can unlock a phone or gain access to functionality normally restricted while the phone is locked without the user authentication factor or biometric. While Google's Smart Lock feature~\cite{androidsupport_smartlock} bypasses the lock screen in some cases, this is by design. We did not find evidence that unauthorized UI-driven lock screen bypasses are commonplace, or relevant to Android forensics.

\subsection{Local Device Data Extraction}\label{sec:android_local_extract}

Once an Android device is seized, an examiner can proceed with extracting the data on it. If a device is seized while it was in use, it should have all of its encryption keys in memory already: in full-disk mode~\cite{aosp_fulldisk}, the system's user data partitions would already be unlocked, and in file-based mode~\cite{aosp_filebased}, the credential-encrypted (CE) keys would have to be in use. 
Of course, in the best case (for law enforcement), the user could have their PIN or password written nearby. With this, a forensics examiner can simply unlock the phone, and use the Android Debug Bridge~\cite{man_adb} to pull a backup off the device via \Gls{USB}, obviating the need for any cracking of the Android device. Also, if the device has an SD card (that is not adopted or encrypted by the system~\cite{aosp_adoptable}), an examiner can simply pull it out of the system and analyze it directly for data. 

\begin{figure}[H]
    \centering
    \includegraphics[width=0.9\linewidth]{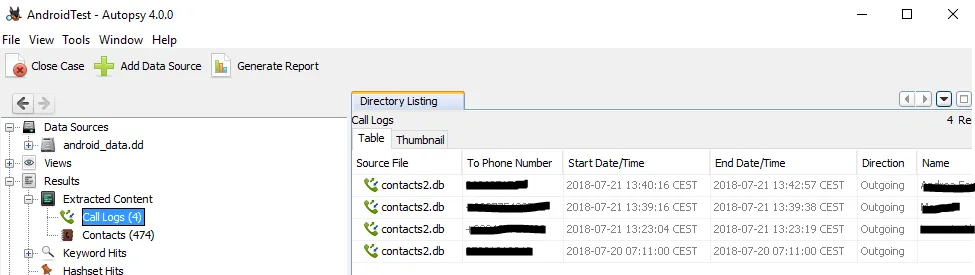}
    \caption{Autopsy Forensic Analysis for an Android Disk Image}\label{img:android_autopsy}
    \caption*{Source: Andrea Fortuna~\cite{android_self_forensics}}
\end{figure}

\begin{figure}
    \centering
    \includegraphics[width=0.7\linewidth]{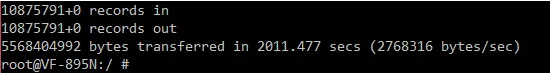}
    \caption{Extracting Data on a Rooted Android Device Using \texttt{dd}}\label{img:android_root_dd}
    \caption*{Source: Andrea Fortuna~\cite{android_self_forensics}}
\end{figure}

If the device is indeed protected, an examiner has several options. The end goal is to extract all of the information off of a device, typically by using some kind of forensic explorer to read organize information. If the device is not secured or has an easily-exploitable vulnerability, (and if the agency has sufficient expertise), they may run exploits directly against the target, and run a tool (such as the open-source Autopsy~\cite{sleuthkit_autopsy}, Figure~\ref{img:android_autopsy}) against the device. Many of the roots discussed in Section~\ref{sec:android_root} were created by members of the Android development community for personal use~\cite{magisk_xda,firmware_xda}, but can be re-purposed to gain access onto a device and attain a root shell~\cite{android_malware_roots}. The shell can then be used to clone the system (using \texttt{dd}, Figure~\ref{img:android_root_dd}) for analysis off the phone~\cite{android_self_forensics}. 

\begin{figure}
    \centering
    \includegraphics[width=0.8\linewidth]{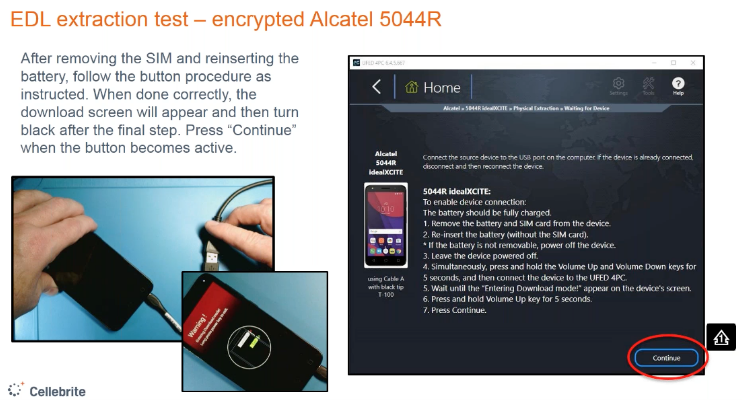}
    \caption{Cellebrite EDL Instructions for an Encrypted Alcatel Android Device}\label{img:android_cellebrite_edl}
    \caption*{Source: Privacy International~\cite{pi_deepdive}}
\end{figure}

\begin{figure}
    \centering
    \includegraphics[height=0.7\textheight]{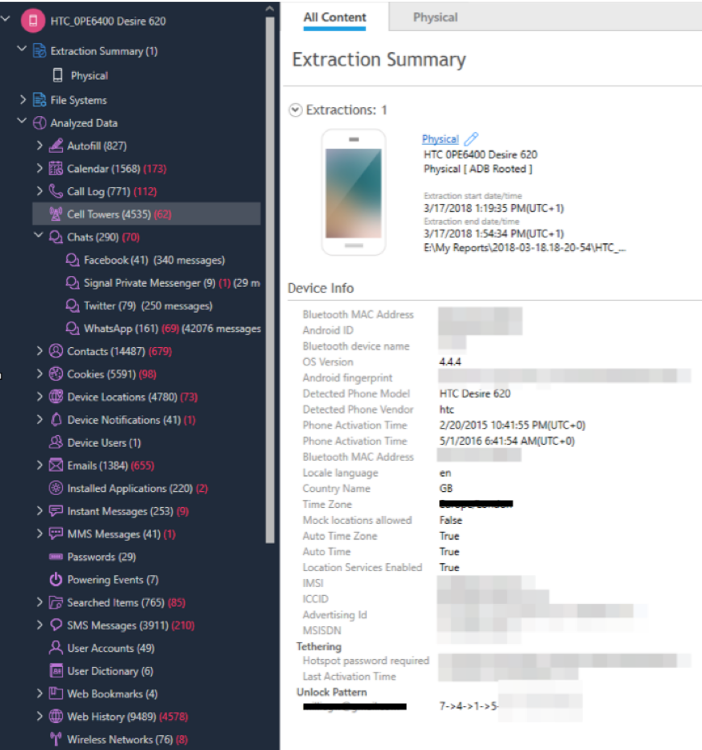}
    \caption{Cellebrite UFED Interface During Extraction of an HTC Desire Android Device}\label{img:android_cellebrite_ufed}
    \caption*{Source: Privacy International~\cite{pi_deepdive}}
\end{figure}

If additional assistance is required, law enforcement examiners can hire a forensics company, like Cellebrite, Oxygen, or Magnet. These organizations are secretive about their proprietary methodologies, but do leave hints about their methods in marketing documentation~\cite{cellebrite_advanced_services,oxygen_latest,magnet_axiom}. Often, these forensics tools draw from a library of exploits. For example, Cellebrite, Oxygen, and Magnet all employ Qualcomm EDL programmers (described in Section~\ref{sec:android_root}) to extract data~\cite{pi_deepdive}. A procedure to put an Alcatel 5044R device in EDL mode for Cellebrite's UFED software can be found in Figure~\ref{img:android_cellebrite_edl}. 
EDL programmers must be signed by the vendor, but many of them have been leaked online~\cite{aleph_qcomm_edl_1}. It is unknown if forensics organizations are legally allowed to use these binaries for this purpose. Once the bypass is run, forensics software typically analyze and exfiltrate data; example output from Cellebrite UFED is found in Figure~\ref{img:android_cellebrite_ufed}.

\subsection{Cloud Data Extraction}\label{sec:android_cloud_extract}
Android has tight integration with Google services, and as a result Android devices store significant amount of information on Google's servers. GMS apps such as Gmail, Google Drive, and Google Photos store data on the Google Cloud backend with no guarantees of end-to-end encryption. This means that Google has unfettered access to this information, and is able give this information to law enforcement upon request. The only exception to this (according to available information) would be Key/Value Backup data stored in the Android Backup Service~\cite{kensinger2018google}, which is end-to-end encrypted.

Google complies with law enforcement requests for information, processing 20,000 search warrants for information in the United States in 2019~\cite{google_transp_2019}. Unlike Apple's iCloud, 
Google's cloud offerings are platform agnostic, and as such Google has information about iPhone users in addition to Android users. Law enforcement has expressed particular interest in Google's store of location data for users~\cite{nyt_sensorvault}. If law enforcement does not wish to go through Google (and potentially have to pay to service their request~\cite{nyt_pay_to_win}), forensic software can attempt to recover cloud authentication tokens on a device and then use it to manually download data from an individual's Google account~\cite{cellebrite_cloud,pi_cloud_extraction}.

\subsection{Conclusions from Bypasses}

The primary takeaway from this discussion is that there are many techniques to bypass user data protections on Android. Lacking an analogue to iOS Complete Protection, decryption keys for user data remain available in memory at all time after the first unlock of the device; live extraction then becomes a question of breaking security controls instead of breaking cryptography or hardware. Additionally, the extent of Google's data collection affords law enforcement and rogue actors alike considerable user data, acquirable either through the legal system or through a device bypass. We discuss further improvements to Android in greater detail in Section~\ref{sec:android_improve}.

\section{Forensic Software for Android}

As Android phones have gained market share, so have Android forensics tools. As with iOS, it is critical to realize the accessibility of professional forensic tools such as Cellebrite's UFED~\cite{cellebrite_ufed}, and even of individualized consulting services such as Cellebrite's Advanced Services~\cite{cellebrite_advanced_services} for unlocking phones. Law enforcement agencies, including local departments, can unlock devices with Advanced Services for as cheap as \$2,000 USD per phone, and even less in bulk~\cite{upturn_mass_extraction}, and commonly do so~\cite{upturn_mass_extraction,vice_db_news,pi_deepdive}.

For a complete list of the forensic tools tested by \Gls{DHS}, as determined by publicly-available reports as of this writing, and the data forensically extracted by those tools, see Appendix~\ref{app:forensic_tools} and archived \Gls{DHS} reports~\cite{dhs_forensics}. Again unfortunately, the NIST standard for testing devices is unclear as to whether the device should be in a locked state during testing~\cite{nist_forensics_spec}; we note, however, that certain categories of data seem inaccessible to forensic software in various cases, which we attribute to Android security controls. It is possible that these failures occur due to the forensic tools inability to correctly extract and display data they should have access to, but based on the patterns of inaccessible data in the reports we consider this unlikely. If this assumption holds, in summary these reports imply the following:

\medskip \noindent
{\em Successful extraction.} In most tests, most or all of the targeted data (see Figure~\ref{list:nist_forensics}) is successfully extracted against the latest Android devices. One notable omission, however, is location data: while forensic tools were largely able to extract data from Android phones, we speculate that since Maps and Location is a GMS component, this data may be stored exclusively in the cloud, leaving none for local extraction.

\medskip \noindent
{\em Limited software diversity.} A small number of forensic software companies frequently iterate their products. This is demonstrated in the \Gls{DHS} tests as many Cellebrite, XRY, Oxygen, MOBILedit, Secure View, and Lantern devices being tested between 2009 and 2019 (see Appendix~\ref{app:forensic_tools}), and each generally successfully extracting data from contemporaneous generations of Android devices over time~\cite{dhs_forensics}.

\medskip \noindent
\emph{Slow adoption of new test devices.} The tests for Android show results for a number of different Android versions, with subsequent tests either remaining on an old version, or skipping over several major versions. Studies would continue to use older Android versions years after they were released. Even in 2019, the latest version tested in one report was 7.1.1 -- a version of Android first tested in 2017 -- even though less than 9\% of U.S. Android users were on Android 7 at the time~\cite{android_market_share_us}. 

\medskip \noindent
{\em Reporting inconsistencies.} Beginning in February 2016, the quality of the reports degrades notably, and it is unclear in many cases whether data was extracted from Android and not displayed properly by the forensic software, or simply not extracted at all. Some reports\footnote{For example see April 2017 ``Electronic Evidence Examiner - Device Seizure v1.0.9466.18457.''} showed inconsistencies between the analysts notes on forensic performance and the summary tables in the final report (e.g. claimed extracted data in the notes, but this success not indicated in the summary table).

Regardless of irregularities in methodology, it is likely given these reports that agencies such as \Gls{FBI} and \Gls{DHS} that have access to these tools are largely able to bypass Android protections and exfiltrate data from Android devices. 

\section{Proposed Improvements to Android}\label{sec:android_improve}

Android security has improved significantly over the years, with modern Android using more cryptography and security controls than ever before. However, forensics experts are still able to compromise the Android system and bypass its protections. Below, we briefly describe some potential improvements that Google can make to Android to make it more resilient against bypass.



\paragraph{Encrypt user data on screen lock} The most important improvement Android can make would be to encrypt user data with a key that is evicted in memory after the screen is locked. File-based encryption with device and credential-encrypted storage types was a step in the right direction, but adding this protection will ensure cryptographically that personal data is unavailable while the user is not using the device. This is not a trivial change, as it would require modifications to the Android process lifecycle~\cite{android_dev_lifecycle} to notify apps that the screen was locked and that their processes no longer have access to their data. This, in turn, would force app developers to adapt their application to the new lifecycle, or risk losing users due to app failures. However, if implemented correctly (and throughout the Android system), such a file encryption mechanism would significantly improve the privacy guarantees of the Android system, and make root exploits less effective.

\paragraph{Implement end-to-end encryption for messaging} Currently, Google is
deploying \Gls{RCS} as a part of its Google Messenger app on
Android~\cite{verge_google_rcs}. Adding end-to-end encryption (as they plan
to~\cite{google_rcs_e2e}) will
improve security for Android users, and using a platform-independent protocol may
help increase end-to-end encryption adoption worldwide. This will also allow
Google to provide stronger guarantees of security for the Messages web application,
as end-to-end encryption would ensure message security between the web client
and the mobile device using encryption rather than relying on a retention policy.

\paragraph{Provide end-to-end security for more Google products} The only offering that offers end-to-end security in the Google service ecosystem is Duo for real-time video calls. Google should add end-to-end encryption for more information in its cloud, especially for data generated or stored by GMS apps on Android. This is a harder ask for Google, as its services run sophisticated machine learning over user data to improve the quality of its service~\cite{google_ml_good}. However, at the very least, Google should extend its end-to-end encryption guarantees from the Android Backup Service to the Android Auto-Backup service as well. Automatic end-to-end security for users and their app data may create tension with law enforcement agencies~\cite{ApplevFBI_apple,ApplevFBI_fbi} but would be a boon for the privacy of users.

\paragraph{Secure the low-level firmware programming interface} Low-level exploits bypass the careful Android security model Google has developed. Firmware programmers are widely available; this makes attacks practical~\cite{aleph_qcomm_edl_1,aleph_qcomm_edl_2}, with forensics software~\cite{pi_deepdive} employing these methods to bypass security controls. As such, SoC vendors and device OEMs should work on securing access to firm. One option is to distribute programmers as hardware devices with tamper-resistant components. In this way, vendors can keep track of programmers and ensure they are not easily leaked. Another option is to consider revoking the certificates of compromised programmers. Phones would flash their firmware with these revocations, and would not respond to firmware mode if it detects a programmer on its revocation list.

\paragraph{Leverage strong trusted hardware} Secure element co-processors add additional security over shared-processor designs such as TrustZone. More parts of Android should tie itself to a secure element, as the separate hardware is more resistant to hardware and software-bound attacks. Android is a worldwide platform, however, and there may be situations where a secure processor cannot be added to a device for price constraints. TrustZone, while not as robust as a secure element, should be used as a fallback. Recommendations like reducing TEE size and bolstering hardware against side-channel attacks \cite{cerdeira2020sok} are sensible improvements to improve TrustZone security for devices that do not support a secure element. 

\paragraph{Expand update efforts} The latest Android devices seem to employ relatively strong data protection techniques. However, Android phones run older software versions in massive numbers~\cite{statista_android_versions}. For devices which are capable of receiving the latest updates, increased efforts for user education and incentivization could improve update numbers, especially paired with improved manufacturer outreach. Devices which have reached end-of-life and cannot receive the latest updates must be clear to users about the risks entailed.

\paragraph{Derive mutual benefit from shared code} As we discuss, Android is based on the Linux \gls{kernel}, and as such derives both the benefits and drawbacks of such sharing. The open-source Linux \gls{kernel} is a widely used, deeply analyzed operating system, and Google is a powerful force for vulnerability discovery and mitigation~\cite{project_zero}. By dedicating time, resources, and engineering effort toward the betterment of the Linux \gls{kernel}, Google immediately reaps the benefits of such efforts through improvements to Android. Similarly, Google may benefit immediately from the open-source efforts of improving the security and performance of the Linux \gls{kernel}.

\paragraph{Embrace the federated nature of Android} One of the greatest strengths of Android is that it is an open ecosystem of several manufacturers. While this can create fragmentation, this also allows for the opportunity to design secure systems that do not rely on a single party (Google) to function. Google should use Android to champion open protocols and manufacturer interoperability through common standards. This will provide end users with thoughtfully and securely designed systems while still maintaining the level of choice they come to expect from the Android ecosystem.

\chapter{Conclusion}\label{sec:conclusion}

Privacy is critical to a free and open society. For one, people behave differently when they know they might be surveilled \cite{chilling_effects1,chilling_effects2,chilling_effects3,chilling_effects4}, and so privacy-preserving communication channels can alone enable truly democratic discussion of ideas. As more of our data is gathered and our interactions occur on (particularly mobile) devices, these considerations become more important. Worse, harms from violations of privacy are concentrated among some of the most vulnerable, chronically disenfranchised populations in our societies such as the Uighur population in China~\cite{uyghurs} and Indigenous protesters in the United States~\cite{surveillance_of_protesters}.

Smartphones have the potential to fundamentally change the balance of privacy in our lives. Not only do they contain our daily schedules and locations, they store and deliver our communications, rich media content documenting our activities, and are a gateway to the Internet. Our devices also contain information about our families and peers, meaning that compromise of their privacy affects our entire interpersonal network. With the rapid advances of data science, machine learning, and the industry of data aggregation, the potential privacy loss due to phone compromise is difficult to overstate. When we fail to protect our data, we are making privacy decisions not only for ourselves but on behalf of anyone with whom we communicate and interact.

The questions we raise in this work are primarily technical. But they stand for a larger question: how do we improve privacy for users of mobile devices? In this work we demonstrate significant limitations in existing systems, and show the existence of bypasses to security controls which protect our. We also present ideas which we hope can be adopted, extended, and improved towards the overall goal of improving privacy. Policy and legislative solutions are also very relevant toward this end; we leave this analysis for experts in the respective fields~\cite{upturn_mass_extraction}, but to summarize their work, meaningful opportunities for change include limiting law enforcement rights to search devices, requiring robust data sealing and deletion standards, and increasing transparency through audits and logging.

Many of these limitations are centered around what data is encrypted when, and where encryption keys are stored. Encryption, unlike any operating systems security control, lacks the complex state space which contributes to vulnerabilities in software. Whereas the operating system must contend with and manage various user and system contexts, and correctly provide access control and handling from even potentially malicious sources, encryption can be summarized in brief: is the data encrypted using a strong cipher, and where are the keys? We find that while much data on iOS and Android is stored encrypted, the keys are often available in memory. This creates an opportunity for a compromised OS kernel to exfiltrate data as we see in various forensic tools and bypasses. Further, in Android we find that many widely-used but outdated versions of Android offer even more limited coverage of encryption, up to as weak as only encrypting data when the device is off. While modern versions offer strong and more granular file-based encryption, older models are relegated to disk encryption; disk encryption is wholly unprepared for the stronger adversaries we consider in our threat model, where running devices may be seized at any time. In the cloud, both platforms extensively store user data on behalf of devices, and while there are options for end-to-end encrypted content such as app developer opt-in backups on Android and certain data categories on iOS, this coverage is limited due to design decisions by Apple and Google.

Secure hardware offers compelling security benefits on both mobile platforms, particularly by giving the devices a place to store encryption keys without risking their immediate compromise. It is these components, whether the Secure Enclave on iOS or TrustZone on Android, which allow mobile devices to contend in our stronger threat model, far beyond what for example most laptop computers could hope to. Secure hardware is the only reliable method of storing encryption keys on-device and protecting them that we find in the industry, modulo potential bypasses of Secure Enclave technology (e.g. by GrayKey) and vulnerabilities in TrustZone.

Going beyond the current state of the art, there are exciting possibilities to create novel methods for increasing the coverage of encryption without limiting performance. This is a deeply technical challenge which may draw in research from cryptography, systems security, and even machine learning in the example of creating prediction systems for limiting available decryption keys. There are also compelling questions about update distribution and synchronization to be answered. Both platforms could benefit greatly from the work of secure multi-party computation researchers in developing privacy-preserving replacements for sensitive but desirable cloud services such as backup, data aggregation, and predictive systems. Additionally, extensive usable security research and other sociological studies can answer questions about how people expect security and privacy controls to work, and compare any gaps with how they are truly implemented. User studies are difficult and expensive, but also have great potential to aid development research by informing them with a human perspective. Follow-up analysis of the effectiveness of security awareness and education works are also critical in this regard.\footnote{A relevant example can be found in~\cite{van2014studying}.}

Privacy lies at the crossroads of technology and policy, of academic and engineering interest, and of producer and consumer effort. Effective privacy controls can be mandated through legislative efforts~\cite{gdpr} as well as through technical design considerations~\cite{apple_privacy,google_titanm}. Thus there is much opportunity for improvement, but these intersections complicate solutions due to differences in understanding, language, and motivations amongst policymakers and technology designers. The challenges we highlight in this work, then, can largely be solved through mutually informed efforts in the policy and technical domains. We urge researchers and engineers to collaborate with policymakers, advocates, and experts in these efforts. Towards improved usage norms, towards efficient secure protocols to enable sensitive use-cases, and towards mobile device privacy \textit{by default} for all people.

\section*{Acknowledgments}\label{sec:ack}

The authors would like to thank Upturn, particularly Emma Weil and Harlan Yu, for their work on the recently released document \textit{Mass Extraction}~\cite{upturn_mass_extraction} and for their excellent and helpful feedback on this work.

The authors would also like to thank the informal reviewers of this work for their technical contributions.

\newpage

{\footnotesize 
\bibliographystyle{unsrt}
\bibliography{main}}

\appendix
\chapter{History of iOS Security Features}
\label{app:ios_history}

\section{iOS Security Features Over Time}

\subsection{Encryption and Data Protection over time} In 2009, iOS 3 was released on the iPhone 3GS with the first iteration of Data Protection, which entailed encrypting the flash storage of the device when powered off, primarily to facilitate rapid erasure of the device. In this initial release, the user passcode was \textit{not} mixed into the encryption key and thus was not necessary for decryption. In 2014, iOS 8 (released along with the iPhone 6 and 6+) increased the coverage of data encrypted using a key derived from the user's passcode, protecting significantly more data with keys unavailable without the user's consent. This decision drew significant criticism from law enforcement, most notably in an early speech by former FBI Director James Comey~\cite{comey14}. Since its inception, Data Protection classes have been introduced and applied to various data as summarized in Table~\ref{tbl:ios_history}. With Data Protection, when files are deleted, their keys are evicted and as such this data is unrecoverable, however, in some cases user-initiated delete options simply move data to a ``deleted'' section of a SQL database on-device which may remain recoverable~\cite{nist_forensics}.

\subsection{Progression of passcodes to biometrics} In early versions of iOS, passcodes defaulted to four digits and were enforced by software running on the application processor. This provided a superficial layer of security, which could potentially be bypassed through a software exploit against the operating system. With the iPhone 5S and iOS 9 in 2015, Apple increased the default for new passcodes to 6 digits while also introducing TouchID, a capacitive fingerprint sensor. In 2017 with the iPhone X, some models of iPhone replaced TouchID with FaceID, a facial recognition biometric, similarly authenticated through the SEP. Apple argues that these ease-of-use features improve security by enabling users to employ higher-quality passcodes (as the biometric sensor reduces the frequency at which they must be entered)~\cite{apple_faceid_security}; the efficacy of this decision has questioned in the scientific literature~\cite{touchidimpact15}.

\subsection{Introduction of the Secure Enclave} Early iOS devices performed most security functions solely using the application processor, which is the same processor that runs the operating system and all application software. As of the iPhone 5S in 2013, Apple began shipping an additional component in iOS devices: the Secure Enclave Processor, or SEP~\cite{apple_security_guides,apple_platform_security}. The SEP architecture and hardware has undergone a few improvements over time as documented in Tables~\ref{tbl:ios_history} and~\ref{tbl:ios_history_hw}.

\subsection{Hardware changes for security} Just as the software, iOS, has evolved year to year, so too has iPhone hardware. Early generations integrated commodity components. However, recent iPhone and iOS devices employ custom hardware and processor designs~\cite{ifixit_a13}. Table~\ref{tbl:ios_history_hw} summarizes changes to iPhone hardware over time which are directly security-relevant, and the remainder of this section covers these changes in detail.

From iPhone 2G to 3GS ({2007--2009}), Apple sourced hardware from Samsung, LG, and other manufacturers to make their phone. Few if any security-specific hardware features were present in these early phones, and this remained true even as Apple transitioned to in-house designed System-on-Chips (SoCs) with the A4 and continued with the A5 and A6~\cite{ifixit_a4,ifixit_a5,ifixit_a6}. These early iterations encrypted the boot NOR memory with a hardware-fused AES key (UID key), driven by a cryptographic accelerator known as the Crypto Engine.\footnote{The Crypto Engine is documented as early as the A5 SoC, but may have been included earlier} The A7 in iPhone 5S introduced the Secure Enclave Processor (SEP) and a new fingerprint sensor in the home button to enable TouchID. The SEP at this iteration included its own UID key used for encryption. Enclave memory is encrypted with a key derived from both the UID key and an ephemeral key stored in the enclave, and enclave memory is authenticated on A8 and later. The A9 SoC in the iPhone 6S/6S+ connected the path between flash storage and memory directly via the Crypto Engine, in order to improve file encryption security and performance, and allows the SEP to generate its UID key rather than be fused with it. SEP key generation also has the advantage that not even the manufacturers know the SEP UID key. The A10 SoC uses the SEP to protect filesystem class keys in Recovery Mode. The A11 SoC present in the iPhone X added the Neural Engine, a CPU component which accelerates neural network computation, and enables FaceID by allowing facial recognition authentication in the SEP to directly access the Neural Engine~\cite{apple_faceid_security}. A11 also introduces the ``integrity tree'' for stronger authentication of enclave memory. The A12 in XS, XS Max, and XR adopted a version of Pointer Authentication Codes (PAC) from \Gls{ARM} v8.3~\cite{arm_pac} with the relevant hardware support as an additional defense against memory corruption exploits, and added a ``secure storage integrated circuit'' for the SEP to use for replay counters, tamper detection, and to aid cryptography and random number generation. Apple notes that the enclave communicated with the secure storage IC via a ``secure protocol.'' The A12 also extended Recovery Mode protections to DFU mode.

\subsection{Moving secrets to the cloud} iCloud Keychain was announced in 2016~\cite{apple_bh_2016}, and then in 2017 with iOS 11 on iPhones 8, 8 Plus, and X CloudKit containers extended an \Gls{API} to third-party developers for them to store arbitrary data encrypted with keys in the user iCloud Keychain (thus not decrypt-able by Apple) called CloudKit~\cite{apple_developer_docs}. Apple additionally introduced the use of CloudKit for iMessage in this time range~\cite{apple_platform_security,apple_icloud_security}.\footnote{With the aforementioned caveat regarding iCloud Backup} The historical development of iCloud security is not extensively documented by Apple, but \S\ref{sec:ios_security} provides an overview of the current documented architecture in depth.

\chapter{History of Android Security Features}\label{app:android_history}

\section{Android Security Features Over Time}

\subsection{Application sandboxing} The original sandbox on Android was filesystem-based, with applications sandboxed from each other via user IDs and permissions. The filesystem on Android has historically been divided into ``internal'' (device-specific) and ``external'' (non-device-specific). The external storage, also known by its directory name of \texttt{/sdcard}, holds arbitrary files, such as music, photos, and documents, that the user can use across several apps or even off the device itself. Internal storage is controlled closely by the Android system -- each app gets access to its own part of internal storage -- but external storage has fewer controls, because of its intention as a shared resource. Traditionally, external storage was a removable flash memory device (such as an SD Card), but the term has evolved to encompass all shared storage on Android, even for those devices without a removable memory slot~\cite{android_dev_external}.

Android 7.0 began locking down access to {\tt /sdcard}, changing directory access control (DAC) permissions from 751 (Unix {\tt rwxr-x-{}-x}) to 700 (Unix {\tt rwx-{}-{}-{}-{}-{}-}) and removing the ability for apps to request explicit URIs to other app's files. Android 10 took this a step further, deprecating raw access to {\texttt /sdcard} and only allowing the system explicit access to \emph{only} files that it has created. All other views of the file system are mediated by the operating system, and require user intervention~\cite{aosp_appsandbox}. Android 11 enforces scoped storage, eliminating {\tt /sdcard} access entirely~\cite{aosp_scoped}. 

An augmentation to the existing filesystem-based sandbox came in the form of SELinux, which was first added to the Linux kernel with Android 4.3 in permissive (logging) mode, and Android 4.4 in enforcing (blocking) mode. Originally, only a subset of domains (SELinux protection types) were utilized for critical system functions; Android 5.0 pushed SELinux enforcement to the rest of the system. Android 6.0 and 7.0 continued to strengthen the SELinux sandbox, tightening the scope of domains and breaking up services to isolate permissions to specific subsystems. Android 8.0 adds updates for Android's Project Treble hardware abstraction initiative, and 9.0 requires that all apps run in individual SELinux sandboxes (and thus preventing apps from making data world-accessible)~\cite{aosp_selinux, aosp_appsandbox}.  

Dovetailing with SELinux is \texttt{seccomp-bpf}, another sandboxing feature Android inherits from the upstream Linux kernel. This facility, added to Android in version 8.0, prevents applications from making syscalls explicitly whitelisted. This whitelist was built from syscalls available to applications (via the Bionic C interface), syscalls required for boot, and syscalls required for popular apps~\cite{lawrence2017seccomp}. Android 9.0 further improves on the filter, paring down the whitelist even further~\cite{android_dev_piechange}.

Application permissions to media, dialer, messages, \emph{etc}., are important to the security of the Android system, but not relevant to the discussion of this work. 


\subsection{Data storage \& encryption} On release, Android did not employ any encryption mechanisms to protect user data. Android 4.4 introduced full-disk encryption, but it would not be until Android 5.0 that it would become the default. Full-disk encryption uses the PIN/password (4.4) or pattern (5.0) to encrypt the encryption key, and stores it in a secure hardware location if available (5.0). Note that this encryption provides protection only for lost or stolen devices, as once the device is powered on, the device remains decrypted~\cite{aosp_fulldisk}. Android 7.0 introduced file-based encryption, discussed above, and Android 10 made it mandatory for all devices going forward~\cite{aosp_filebased}. File-based encryption allows devices to Direct Boot, allowing critical system functions to run without the phone unlocking~\cite{android_dev_directboot}. The prior full-disk encryption used the Linux kernel's \texttt{dm-crypt} block-level encryption, meaning that the user's passcode was required before any data could be unlocked. Encryption of file metadata was integrated into Android 9,0. The combination of file-based encryption and metadata encryption is thus equivalent to full-disk encryption~\cite{aosp_metadata}.

\subsection{Integrity checking} Starting in version 4.4, Android validated boot code via the \texttt{dm-verity} Linux kernel module, which uses a secure hardware-backed hash tree to validate data on a partition. Any device corruption would be detected and presented to the user as a warning, but the device would still be allowed to boot. It would not be until Android 7.0 that Verified Boot would be strictly enforced, preventing boot if the {\tt dm-verity} check fails. A new version of Verified Boot, known as Android Verified Boot, was introduced in version 8.0 alongside the changes for Project Treble. \Gls{AVB} also added rollback protection via features in the TEE~\cite{aosp_verified}. Note that this integrity check is disabled when a user roots their device via an unlocked bootloader.

In addition to the checks performed at boot time, the Android system validates the APK (Android Package) of an app before installation; if the signature on the APK does not verify with its public key, Android refuses to install it. Android 2.1 introduced the first version APK verification, using the basic Java JAR verification mechanism. Android 7.0 introduced an APK-specific signing scheme, and Android 9.0 improved upon it by allowing for key rotation on APK updates~\cite{aosp_appsigning}. Note that this is separate from the installation of apps from trusted sources (such as Google Play), which is a separate mechanism \cite{android_dev_sign}.

Google has since introduced App Bundles, which consist of pre-compilation code and resources. Google has mandated that all new Android apps on the Play Store use App Bundles some time in 2021~\cite{androidpolice_appbundles}. Before an app is published on Google Play, Google builds and signs the final APK bundle. Rather than allowing developers to directly sign APKs, App Bundles shift control of final signing keys to Google. This move has raised questions from app developers who protest that Google is now empowered to modify apps before distributing them, potentially to inject Google advertising, or at the request of overreaching governments~\cite{commonsware}.

\subsection{Trusted hardware} The history of trusted hardware on Android is inherently more complex, as the Open Handset Alliance does not mandate the use of any specific hardware for Android~\cite{android_compat}, and there is no requirement in Android that the TEE be an \Gls{ARM} TrustZone OS, just that it supports the relevant hardware features~\cite{android_compat}. However, Qualcomm's QSEE environment meets Android requirements~\cite{qsee_info}, and Google provides a reference TEE, Trusty, for general use~\cite{aosp_trusty}.

One of the early uses of TEEs in Android was for key material. Starting in version 4.1, Android provided a Keymaster Hardware Abstraction Layer (HAL) to store keys and perform signing and verifying operations in a trusted application (Keymaster TA). Successive Android versions increased the capabilities of the Keymaster TA. Android 6.0 added symmetric features (AES and HMAC) and key usage control. Keymaster 2, released with Android 7.0, added key attestation, allowing remote clients to verify that a key was stored in a Keymaster TEE, as well as version bindings, preventing version rollbacks. Keymaster 3 (Android 8.0) extended the attestation features provided by the underlying TEE to also provide ID attestation using the phone's IMEI and hardware serial number. Keymaster is not the only Android system that uses the TEE: the Gatekeeper and Fingerprint user authenticator, mentioned previously, are also implemented in trusted applications. The authentication components use TEE features to store passwords and authentication data~\cite{aosp_keystore}.

Recent years have seen Android manufacturers adding additional hardware security modules to their devices. This module, known as a \emph{secure element}, is an analogue to the Secure Enclave in iOS: a dedicated cryptographic module, separate from the primary processor, for computation on secret data. The Android OS view of the use of secure elements is through StrongBox Keymaster interface, added in Android 9. This extends the Keymaster TA functions to support cryptographic operations off the main system processor~\cite{aosp_release9}.
\chapter{History of Forensic Tools}\label{app:forensic_tools}

The tables in Appendix~\ref{sec:fs_access} summarize the \Gls{DHS} forensic software tests~\cite{dhs_forensics}. They are organized by year, mirroring the tables in Appendix~\ref{sec:fs_details} which provide additional details including full software names and versions.

\section{Forensic Tools Access}\label{sec:fs_access}

\begin{tcolorbox}
\begin{figure}[H]
    \centering
    \caption{Legend for Forensic Access Tables}
    \label{fig:legend_fs_access}
\begin{itemize}
    \item[] \checkmark~successful extraction: data was accessed and displayed to the investigator
    \item[] \dmark~partial extraction
    \item[] \xmark~extraction failure
    \item[] - not applicable
\end{itemize}
\end{figure}
\end{tcolorbox}

For these tables, we display results for the latest available hardware and software iteration of a flagship (iPhone, Google Pixel, Google Nexus, Samsung Galaxy) device. Note that this is commonly not the latest available device/running the latest software.

Importantly, it is not always clear from the reports whether `extraction failure' implies the inaccessibility of data due to encryption; it seems that often this implies a simple failure of the forensic tool to clearly document and display the accessed data to the investigator. Commonly, `not applicable' implies that the forensic tool does not advertise or claim to be able to access the data in question, although in some cases it indicates that circumstances inhibited testing of the data category. The rows labeled \textbf{iOS}
and \textbf{Android} aggregate access across the table. That is, if any tool was able to access a particular data category on the platform, this row will contain \checkmark. If no tool achieved better than partial access, \dmark. If all tools failed or were not applicable, \xmark~and -, respectively.
\bigskip

\begin{tcolorbox}
\begin{table}[H]
    \centering
    \caption{iOS and Android Forensic Tool Access (2019)}
    \label{tab:access_19}
\begin{tabular}{r|c|c|c|c|c|c|c|c|c|c|c|c|c|c|c|}
\textbf{Forensic Tool}&
\rot{IMEI} &
\rot{MEID/ESN} &
\rot{MSISDN} &
\rot{Contacts} &
\rot{Calendars} &
\rot{Memos/Notes} &
\rot{Call Logs} &
\rot{SMS/MMS} &
\rot{Files} &
\rot{App Data} &
\rot{Social Media} &
\rot{Bookmarks} &
\rot{Browser History} &
\rot{Email} &
\rot{GPS Data}
    \\ \hline
\textbf{iOS}&\checkmark&\checkmark&\checkmark&\checkmark&\checkmark&\checkmark&\checkmark&\checkmark&\checkmark&\checkmark&\checkmark&\checkmark&\checkmark&\checkmark&\checkmark\\
UFED Kiosk&\checkmark&-&\checkmark&\dmark&\dmark&\checkmark&\checkmark&\dmark&\checkmark&-&\dmark&-&-&-&\checkmark\\
Device Seizure&\checkmark&-&\checkmark&\checkmark&\checkmark&\checkmark&\checkmark&\checkmark&\checkmark&\checkmark&\checkmark&\dmark&\checkmark&\checkmark&\xmark\\
GrayKey&\checkmark&\checkmark&\checkmark&\checkmark&\checkmark&\checkmark&\checkmark&\checkmark&\checkmark&\checkmark&\checkmark&\checkmark&\checkmark&\checkmark&\checkmark\\
UFED 4PC&\checkmark&\checkmark&\checkmark&\checkmark&\checkmark&\checkmark&\checkmark&\checkmark&\checkmark&\checkmark&\dmark&\checkmark&\checkmark&-&\checkmark\\
&&&&&&&&&&&&&&&\\
\textbf{Android}&\checkmark&-&\checkmark&\checkmark&\checkmark&\checkmark&\checkmark&\checkmark&\checkmark&\checkmark&\xmark&\xmark&\xmark&-&\checkmark\\
UFED Kiosk&\checkmark&-&\checkmark&\dmark&\checkmark&\checkmark&\checkmark&\checkmark&\checkmark&\checkmark&-&-&-&-&\checkmark\\
Device Seizure&\checkmark&-&\checkmark&\checkmark&\checkmark&-&\checkmark&\checkmark&\checkmark&\checkmark&\xmark&\xmark&\xmark&-&\xmark\\
GrayKey&-&-&-&-&-&-&-&-&-&-&-&-&-&-&-\\
UFED 4PC&\checkmark&-&\checkmark&\dmark&\checkmark&\xmark&\checkmark&\checkmark&\checkmark&\xmark&-&-&-&-&\xmark\\
\end{tabular}
\end{table}
\end{tcolorbox}

\begin{tcolorbox}
\begin{table}[H]
    \centering
    \caption{iOS and Android Forensic Tool Access (2018)}
    \label{tab:access_18}
\begin{tabular}{r|c|c|c|c|c|c|c|c|c|c|c|c|c|c|c|}
\textbf{Forensic Tool}&
\rot{IMEI} &
\rot{MEID/ESN} &
\rot{MSISDN} &
\rot{Contacts} &
\rot{Calendars} &
\rot{Memos/Notes} &
\rot{Call Logs} &
\rot{SMS/MMS} &
\rot{Files} &
\rot{App Data} &
\rot{Social Media} &
\rot{Bookmarks} &
\rot{Browser History} &
\rot{Email} &
\rot{GPS Data}
    \\ \hline
\textbf{iOS}&\checkmark&-&\checkmark&\checkmark&\checkmark&\checkmark&\checkmark&\checkmark&\checkmark&\checkmark&\checkmark&\checkmark&\checkmark&\xmark&\checkmark\\
Blacklight&\checkmark&-&\checkmark&\checkmark&\checkmark&\checkmark&\checkmark&\checkmark&\dmark&\checkmark&\dmark&\checkmark&\checkmark&-&\xmark\\
Mobilyze&\checkmark&-&\checkmark&\dmark&\checkmark&\xmark&\checkmark&\dmark&\dmark&\checkmark&\dmark&\checkmark&\checkmark&-&\checkmark\\
XRY&\checkmark&-&\checkmark&\checkmark&\checkmark&\checkmark&\checkmark&\checkmark&\checkmark&\checkmark&\dmark&\checkmark&\checkmark&-&\checkmark\\
XRY Kiosk&\checkmark&-&\checkmark&\checkmark&\checkmark&\checkmark&\checkmark&\checkmark&\checkmark&\checkmark&\dmark&\checkmark&\checkmark&-&\checkmark\\
Magnet AXIOM&\xmark&-&\xmark&\dmark&\dmark&\checkmark&\checkmark&\dmark&\checkmark&\checkmark&\dmark&\checkmark&\checkmark&-&\checkmark\\
Final&\checkmark&-&\checkmark&\dmark&\dmark&\checkmark&\checkmark&\checkmark&\checkmark&-&\dmark&\checkmark&\checkmark&-&\xmark\\
Device Seizure&\checkmark&-&\checkmark&\dmark&\checkmark&\checkmark&\checkmark&\checkmark&\checkmark&-&\dmark&\checkmark&\checkmark&-&\checkmark\\
Katana Triage&\checkmark&-&\checkmark&\dmark&\checkmark&\xmark&\checkmark&\checkmark&\checkmark&-&-&\checkmark&\checkmark&-&\xmark\\
MD-NEXT/RED&\checkmark&-&\checkmark&\dmark&\checkmark&\checkmark&\checkmark&\checkmark&\checkmark&\xmark&\dmark&\checkmark&\checkmark&-&\checkmark\\
Oxygen&\checkmark&-&\checkmark&\checkmark&\checkmark&\dmark&\checkmark&\checkmark&\dmark&\xmark&\checkmark&\checkmark&\checkmark&-&\xmark\\
MOBILEdit&\checkmark&-&\checkmark&\checkmark&\dmark&\checkmark&\checkmark&\checkmark&\checkmark&\checkmark&\dmark&\checkmark&\checkmark&-&\checkmark\\
UFED Touch&\checkmark&-&\checkmark&\checkmark&\checkmark&\checkmark&\checkmark&\checkmark&\checkmark&\checkmark&\dmark&\checkmark&\checkmark&\xmark&\checkmark\\
&&&&&&&&&&&&&&&\\
\textbf{Android}&\checkmark&-&\checkmark&\checkmark&\checkmark&\checkmark&\checkmark&\checkmark&\checkmark&\checkmark&\checkmark&\checkmark&\checkmark&\checkmark&\checkmark\\
Blacklight&\checkmark&-&\checkmark&\checkmark&\checkmark&\checkmark&\checkmark&\dmark&\checkmark&\checkmark&\dmark&\xmark&\xmark&\xmark&\xmark\\
Mobilyze&\checkmark&-&\checkmark&\dmark&\xmark&\xmark&\checkmark&\dmark&\dmark&\xmark&\xmark&\xmark&\xmark&\xmark&\xmark\\
XRY&\checkmark&-&\checkmark&\checkmark&\xmark&\checkmark&\checkmark&\checkmark&\checkmark&\checkmark&\dmark&\xmark&\xmark&\xmark&\checkmark\\
XRY Kiosk&\checkmark&-&\checkmark&\checkmark&\xmark&\checkmark&\checkmark&\checkmark&\checkmark&\checkmark&\dmark&\xmark&\xmark&\xmark&\checkmark\\
Magnet AXIOM&\xmark&-&\xmark&\xmark&\checkmark&\xmark&\xmark&\xmark&\dmark&\xmark&\xmark&\xmark&\xmark&\xmark&\xmark\\
Final&\xmark&-&\checkmark&\dmark&\checkmark&\xmark&\checkmark&\dmark&\checkmark&\checkmark&\xmark&\checkmark&\checkmark&\checkmark&\xmark\\
Device Seizure&\checkmark&-&\checkmark&\checkmark&\checkmark&-&\checkmark&\dmark&\checkmark&\checkmark&\xmark&\xmark&\xmark&-&\xmark\\
Katana Triage&\xmark&-&\xmark&\dmark&\checkmark&-&\checkmark&\checkmark&\checkmark&\checkmark&-&-&-&-&\xmark\\
MD-NEXT/RED&\checkmark&-&\checkmark&\checkmark&\checkmark&\checkmark&\checkmark&\checkmark&\checkmark&\checkmark&\checkmark&\checkmark&\checkmark&\checkmark&\checkmark\\
Oxygen&\checkmark&-&\checkmark&\checkmark&\xmark&\checkmark&\checkmark&\dmark&\checkmark&\checkmark&\checkmark&\xmark&\xmark&\xmark&\xmark\\
MOBILEdit&\checkmark&-&\checkmark&\checkmark&\checkmark&\xmark&\checkmark&\dmark&\checkmark&\checkmark&\xmark&\xmark&\xmark&\xmark&\xmark\\
UFED Touch&\checkmark&-&\checkmark&\dmark&\checkmark&\dmark&\checkmark&\checkmark&\checkmark&\checkmark&\xmark&\xmark&\xmark&\xmark&\checkmark\\
\end{tabular}
\end{table}
\end{tcolorbox}

\begin{tcolorbox}
\begin{table}[H]
    \centering
    \caption{iOS and Android Forensic Tool Access (2017)}
    \label{tab:access_17}
\begin{tabular}{r|c|c|c|c|c|c|c|c|c|c|c|c|c|c|c|}
\textbf{Forensic Tool}&
\rot{IMEI} &
\rot{MEID/ESN} &
\rot{MSISDN} &
\rot{Contacts} &
\rot{Calendars} &
\rot{Memos/Notes} &
\rot{Call Logs} &
\rot{SMS/MMS} &
\rot{Files} &
\rot{App Data} &
\rot{Social Media} &
\rot{Bookmarks} &
\rot{Browser History} &
\rot{Email} &
\rot{GPS Data}
    \\ \hline
\textbf{iOS}&\checkmark&-&\checkmark&\checkmark&\checkmark&\checkmark&\checkmark&\checkmark&\checkmark&\checkmark&\dmark&\checkmark&\checkmark&\checkmark&\checkmark\\
Blacklight&\checkmark&-&\checkmark&\checkmark&\checkmark&\checkmark&\checkmark&\dmark&\dmark&\xmark&\dmark&\checkmark&\checkmark&\xmark&\checkmark\\
Mobilyze&\checkmark&-&\checkmark&\dmark&\checkmark&\xmark&\checkmark&\dmark&\dmark&\xmark&\xmark&\checkmark&\checkmark&\xmark&\xmark\\
Secure View&\checkmark&-&\checkmark&\dmark&\checkmark&\dmark&\checkmark&\dmark&\checkmark&\xmark&\dmark&\checkmark&\checkmark&-&\checkmark\\
MOBILEdit&\checkmark&-&\checkmark&\checkmark&\checkmark&\dmark&\checkmark&\dmark&\checkmark&\checkmark&\dmark&\checkmark&\checkmark&-&\xmark\\
UFED 4PC&\checkmark&-&\checkmark&\dmark&\checkmark&\dmark&\checkmark&\dmark&\checkmark&\checkmark&\dmark&\checkmark&\checkmark&\checkmark&\checkmark\\
XRY Kiosk&\checkmark&-&\checkmark&\checkmark&\checkmark&\checkmark&\checkmark&\checkmark&\checkmark&\xmark&\dmark&\checkmark&\checkmark&-&\checkmark\\
XRY&\checkmark&-&\checkmark&\checkmark&\checkmark&\checkmark&\checkmark&\checkmark&\checkmark&\xmark&\dmark&\checkmark&\checkmark&-&\checkmark\\
Lantern&\xmark&-&\checkmark&\dmark&\checkmark&\checkmark&\checkmark&\dmark&\checkmark&-&-&\checkmark&\checkmark&-&\checkmark\\
Final&\checkmark&-&\checkmark&\dmark&\checkmark&\dmark&\checkmark&\checkmark&\dmark&\xmark&\dmark&\dmark&\checkmark&\xmark&\checkmark\\
Device Seizure&\checkmark&-&\checkmark&\dmark&\checkmark&\dmark&\checkmark&\checkmark&\checkmark&\checkmark&\dmark&\checkmark&\checkmark&-&\checkmark\\
MPE+&\checkmark&-&\xmark&\checkmark&\checkmark&\dmark&\checkmark&\checkmark&\checkmark&\xmark&\dmark&\checkmark&\checkmark&\xmark&\checkmark\\
MOBILEdit&\checkmark&-&\xmark&\checkmark&\checkmark&\dmark&\checkmark&\checkmark&\checkmark&\xmark&\dmark&\checkmark&\checkmark&\xmark&\checkmark\\
XRY Kiosk&\checkmark&-&\checkmark&\checkmark&\checkmark&\checkmark&\checkmark&\checkmark&\checkmark&-&\dmark&\checkmark&\checkmark&-&\checkmark\\
&&&&&&&&&&&&&&&\\
\textbf{Android}&\checkmark&-&\checkmark&\checkmark&\checkmark&\checkmark&\checkmark&\checkmark&\checkmark&\checkmark&\dmark&\checkmark&\checkmark&\xmark&\checkmark\\
Blacklight&\checkmark&-&\checkmark&\dmark&\dmark&\dmark&\checkmark&\dmark&\dmark&\dmark&\dmark&\xmark&\xmark&\xmark&\xmark\\
Mobilyze&\checkmark&-&\checkmark&\dmark&\xmark&\xmark&\checkmark&\dmark&\dmark&\xmark&\xmark&\checkmark&\checkmark&\xmark&\xmark\\
Secure View&\xmark&-&\xmark&\dmark&\checkmark&\xmark&\checkmark&\dmark&\checkmark&\checkmark&\dmark&\xmark&\xmark&\xmark&\xmark\\
MOBILEdit&\checkmark&-&\checkmark&\checkmark&\checkmark&\xmark&\checkmark&\dmark&\checkmark&\checkmark&\dmark&\xmark&\xmark&\xmark&\checkmark\\
UFED 4PC&\checkmark&-&\checkmark&\dmark&\checkmark&\checkmark&\checkmark&\checkmark&\checkmark&\checkmark&\dmark&\xmark&\xmark&\xmark&\checkmark\\
XRY Kiosk&\checkmark&-&\checkmark&\dmark&\checkmark&\checkmark&\checkmark&\checkmark&\checkmark&\checkmark&\xmark&\xmark&\xmark&\xmark&\checkmark\\
XRY&\checkmark&-&\checkmark&\dmark&\checkmark&\checkmark&\checkmark&\checkmark&\checkmark&\checkmark&\xmark&\xmark&\xmark&\xmark&\checkmark\\
Lantern&\xmark&-&\xmark&\dmark&\checkmark&-&\checkmark&\checkmark&\checkmark&\checkmark&-&\xmark&\xmark&-&\checkmark\\
Final&\xmark&-&\xmark&\dmark&\checkmark&\xmark&\checkmark&\dmark&\checkmark&\checkmark&\xmark&\xmark&\xmark&\xmark&\xmark\\
Device Seizure&\checkmark&-&\checkmark&\checkmark&\checkmark&\xmark&\checkmark&\dmark&\checkmark&\checkmark&\dmark&\checkmark&\checkmark&-&\checkmark\\
MPE+&\checkmark&-&\checkmark&\dmark&\xmark&\xmark&\checkmark&\dmark&\checkmark&\xmark&\xmark&\checkmark&\xmark&\xmark&\xmark\\
MOBILEdit&\checkmark&-&\xmark&\checkmark&\checkmark&\xmark&\checkmark&\checkmark&\checkmark&\checkmark&\dmark&\checkmark&\xmark&\xmark&\checkmark\\
XRY Kiosk&\checkmark&-&\checkmark&\checkmark&\checkmark&-&\checkmark&\checkmark&\checkmark&\checkmark&\dmark&\xmark&\xmark&\xmark&\xmark\\
\end{tabular}
\end{table}
\end{tcolorbox}

\begin{tcolorbox}
\begin{table}[H]
    \centering
    \caption{iOS Forensic Tool Access (2010--2016)}
    \label{tab:ios_access_to16}
\begin{tabular}{r|c|c|c|c|c|c|c|c|c|c|c|c|c|c|c|}
\textbf{Forensic Tool}&
\rot{IMEI} &
\rot{MEID/ESN} &
\rot{MSISDN} &
\rot{Contacts} &
\rot{Calendars} &
\rot{Memos/Notes} &
\rot{Call Logs} &
\rot{SMS/MMS} &
\rot{Files} &
\rot{App Data} &
\rot{Social Media} &
\rot{Bookmarks} &
\rot{Browser History} &
\rot{Email} &
\rot{GPS Data}
    \\ \hline
\textbf{iOS}&\checkmark&\checkmark&\checkmark&\checkmark&\checkmark&\checkmark&\checkmark&\checkmark&\checkmark&\checkmark&\checkmark&\checkmark&\checkmark&\xmark&\checkmark\\
MOBILEdit&\checkmark&-&\xmark&\checkmark&\checkmark&\checkmark&\checkmark&\checkmark&\checkmark&\xmark&\dmark&\checkmark&\checkmark&\xmark&\checkmark\\
XRY&\checkmark&-&\checkmark&\checkmark&\checkmark&\checkmark&\checkmark&\checkmark&\checkmark&-&\dmark&\checkmark&\checkmark&-&\checkmark\\
Oxygen&\checkmark&-&\checkmark&\checkmark&\checkmark&\checkmark&\checkmark&\checkmark&\dmark&\xmark&\dmark&\checkmark&\checkmark&-&\checkmark\\
Secure View&\xmark&-&\xmark&\xmark&\checkmark&\xmark&\checkmark&\checkmark&\dmark&\xmark&\dmark&\xmark&\xmark&-&\xmark\\
UFED Touch&\checkmark&-&\checkmark&\dmark&\checkmark&\checkmark&\dmark&\checkmark&\checkmark&\checkmark&\dmark&\checkmark&\checkmark&-&-\\
Device Seizure&\xmark&-&\checkmark&\dmark&\checkmark&\checkmark&\checkmark&\checkmark&\dmark&\xmark&\dmark&\xmark&\xmark&\xmark&\xmark\\
BlackLight&\xmark&-&\checkmark&\xmark&\xmark&\xmark&\xmark&\xmark&\dmark&\xmark&\xmark&\xmark&\xmark&\xmark&\xmark\\
UFED 4PC&-&\checkmark&\checkmark&\checkmark&\checkmark&\checkmark&\checkmark&\checkmark&\checkmark&\xmark&\checkmark&\checkmark&\checkmark&-&-\\
MOBILEdit&\checkmark&-&\xmark&\checkmark&\checkmark&\checkmark&\checkmark&\dmark&\checkmark&\xmark&\checkmark&\xmark&\xmark&-&-\\
PF Express&\checkmark&-&\xmark&\xmark&\xmark&\xmark&\xmark&\dmark&\checkmark&\xmark&\dmark&\xmark&\xmark&-&-\\
Device Seizure&\checkmark&-&\checkmark&\dmark&\checkmark&\dmark&\dmark&\dmark&\dmark&\checkmark&\dmark&\checkmark&\checkmark&-&-\\
EnCase&-&\xmark&\xmark&\checkmark&\checkmark&\checkmark&\checkmark&\checkmark&\checkmark&\checkmark&\dmark&\checkmark&\checkmark&-&-\\
Lantern&-&\checkmark&\checkmark&\xmark&\xmark&\xmark&\xmark&\xmark&\checkmark&\xmark&\dmark&\xmark&\xmark&-&-\\
Oxygen&-&\checkmark&\checkmark&\checkmark&\checkmark&\checkmark&\checkmark&\checkmark&\checkmark&\checkmark&\checkmark&\checkmark&\checkmark&-&-\\
Secure View&-&-&-&\dmark&\checkmark&\xmark&\dmark&\checkmark&\checkmark&\checkmark&\xmark&\checkmark&\checkmark&-&-\\
Crime Lab&-&\checkmark&\checkmark&\checkmark&\checkmark&\checkmark&\checkmark&\checkmark&\checkmark&\checkmark&\checkmark&\checkmark&\checkmark&-&-\\
MPE+&-&\checkmark&\checkmark&\dmark&\xmark&\checkmark&\xmark&\checkmark&\checkmark&\xmark&\checkmark&\checkmark&\checkmark&-&-\\
UFED PA&-&\checkmark&\checkmark&\dmark&\dmark&\checkmark&\checkmark&\dmark&\checkmark&\dmark&\xmark&\xmark&\xmark&-&-\\
XRY/XACT&-&\checkmark&\checkmark&\checkmark&\checkmark&\checkmark&\checkmark&\dmark&\checkmark&\checkmark&\checkmark&\checkmark&\checkmark&-&-\\
EnCase&-&-&-&\checkmark&\checkmark&\checkmark&\checkmark&\checkmark&\checkmark&-&-&\checkmark&\checkmark&-&-\\
Device Seizure&\xmark&\xmark&\dmark&\dmark&\checkmark&\checkmark&\dmark&\dmark&\dmark&-&-&\checkmark&\checkmark&-&-\\
XRY&\xmark&\xmark&-&\checkmark&\checkmark&\checkmark&\checkmark&\checkmark&\checkmark&\checkmark&-&\checkmark&\checkmark&-&-\\
Lantern&\xmark&\xmark&-&\dmark&\checkmark&\checkmark&\checkmark&\checkmark&\checkmark&-&-&\checkmark&\checkmark&-&-\\
Secure View&-&-&-&\dmark&\checkmark&\checkmark&\checkmark&\checkmark&\checkmark&-&-&\checkmark&\checkmark&-&-\\
MPE+&\xmark&-&\xmark&\checkmark&\checkmark&\checkmark&\checkmark&\checkmark&\checkmark&-&-&\checkmark&\checkmark&-&-\\
UFED&\checkmark&\checkmark&\checkmark&\dmark&\checkmark&\checkmark&\checkmark&\checkmark&\checkmark&-&-&\checkmark&\checkmark&-&-\\
Mobilyze&\checkmark&\checkmark&\checkmark&\dmark&\checkmark&\checkmark&\checkmark&\dmark&\checkmark&-&-&-&-&-&-\\
Zdziarski&\checkmark&\checkmark&\checkmark&\checkmark&\checkmark&\checkmark&\checkmark&\checkmark&\checkmark&\checkmark&-&\checkmark&\checkmark&-&-\\
iXAM&\checkmark&\checkmark&\checkmark&\checkmark&\checkmark&\checkmark&\checkmark&\checkmark&\checkmark&\checkmark&-&\checkmark&\checkmark&-&-\\
Secure View&\checkmark&\checkmark&\checkmark&-&-&-&\checkmark&\checkmark&-&-&-&-&-&-&\checkmark\\
XRY&\checkmark&\checkmark&\checkmark&\checkmark&\checkmark&\checkmark&\checkmark&\checkmark&\checkmark&-&-&\checkmark&\checkmark&-&-\\
\end{tabular}
\end{table}
\end{tcolorbox}

\begin{tcolorbox}
\begin{table}[H]
    \centering
    \caption{Android Forensic Tool Access (2010--2016)}
    \label{tab:android_access_to16}
\begin{tabular}{r|c|c|c|c|c|c|c|c|c|c|c|c|c|c|c|}
\textbf{Forensic Tool}&
\rot{IMEI} &
\rot{MEID/ESN} &
\rot{MSISDN} &
\rot{Contacts} &
\rot{Calendars} &
\rot{Memos/Notes} &
\rot{Call Logs} &
\rot{SMS/MMS} &
\rot{Files} &
\rot{App Data} &
\rot{Social Media} &
\rot{Bookmarks} &
\rot{Browser History} &
\rot{Email} &
\rot{GPS Data}
    \\ \hline
\textbf{Android}&\checkmark&\checkmark&\checkmark&\checkmark&\checkmark&\checkmark&\checkmark&\checkmark&\checkmark&\checkmark&\checkmark&\checkmark&\checkmark&\xmark&\checkmark\\
MOBILEdit&\checkmark&-&\xmark&\checkmark&\checkmark&\xmark&\checkmark&\dmark&\checkmark&\checkmark&\xmark&\xmark&\xmark&\xmark&\xmark\\
XRY&\checkmark&-&\checkmark&\checkmark&\checkmark&-&\checkmark&\checkmark&\checkmark&\checkmark&\dmark&\xmark&\xmark&\xmark&\xmark\\
Oxygen&\checkmark&-&\xmark&\checkmark&\checkmark&\xmark&\checkmark&\dmark&\checkmark&\checkmark&\dmark&\xmark&\xmark&\xmark&\xmark\\
Secure View&\checkmark&-&\checkmark&\dmark&\checkmark&\xmark&\checkmark&\dmark&\checkmark&\checkmark&\xmark&\xmark&\xmark&\xmark&\xmark\\
UFED Touch&\checkmark&-&\checkmark&\checkmark&\checkmark&\xmark&\checkmark&\checkmark&\checkmark&\checkmark&\dmark&\checkmark&\checkmark&-&-\\
Device Seizure&\checkmark&-&\checkmark&\checkmark&\checkmark&\xmark&\checkmark&\dmark&\checkmark&\checkmark&\xmark&\xmark&\xmark&\xmark&\xmark\\
BlackLight&\checkmark&-&\checkmark&\checkmark&\xmark&\xmark&\checkmark&\dmark&\checkmark&\checkmark&\dmark&\checkmark&\checkmark&\xmark&\checkmark\\
UFED 4PC&\xmark&-&\xmark&\checkmark&\checkmark&\xmark&\checkmark&\checkmark&\checkmark&\checkmark&\checkmark&\checkmark&\checkmark&-&-\\
MOBILEdit&-&\checkmark&\xmark&\checkmark&\checkmark&\xmark&\checkmark&\dmark&\checkmark&\checkmark&\dmark&\checkmark&\checkmark&-&-\\
PF Express&\checkmark&-&\xmark&\checkmark&\checkmark&\xmark&\checkmark&\checkmark&\checkmark&\checkmark&\dmark&\dmark&\xmark&-&-\\
Device Seizure&\checkmark&-&\xmark&\checkmark&\checkmark&\xmark&\checkmark&\checkmark&\checkmark&\checkmark&\dmark&\checkmark&\checkmark&-&-\\
EnCase&\checkmark&-&\xmark&\checkmark&\checkmark&\xmark&\checkmark&\checkmark&\checkmark&\checkmark&\xmark&\xmark&\checkmark&-&-\\
Lantern&-&-&-&-&-&-&-&-&-&-&-&-&-&-&-\\
Oxygen&-&\checkmark&\xmark&\dmark&\checkmark&\dmark&\checkmark&\dmark&\checkmark&\checkmark&\dmark&\checkmark&\checkmark&-&-\\
Secure View&-&-&-&\dmark&-&\xmark&\checkmark&\checkmark&\checkmark&\checkmark&\checkmark&\dmark&\checkmark&-&-\\
Crime Lab&-&-&-&-&-&-&-&-&-&-&-&-&-&-&-\\
MPE+&\checkmark&-&\checkmark&\dmark&-&-&\checkmark&\checkmark&\xmark&\xmark&-&\dmark&\checkmark&-&-\\
UFED PA&\checkmark&-&\checkmark&\dmark&\checkmark&\xmark&\checkmark&\checkmark&\checkmark&\xmark&\xmark&\xmark&\xmark&-&-\\
XRY/XACT&\checkmark&-&\xmark&\dmark&\checkmark&\xmark&\checkmark&\checkmark&\checkmark&\xmark&\checkmark&\checkmark&\checkmark&-&-\\
EnCase&\checkmark&\checkmark&\checkmark&\checkmark&\checkmark&\checkmark&\checkmark&\checkmark&-&\checkmark&-&\checkmark&\checkmark&-&-\\
Device Seizure&\checkmark&\checkmark&-&\checkmark&\dmark&\dmark&\checkmark&\checkmark&\xmark&\xmark&-&\checkmark&\checkmark&-&-\\
XRY&\checkmark&\checkmark&-&\checkmark&\checkmark&\checkmark&\checkmark&\checkmark&\checkmark&\checkmark&-&\checkmark&\checkmark&-&-\\
Lantern&-&-&-&-&-&-&-&-&-&-&-&-&-&-&-\\
Secure View&-&-&-&\dmark&\checkmark&\checkmark&\checkmark&\checkmark&\xmark&-&-&\checkmark&\checkmark&-&-\\
MPE+&\xmark&\xmark&\xmark&\checkmark&\checkmark&\checkmark&\checkmark&\checkmark&-&\checkmark&-&-&-&-&-\\
UFED&\checkmark&\checkmark&-&\dmark&\checkmark&\checkmark&\checkmark&\dmark&\checkmark&-&-&-&-&-&-\\
Mobilyze&-&-&-&-&-&-&-&-&-&-&-&-&-&-&-\\
Zdziarski&-&-&-&-&-&-&-&-&-&-&-&-&-&-&-\\
iXAM&-&-&-&-&-&-&-&-&-&-&-&-&-&-&-\\
Secure View&-&-&-&\checkmark&\xmark&\xmark&\checkmark&\checkmark&-&-&-&-&-&-&-\\
XRY&-&-&-&-&-&-&-&-&-&-&-&-&-&-&-\\
\end{tabular}
\end{table}
\end{tcolorbox}

\section{Forensic Tools Listing}\label{sec:fs_details}
In the following tables~\ref{tab:forensics_19},~\ref{tab:forensics_18},~\ref{tab:forensics_17}, and~\ref{tab:forensics_to16}, we enumerate the forensic software tools evaluated by \Gls{DHS} as of this writing. These tables are divided by year for readability.
\bigskip
\begin{tcolorbox}
\begin{table}[H]
    \centering
    \begin{tabular}{p{0.09\linewidth}|p{0.15\linewidth}|p{0.12\linewidth}|p{0.32\linewidth}|p{0.18\linewidth}}
     \textbf{Date} & \textbf{iPhone (iOS)}\footnote{Latest model/version included in test.} & \textbf{Android}\footnote{Latest version included in test.} & \textbf{Forensic Tool} & \textbf{Version} \\\hline
      Sep '19 & 7 Plus (10.2) & 7.1.1 & UFED InField Kiosk & 7.5.0.875 \\\hline 
      Jun '19 & X and 8 Plus (11)\footnote{The iPhone X ran iOS 11.3.1, the iPhone 8 ran 11.4.1.} & - & Paraben's Electronic Evidence Examiner – Device Seizure & 2.2.11812.15844 \\\hline
      Jun '19 & X and 8 Plus (11)\footnote{X (11.3.1), 8 (11.4.1).} & N/A\footnote{iOS-only tool.} & GrayKey OS/App Bundle & 1.4.2/1.11.2.5 \\\hline
     Apr '19 & X and 8 Plus (11)\footnote{X (11.3.1), 8 (11.4.1).} & 8.1.0 & UFED 4PC/Physical Analyzer & 7.8.0.942 \\\hline
    \end{tabular}
    \caption{History of Forensic Tools (2019)}\label{tab:forensics_19}
    \caption*{Source: \Gls{DHS}~\cite{dhs_forensics}}
\end{table}
\end{tcolorbox}     
\begin{tcolorbox}
\begin{table}[H]
    \centering
    \begin{tabular}{p{0.09\linewidth}|p{0.15\linewidth}|p{0.12\linewidth}|p{0.36\linewidth}|p{0.14\linewidth}}
     \textbf{Date} & \textbf{iPhone (iOS)}\footnote{Latest model/version included in test.} & \textbf{Android}\footnote{Latest version included in test.} & \textbf{Forensic Tool} & \textbf{Version} \\\hline
     Nov '18 & 7 Plus (10.2) & 7.1.1 & Blacklight 2018 & 1.1 \\\hline
     Nov '18 & 7 Plus (10.2) & 7.1.1 & Mobilyze & 2018.1 \\\hline
     Nov '18 & 7 Plus (10.2) & 7.1.1 & XRY & 7.8.0 \\\hline
     Nov '18 & 7 Plus (10.2) & - & XRY Kiosk & 7.8.0 \\\hline 
     Oct '18 & 7 Plus (10.2) & 7.0 & Magnet AXIOM & 1.2.1.6994 \\\hline
     Jul '18 & 7 Plus (10.2) & 6.0.1 & Final Mobile Forensics & 2018.02.07 \\\hline
     Jun '18 & 7 Plus (10.2) & 7.1.1 & Electronic Evidence Examiner - Device Seizure & 1.7 \\\hline
     May '18 & 7 (10.2) & 7.1.1 & Katana Forensics Triage & 1.1802.220 \\\hline
     Apr '18 & 7 Plus (10.2) & 7.1.1 & MD-NEXT/MD-RED & 1.75/2.3 \\\hline
     Apr '18 & 7 Plus (10.2) & 7.1.1 & Oxygen Forensics & 10.0.0.81 \\\hline
     Jan '18 & 7 Plus (10.2) & 7.1.1 & MOBILedit Forensics & 9.1.0.22420 \\\hline
     Jan '18 & 7 Plus (10.2) & 7.1.1 & UFED Touch/Physical Analyzer & 6.2.1.17 \\\hline
         \end{tabular}
    \caption{History of Forensic Tools (2018)}\label{tab:forensics_18}
    \caption*{Source: \Gls{DHS}~\cite{dhs_forensics}}
\end{table}
\end{tcolorbox}     
\begin{tcolorbox}
\begin{table}[H]
    \centering
    \begin{tabular}{p{0.09\linewidth}|p{0.15\linewidth}|p{0.12\linewidth}|p{0.36\linewidth}|p{0.14\linewidth}}
     \textbf{Date} & \textbf{iPhone (iOS)}\footnote{Latest model/version included in test.} & \textbf{Android}\footnote{Latest version included in test.} & \textbf{Forensic Tool} & \textbf{Version} \\\hline
     Dec '17 & 7 Plus (10.2) & 7.1.1 & Blacklight & 2016.3.1 \\\hline
     Dec '17 & 7 Plus (10.2) & 7.1.1 & Mobilyze & 2017.1 \\\hline
     Nov '17 & 7 (10.2) & 7.1.1 & Secure View & 4.3.1 \\\hline
     Nov '17 & 7 Plus (10.2) & 7.1.1 & MOBILedit Forensics Express & 4.2.1.11207 \\\hline
     Sep '17 & 7 Plus (10.2) & 7.1.1 & UFED 4PC/Physical Analyzer & 6.2.1 \\\hline
     Aug '17 & 7 Plus (10.2) & 7.1.1 & XRY Kiosk & 7.3.0 \\\hline
     Aug '17 & 7 Plus (10.2) & 7.1.1 & XRY & 7.3.1 \\\hline
     Jul '17 & 7 (10.2) & 7.1.1 & Lantern & 4.6.8 \\\hline
     Jun '17 & 7 (10.2) & 7.1.1 & Final Mobile Forensics & 2017.02.06 \\\hline
     Apr '17 & 6S (9.2.1) & 5.1.1 & Electronic Evidence Examiner Device Seizure & 1.0.9466.18457 \\\hline
     Mar '17 & 6S (9.2.1) & 5.1.1 & Mobile Phone Examiner Plus & 5.6.0 \\\hline
     Mar '17 & 6S (9.2.1) & 5.1.1 & MOBILedit Forensic Express & 3.5.2.7047 \\\hline
     Jan '17 & 6S (9.2.1) & 5.1.1 & XRY Kiosk & 7.0.0.36568 \\\hline
    \end{tabular}
    \caption{History of Forensic Tools (2017)}\label{tab:forensics_17}
    \caption*{Source: \Gls{DHS}~\cite{dhs_forensics}}
\end{table}
\end{tcolorbox}
\begin{tcolorbox}
\begin{table}[H]
    \centering
    \begin{tabular}{p{0.09\linewidth}|p{0.19\linewidth}|p{0.11\linewidth}|p{0.35\linewidth}|p{0.11\linewidth}}
     \textbf{Date} & \textbf{iPhone (iOS)}\footnote{Latest model/version included in test.} & \textbf{Android}\footnote{Latest version included in test.} & \textbf{Forensic Tool} & \textbf{Version} \\\hline
     Dec '16 & 6S (9.2.1) & - & MOBILedit Forensic & 8.6.0.20354 \\\hline
     Nov '16 & 6S (9.2.1) & 5.1.1 & MOBILedit Forensic & 8.6.0.20354 \\\hline
     Nov '16 & 6S (9.2.1) & - & XRY & 7.0.1.37853 \\\hline
     Aug '16 & 6S (9.2.1) & 5.1.1 & Oxygen Forensics & 8.3.1.105 \\\hline
     Jul '16 & 6S (9.2.1) & - & Secure View & 4.1.9 \\ \hline
     Jul '16 & 5S (7.1) & - & UFED Touch & 4.4.0.1 \\\hline
     May '16 & 6S (9.2.1) & - & Device Seizure & 7.4 \\\hline
     May '16 & 6S (9.2.1) & 5.1.1 & BlackLight & 2016.1 \\\hline
     Jan '16 & 5S (7.1) & - & UFED 4PC/Physical Analyzer & 4.2.6.4-5 \\\hline
     Dec '15 & 5S (7.1) & - & MOBILedit Forensic & 7.8.3.6085 \\\hline
     Dec '15 & 5S (7.1) & - & Phone Forensics Express & 2.1.2.2761 \\ \hline
     Jun '15 & 5S (7.1) & - & Device Seizure & 6.8 \\\hline
     Apr '15 & 5S (7.1) & - & EnCase Smartphone Examiner & 7.10.00.103 \\\hline
     Jun '15 & 5S (7.1) & - & Lantern & 4.5.6 \\\hline
     Mar '15 & 5S (7.1) & - & Oxygen Forensic Analyst & 7.0.0.408 \\ \hline
     Feb '15 & 5S (7.1) & - & Secure View & 3.16.4 \\ \hline
     Dec '14 & 5S (7.1) & N/A & iOS Crime Lab & 1.0.1 \\\hline
     Dec '14 & 5S (7.1) & - & Mobile Phone Examiner Plus & 5.3.73 \\ \hline
     Oct '14 & 5S (7.1) & - & UFED Physical Analyzer & 3.9.6.7 \\ \hline
     Sep '14 & 5S (7.1) & - & XRY/XACT & 6.10.1 \\ \hline
     Apr '13 & 4 (5.0.1) & - & EnCase Smartphone Examiner & 7.0.3 \\ \hline
     Feb '13 & 4 (5.0.1) & 2.2.1 & Device Seizure & 5.0 \\ \hline
     Feb '13 & 4 (5.0.1) & - & Micro Systemation XRY & 6.3.1 \\ \hline
     Feb '13 & 4 (5.0.1) & N/A & Lantern & 2.3 \\ \hline
     Feb '13 & 4 (4.3.3) & 2.1.1 & Secure View 3 & 3.8.0 \\\hline
     Sep '12 & 4 (4.3.3) & 2.1.1 & Mobile Phone Examiner Plus & 4.6.0.2 \\ \hline
     Sep '12 & 4 (4.3.3) & 2.1.1 & CelleBrite UFED & 1.1.8.6 \\\hline
     Jan '11 & 3GS (3.0) & N/A & Mobilyze & 1.1 \\ \hline
     Dec '10 & 3GS (3.0) & N/A & Zdziarski’s Method & N/A \\ \hline
     Dec '10 & 3GS (3.0) & N/A & iXAM & 1.5.6 \\ \hline
     Nov '10 & 3GS (3.0) & 1.5 & Secure View & 2.1.0 \\\hline
     Nov '10 & 3GS (3.0) & 1.5 & XRY & 5.0.2 \\\hline
    \end{tabular}
    \caption{History of Forensic Tools (2010--2016)}\label{tab:forensics_to16}
    \caption*{Source: \Gls{DHS}~\cite{dhs_forensics}}
\end{table}
\end{tcolorbox}

\printglossaries

\end{document}